\documentclass[aps,pra,showpacs,twocolumn,floatfix,eqsecnum,superscriptaddress]{revtex4-1}

\usepackage{graphicx}
\usepackage{amssymb,amsmath,bbm,braket,indentfirst,bm,setspace,amsthm,theorem,pifont}
\usepackage[colorlinks=true,linktoc=page,linkcolor=blue,citecolor=magenta]{hyperref}
\usepackage[normalem]{ulem}   % to strike things out
\usepackage{cancel,soul}
\usepackage[pdftex]{color}% needed for pdf colors

\allowdisplaybreaks 
\makeatletter
\renewcommand*\env@matrix[1][*\c@MaxMatrixCols c]{%
  \hskip -\arraycolsep
  \let\@ifnextchar\new@ifnextchar
  \array{#1}}
\makeatother

\newcommand{\pri}[1]{\ensuremath #1^{\prime}}

\newcommand{\sgn}{\ensuremath \text{sgn}}

\renewcommand\({\ensuremath \left(}
\renewcommand\){\ensuremath \right)}
\renewcommand\[{\ensuremath \left[}
\renewcommand\]{\ensuremath \right]}
\def\:={\,\raisebox{0.85pt}{.}\hspace{-2.78pt}\raisebox{2.85pt}{.}\!\!=\,}
\def\=:{\,=\!\!\raisebox{0.85pt}{.}\hspace{-2.78pt}\raisebox{2.85pt}{.}\,}

\newcommand{\ua}{\uparrow}
\newcommand{\da}{\downarrow}

\newcommand{\xmark}{\text{\ding{55}}}

\begin{document}
\title{Ground state degeneracy of non-Abelian topological phases\\
       from coupled wires}

\author{Thomas~Iadecola} \affiliation{Physics Department, Boston
University, Boston, Massachusetts 02215, USA}
\affiliation{Kavli Institute for Theoretical Physics, University of California,
Santa Barbara, California 93106, USA }
\affiliation{Joint Quantum Institute and Condensed Matter Theory Center,
Department of Physics, University of Maryland, College Park, Maryland 20742, USA}

\author{Titus~Neupert} \affiliation{Department of Physics,
University of Zurich, Winterthurerstrasse 190, 8057 Zurich, Switzerland}

\author{Claudio~Chamon} \affiliation{Physics Department, Boston
University, Boston, Massachusetts 02215, USA}

\author{Christopher~Mudry} \affiliation{Condensed Matter Theory Group,
Paul Scherrer Institute, CH-5232 Villigen PSI, Switzerland}
  
\date{\today} % Activate to display a given date or no date
\begin{abstract}
We construct a family of two-dimensional non-Abelian topological
phases from coupled wires using a non-Abelian bosonization approach.
We then demonstrate how to determine the nature of the non-Abelian
topological order (in particular, the anyonic excitations and the
topological degeneracy on the torus) realized in the resulting gapped phases of 
matter. This paper focuses on the detailed case study of a coupled-wire
realization of the bosonic $su(2)^{\,}_{2}$ Moore-Read state, but the approach
we outline here can be extended to general bosonic $su(2)^{\,}_{k}$ topological
phases described by non-Abelian Chern-Simons theories.  We also discuss
possible generalizations of this approach to the construction of three-dimensional
non-Abelian topological phases.
\end{abstract}
\maketitle
\vspace{-.5cm}

\tableofcontents

\section{Introduction}
\label{sec: Introduction}

\subsection{Motivation}
\label{subsec: Motivation}

In recent decades, topological order has emerged as a novel paradigm
for describing states of matter.  Motivated by the study of the
fractional quantum Hall effect and chiral spin liquids, theoretical
investigations uncovered a rich landscape of topologically ordered
phases in two spatial dimensions.  The unifying features common to all
phases in this landscape are 1) the degeneracy of the ground state
when the system is defined on a manifold with nonzero
genus~\cite{Wen89b}, and 2) the (intimately related) existence of
fractionalized excitations in the gapped bulk~\cite{Oshikawa06}.
The theoretical understanding of these topologically ordered phases has
been placed on a firm mathematical footing rooted in the apparatus of
modular tensor categories%
~\cite{Friedan87,Froehlich88,Moore89a,Froehlich90a,Etingof2015}.
While numerous problems remain open to investigation, such as the
inclusion of symmetries%
~\cite{Barkeshli14,Teo15,Barkeshli16,Thorngren16} and the
description of topological phases starting from interacting electrons%
~\cite{Gu14,Cheng15,Kapustin15,Gaiotto16,Ware16,Tarantino16,Williamson16},
this mathematical framework provides an indispensable point of
reference in the ongoing effort to understand strongly interacting
topological states of matter in two spatial dimensions (2D).

Despite this progress, the construction of tractable microscopic models
for topological states of matter starting from local spin or electronic degrees
of freedom remains challenging.  Especially challenging are chiral
(i.e., time-reversal-breaking) topological phases, which cannot be represented by
exactly solvable lattice models whose Hamiltonians consist of local commuting
projectors~\cite{Kapustin18} (in contrast to, e.g., Kitaev's toric code and quantum
double models~\cite{Kitaev03}).  There is, however, an approach that allows
for the development of tractable models even in the case of chiral phases:
the coupled-wire construction. In this approach, a 2D state of matter is constructed
by coupling together many one-dimensional (1D) sub-systems with appropriate
many-body interactions.  These 1D subsystems are typically described by gapless
(1+1)-dimensional effective field theories whose underlying microscopic constituents
are electrons, bosons, or spins. The couplings between these subsystems can lead
to fractionalization and other exotic phenomena. The utility of
this approach lies in the fact that numerous analytical techniques exist
for quantum field theories in (1+1)-dimensional spacetime,
enabling the description of a wide variety of strongly interacting states
of matter in a controlled manner. Coupled-wire constructions have been
used to build a variety of strongly correlated phases in 2D, including non-Fermi
liquids~\cite{Emery00,Mukhopadhyay01,Vishwanath01}
as well as Abelian and non-Abelian fractional quantum Hall states and spin liquids%
~\cite{Poilblanc87,Yakovenko91,Lee94,Sondhi01,Kane02,Teo14,Mong14,Neupert14,
Meng15a,Gorohovsky15,Huang16a,Huang16b,Chen17}.

The subject of this paper is the construction and characterization of non-Abelian
topological phases within the coupled-wire approach.  Previous studies of coupled-wire
constructions of non-Abelian topological phases have inferred the non-Abelian
nature of the topological order from the structure of the edge states (e.g., their chiral
central charge) when the system is studied in a cylindrical geometry (see, e.g.,
Refs.~\onlinecite{Teo14,Meng15a,Huang16a,Chen17}).
However, knowledge of the edge theory alone is insufficient to fully determine the
nature of the topological order in the bulk.  For example, a chiral $su(2)^{\,}_{2}$
topological phase has edge states with central charge $3/2$ that can be described
by three independent chiral Majorana modes, but so does a stack of three decoupled
copies of a noninteracting $p_x+\mathrm i p_y$ superconductor. The former topological
phase has a threefold topological ground state degeneracy on the torus, while the latter
does not. Thus, in order to verify the assumed correspondence between the gapless edge theory
the bulk topological order in these models, it is necessary to study independently the bulk
topological order itself.

In this work we construct a family of $su(2)^{\,}_{k}$ topological phases using a
coupled-wire approach based on non-Abelian bosonization~\cite{Witten84,Huang16a}.
This family of topological phases is putatively described at low energies by the
family of $SU(2)$ Chern-Simons theories at level $k$~\cite{Witten89,Cabra00}.  We aim
to make this connection more concrete by demonstrating how to calculate the
topological degeneracy on the torus of the coupled-wire construction so that
it can be compared with the value $k+1$ expected from the Chern-Simons theory.
Focusing on the $su(2)^{\,}_{2}$ case (which in quantum Hall terminology is known
as the bosonic Moore-Read state), we show in detail how to do this within
the coupled-wire setup and verify that the ground state of this model on the torus
is indeed threefold degenerate.  Our discussion and calculations deal at length
with subtleties encountered elsewhere~\cite{Oshikawa07} in the study of non-Abelian
topological phases, but has the benefit that the coupled-wire construction allows
one to use explicit expressions for the operators that are used to compute the degeneracy.
For a more detailed summary of our results, see Sec.~\ref{subsec: Outline and summary of results}.

Although we study 2D topological phases in this work, another motivation
for the present study is the possibility of using coupled-wire constructions
to study topological phases in three dimensions (3D).
The theoretical proposal~\cite{Fu07,Moore07} and experimental
discovery~\cite{Hsieh08,Chen09,Hsieh09,Hsieh09b} of three-dimensional
topological insulators (TIs) protected by time-reversal symmetry (TRS)
underscores the natural question of what types of topological phases are
possible in 3D, and whether these phases can be classified in a manner
analogous to what has been achieved for 2D topological phases.
Numerous examples of topologically ordered phases in three spatial
dimensions have been studied theoretically.
One example of such phases are so-called fractional TIs (FTIs),
which are defined as gapped 3D phases with TRS whose bulk axion
electromagnetic response is characterized by axion angles
$\theta$ that are rational multiples of $\pi$. Consistency with TRS
then demands the presence of topological order in the bulk%
~\cite{Maciejko10,Swingle11}.
Other more elementary examples include discrete gauge theories and their
twisted counterparts~\cite{Dijkgraaf90,Wang15,Wan15,Else17}.  There also exists
a procedure, the Crane-Yetter/Walker-Wang
construction~\cite{Crane93,Walker12,Wang16,Williamson17}, that can be
used to build certain 3D topological phases.
Despite this progress, the question of what kinds of strongly
interacting topological phases can exist in 3D
is far from settled. This is especially true of non-Abelian topological
orders.

The coupled-wire approach has recently been generalized to
3D, yielding a variety of phases including Weyl
semimetals~\cite{Vazifeh13,Meng15b},
fractional topological insulators~\cite{Sagi15b},
and strongly-correlated phases described by
emergent Abelian gauge theories~\cite{Iadecola16,Fuji19}.
The goal of extending this approach to construct and characterize
new non-Abelian phases in 3D is thus a natural one.  The results
of this paper can be used as a starting point for these investigations.
In Sec.~\ref{sec: Challenges for extensions to three-dimensional space}
we provide an overview of some challenges to overcome in the
extension of the non-Abelian coupled-wire approach to 3D.
Given the paucity of tractable microscopic spin- and/or electron-based
models for non-Abelian topological phases in 3D, we believe that the
coupled-wire approach will be a valuable tool to search for and characterize
candidates for new 3D topological phases of matter.

\subsection{Outline and summary of results}
\label{subsec: Outline and summary of results}

We now provide an overview of the organization of the paper and
summarize the results.

In Sec.~\ref{sec: Non-Abelian bosonization of a single wire}, we
review how to bosonize a multi-flavor fermionic wire in terms of the
currents associated with the non-Abelian internal symmetry group of
the wire~\cite{Witten84}.  This bosonization scheme has been used to
address a wide variety of physical problems in 1D,
including the multichannel Kondo 
effect~\cite{Affleck90,Affleck91a,Affleck91b} and marginally-perturbed
conformal field theories (CFTs)~\cite{Tsvelik14}.
In Ref.~\onlinecite{Huang16a} it was also used as a starting point for the
construction of a series of non-Abelian topological phases in
2D. In Sec.\ \ref{subsec: Intrawire interactions}, we show
how to add intrawire interactions to drive the fermionic wire to a
strong-coupling fixed point described by an $su(2)^{\,}_{k}$ CFT.
This treatment is crucial for what follows, as these CFTs are used as
building blocks for the coupled-wire constructions of the subsequent
sections; the non-Abelian topologically ordered phases that we construct
later in the paper inherit their non-Abelian character from the $su(2)^{\,}_{k}$ CFTs.

In Sec.%
~\ref{sec: Warm-up: Non-Abelian topological order in two dimensions},
we describe how to construct non-Abelian topological
phases of matter in 2D
starting from a one-dimensional array of decoupled
$su(2)^{\,}_{k}$ CFTs and using
current-current interactions to couple channels in neighboring wires
that have opposite chirality.  These couplings can be viewed
as arising from continuum limits of microscopic
interactions between the spin sectors of neighboring
wires (see, e.g., Refs.~\onlinecite{Huang16b} and \onlinecite{Chen17}),
and they are marginally relevant under the renormalization
group (RG). The flow to a strong-coupling fixed point is associated to the
opening of a gap in the bulk of the array of coupled wires, while leaving
chiral $su(2)^{\,}_{k}$ modes on the boundaries when the model is defined on a
cylinder~\cite{Huang16a}.

Once we have shown how to gap the bulk of the array, in
Sec.~\ref{subsec: Case study: su(2)_{2}}
we focus on the specific example of $su(2)^{\,}_{2}$
(which is related to the Moore-Read state for bosons at filling factor $\nu=1$),
and show how to characterize the
bulk topological order within the coupled-wire construction.
The procedure for doing so
hinges on using the primary operators of the unperturbed CFTs in each
wire to construct nonlocal ``string operators''
that commute with the
interaction term and satisfy a nontrivial algebra among themselves.
These string operators can then be used to determine the topological
ground-state degeneracy of the coupled-wire theory on the torus.  More
specifically, these string operators can be used to
construct a representation of the ground-state manifold of the
coupled-wire theory at strong coupling.

In particular, in Sec.~\ref{subsec: String operators and topological degeneracy}
we show that the algebra of these string operators suggests the algebra of Wilson loops in a
$\mathbb{Z}^{\,}_{2}$ gauge theory. Namely, there are four nonlocal string
operators that break into two sets of anticommuting operators.
Naive intuition derived from Abelian gauge theory then suggests that the ground-state
degeneracy on the torus should be fourfold.  However, one finds that
one of these four putative ground states cannot reside in the
ground-state manifold.  The reason for this has deep connections to
the non-Abelian algebra of primary operators in the
CFT~\cite{Moore89a}, and has come up before in less microscopic
studies of related topological phases~\cite{Oshikawa07}.  In this way,
we conclude that the topological degeneracy of the $su(2)^{\,}_{2}$
topological phase in 2D is three, rather than four.
This exclusion of states from the ground-state manifold based on
non-Abelian operator algebras is at the heart of what distinguishes
non-Abelian topological phases from Abelian ones and serves as
a useful operational criterion indicating when a topological
phase constructed from coupled wires is non-Abelian.
We expect that the techniques of
Sec.~\ref{subsec: String operators and topological degeneracy}
can be extended to the other $su(2)^{\,}_{k}$ phases defined in
in Sec.~\ref{subsec: Definition of the class of models 2D}
and used to show that these phases possess
a topological degeneracy on the torus of $k+1$, in agreement
with the value obtained within non-Abelian Chern Simons
theory~\cite{Witten89,Cabra00}.

In Sec.~\ref{sec: Challenges for extensions to three-dimensional space}
we provide an overview of prospects for generalizing the construction
presented in this paper to 3D.  We identify challenges that make such
a generalization a delicate matter, and we propose several possible ways
of overcoming these challenges.  We believe that these observations
will help to define a path forward for the use of coupled-wire constructions
in the construction of new non-Abelian phases of matter in 3D.

\section{Non-Abelian bosonization of a single wire}
\label{sec: Non-Abelian bosonization of a single wire}

\subsection{Free-fermion wire}
\label{subsec: Free-fermion wire}

Consider a one-dimensional wire containing $N^{\,}_{\mathrm{c}}$ 
``colors'' (orbitals) of spinful fermions.  
Its action $S^{\,}_{0,\mathrm{wire}}$
is the integral over time $t$ and the coordinate $z$ along the
wire of the Lagrangian density
\begin{equation}\label{eq: single non-abelian wire}
\begin{split}
\mathcal{L}^{\,}_{0,\mathrm{wire}}\:=
2
\sum_{\sigma=\ua,\da}
\sum^{N^{\,}_{\mathrm{c}}}_{\alpha=1}
\Big(&
\chi^{*}_{\mathrm{L},\sigma,\alpha}\,
\mathrm{i}\partial^{\,}_{\mathrm{L}}
\chi^{\,}_{\mathrm{L},\sigma,\alpha}
\\
&
+ 
\chi^{*}_{\mathrm{R},\sigma,\alpha}\,
\mathrm{i}\partial^{\,}_{\mathrm{R}}
\chi^{\,}_{\mathrm{R},\sigma,\alpha} 
\Big).
\end{split}
\end{equation}
The derivatives $\partial^{\,}_{\mathrm{M}}\equiv\partial^{\,}_{z^{\,}_{\mathrm{M}}}$
($\mathrm{M}=\mathrm{L},\mathrm{R}$) are taken with respect to the chiral
(light-cone) coordinates
\begin{equation}
z^{\,}_{\mathrm{L}}\equiv t+z,
\qquad
z^{\,}_{\mathrm{R}}\equiv t-z.
\label{eq: def z of M}
\end{equation}
We assume periodic boundary conditions along the wire, i.e.,
in the $z$-direction. The four Grassmann-valued fields 
$\chi^{*}_{\mathrm{L},\sigma,\alpha}$,
$\chi^{\,}_{\mathrm{L},\sigma,\alpha}$,
$\chi^{*}_{\mathrm{R},\sigma,\alpha}$,
$\chi^{\,}_{\mathrm{R},\sigma,\alpha}$
are independent of each other.

Such a wire has the internal symmetry
$U(2N^{\,}_{\mathrm{c}})^{\,}_{\mathrm{L}}\times U(2N^{\,}_{\mathrm{c}})^{\,}_{\mathrm{R}}$.
The central idea of the coupled-wire constructions presented in this
paper is to decompose the Lie algebra associated with this symmetry using
the following identity (or ``conformal embedding'')~\cite{DiFrancesco97},
\begin{equation}
u(2N^{\,}_{\mathrm{c}})^{\,}_{1}=
u(1)\oplus
su(2)^{\,}_{N^{\,}_{\mathrm{c}}}
\oplus
su(N^{\,}_{\mathrm{c}})^{\,}_{2},
\label{eq: conformal embedding}
\end{equation}
where we have employed the notation $g^{\,}_{k}$ for the affine Lie algebra
at level $k$ associated with the 
connected, compact, and simple Lie group $G$.
(For a review of affine Lie algebras, see, e.g., Ref.~\onlinecite{DiFrancesco97}.)  
Equation \eqref{eq: conformal embedding} 
tells us that the theory~\eqref{eq: single non-abelian wire}
has three conserved currents 
$j^{\,}_{\mathrm{R}}$,
$J^{a}_{\mathrm{R}}$,
and $\mathsf{J}^{\mathsf{a}}_{\mathrm{R}}$
corresponding to the affine Lie algebras
$u(1)$,
$su(2)^{\,}_{N^{\,}_{\mathrm{c}}}$,
and $su(N^{\,}_{\mathrm{c}})^{\,}_{2}$, respectively.  
(Note that, of course, there are analogous conserved currents 
$j^{\,}_{\mathrm{L}},J^{a}_{\mathrm{L}},$ and $\mathsf{J}^{\mathsf{a}}_{\mathrm{L}}$ 
for the left-handed sector.) We use indices
$a=1,2,3$ to label the generators of $SU(2)$ and $\mathsf{a} =
1,\cdots,N^{2}_{\mathrm{c}}-1$ to label the generators of
$SU(N^{\,}_{\mathrm{c}})$.  In terms of the complex fermions, these
currents are given by
\begin{subequations}
\label{eq: non-Abelian currents}
\begin{align}
& 
j^{\,}_{\mathrm{M}}\:=
\sum_{\sigma=\ua,\da}
\sum^{N^{\,}_{\mathrm{c}}}_{\alpha=1}
\chi^{*}_{\mathrm{M},\sigma,\alpha}\,\chi^{\,}_{\mathrm{M},\sigma,\alpha},
\label{eq: non-Abelian currents a}
\\
&
J^{a}_{\mathrm{M}}\:=
\frac{1}{2}
\sum_{\sigma,\sigma^{\prime}=\ua,\da}
\sum^{N^{\,}_{\mathrm{c}}}_{\alpha=1}
\chi^{*}_{\mathrm{M},\sigma,\alpha}\,
\sigma^{a}_{\sigma\sigma^{\prime}}\, 
\chi^{\,}_{\mathrm{M},\sigma^{\prime},\alpha},
\label{eq: non-Abelian currents b}
\\
&
\mathsf{J}^{\mathsf{a}}_{\mathrm{M}}\:=
\sum_{\sigma=\ua,\da}
\sum^{N^{\,}_{\mathrm{c}}}_{\alpha,\alpha^{\prime}=1}
\chi^{*}_{\mathrm{M},\sigma,\alpha}\,
T^{\mathsf{a}}_{\alpha\alpha^{\prime}}\, 
\chi^{\,}_{\mathrm{M},\sigma,\alpha^{\prime}},
\label{eq: non-Abelian currents c}
\end{align}
\end{subequations}
with $\mathrm{M}=\mathrm{L},\mathrm{R}$.
The $U(1)$ currents $j^{\,}_{\mathrm{M}}$
with $\mathrm{M}=\mathrm{L},\mathrm{R}$
are associated with charge conservation. The $SU(2)$ currents
$J^{a}_{\mathrm{M}}$
with $\mathrm{M}=\mathrm{L},\mathrm{R}$ and $a=1,2,3$
are associated with the spin-rotation symmetry.
The $SU(N^{\,}_{\mathrm{c}})$ currents $\mathsf{J}^{\mathsf{a}}_{\mathrm{M}}$
with $\mathrm{M}=\mathrm{L},\mathrm{R}$
and $\mathsf{a}=1,\cdots,N^{2}_{\mathrm{c}}-1$
are associated with the color isospin-rotation symmetry.  The generators
$\sigma^{a}/2$ of $SU(2)$ and $T^{\mathsf{a}}$ of $SU(N^{\,}_{\mathrm{c}})$ obey
the normalizations and the independent algebras
\begin{subequations}
\begin{align}
\mathrm{tr}\,\left(\sigma^{a}\,\sigma^{b}\right)=
2\delta^{ab},
\qquad
[\sigma^{a},\sigma^{b}]=
2\mathrm{i}\,\epsilon^{abc}\,\sigma^{c},
\\
\mathrm{tr}\,\left(T^{\mathsf{a}}\,T^{\mathsf{b}}\right)=
\frac{1}{2}\delta^{\mathsf{ab}},
\qquad
[T^{\mathsf{a}},T^{\mathsf{b}}]=
\mathrm{i}\,f^{\mathsf{abc}}\,T^{\mathsf{c}},
\end{align}
\end{subequations}
where $\epsilon^{\,}_{abc}$ is the Levi-Civita symbol and
$f^{\,}_{\mathsf{abc}}$ are the structure constants of
$SU(N^{\,}_{\mathrm{c}})$.  With these definitions, one can build the
energy-momentum tensor for the free theory defined by 
the Lagrangian density
\eqref{eq: single non-abelian wire} 
using the Sugawara construction%
~\cite{Sugawara68,Affleck90,Affleck91a,Affleck91b} 
for the energy-momentum tensor
in the M-moving sector,
\begin{subequations}
\label{eq: fermionic conformal embedding energy momentum tensor}
\begin{align} 
T^{\,}_{\mathrm{M}}[u(2N^{\,}_{\mathrm{c}})^{\,}_{1}]= 
T^{\,}_{\mathrm{M}}[u(1)] 
+ 
T^{\,}_{\mathrm{M}}[su(2)^{\,}_{N^{\,}_{\mathrm{c}}}] 
+ 
T^{\,}_{\mathrm{M}}[su(N^{\,}_{\mathrm{c}})^{\,}_{2}].
\end{align}
Here,
\begin{align}
&
T^{\,}_{\mathrm{M}}[u(2N^{\,}_{\mathrm{c}})^{\,}_{1}]\:= 
\frac{1}{\pi}
\sum_{\sigma=\ua,\da}
\sum^{N^{\,}_{\mathrm{c}}}_{\alpha=1}
\chi^{*}_{\mathrm{M},\sigma,\alpha}\, 
\mathrm{i}\partial^{\,}_{\mathrm{M}}\chi^{\,}_{\mathrm{M},\sigma,\alpha},
\\
&
T^{\,}_{\mathrm{M}}[u(1)]\:= 
\frac{1}{4N^{\,}_{\mathrm{c}}}\,  
j^{\,}_{\mathrm{M}}\,j^{\,}_{\mathrm{M}},
\\
& 
T^{\,}_{\mathrm{M}}[su(2)^{\,}_{N^{\,}_{\mathrm{c}}}]\:= 
\frac{1}{N^{\,}_{\mathrm{c}}+2}
\sum^{3}_{a=1} 
J^{a}_{\mathrm{M}}\,J^{a}_{\mathrm{M}},
\\
&
T^{\,}_{\mathrm{M}}[su(N^{\,}_{\mathrm{c}})^{\,}_{2}]\:= 
\frac{1}{2+N^{\,}_{\mathrm{c}}}
\sum^{N^{2}_{\mathrm{c}}-1}_{\mathsf{a}=1} 
\mathsf{J}^{\mathsf{a}}_{\mathrm{M}}\,\mathsf{J}^{\mathsf{a}}_{\mathrm{M}}.
\end{align}
\end{subequations}
With these definitions, it follows that the Hamiltonian density associated with
the free Lagrangian density~\eqref{eq: single non-abelian wire} is given by
\begin{equation}
\mathcal{H}^{\,}_{0,\mathrm{wire}}
\!\!\:=\!\!
2\pi
\!\!\!\!\!
\sum_{\mathrm{M}=\mathrm{L},\mathrm{R}}
\!\!\!\!
\left(
T^{\,}_{\mathrm{M}}[u(1)]
\!+\!
T^{\,}_{\mathrm{M}}[su(2)^{\,}_{N^{\,}_{\mathrm{c}}}]
\!+\!
T^{\,}_{\mathrm{M}}[su(N^{\,}_{\mathrm{c}})^{\,}_{2}]
\right)\!.
\end{equation}
Rewriting the free theory~\eqref{eq: single non-abelian wire}
in terms of the currents~\eqref{eq: non-Abelian currents}
amounts to performing a non-Abelian bosonization of the
free theory.  This rewriting highlights the fact that a theory of
multiple flavors of free fermions can be broken up into independent
charge [$u(1)$], spin [$su(2)^{\,}_{N^{\,}_{\mathrm{c}}}$],
and color (orbital)
[$su(N^{\,}_{\mathrm{c}})^{\,}_{2}$] sectors.

\subsection{Intrawire interactions}
\label{subsec: Intrawire interactions}

Having rewritten the free theory~\eqref{eq: single non-abelian wire}
in terms of the non-Abelian currents~\eqref{eq: non-Abelian currents},
we now wish to isolate the $su(2)^{\,}_{N^{\,}_{\mathrm{c}}}$ spin degrees of freedom
by removing the $u(1)$ charge and $su(N^{\,}_{\mathrm{c}})^{\,}_{2}$
color (orbital) degrees
of freedom from the low-energy sector of the theory. We accomplish this
by adding interactions that gap out the latter pair of degrees of freedom.

To gap out the charge sector, we add to the free 
Lagrangian density~\eqref{eq: single non-abelian wire} 
the interaction term
\begin{subequations}\label{eq: umklapp}
\begin{align}\label{eq: umklapp a}
\mathcal{L}^{\,}_{\mathrm{int}}[u(1)]\:=
-
\lambda^{\,}_{u(1)}
\cos\left(\sqrt{2}\,(\phi^{\,}_{\mathrm{R}}+\phi^{\,}_{\mathrm{L}})\right).
\end{align}
The chiral bosonic fields $\phi^{\,}_{\mathrm{M}}$ are defined by the Abelian
bosonization identity
\begin{align}\label{eq: umklapp b}
j^{\,}_{\mathrm{M}}=
-
\frac{1}{\sqrt{2}\,\pi}\,
\partial^{\,}_{\mathrm{M}}\phi^{\,}_{\mathrm{M}}.
\end{align}
\end{subequations}
In the fermionic language, the interaction~\eqref{eq: umklapp a} is interpreted
as an Umklapp process. It is marginally relevant in the renormalization
group (RG) sense, i.e., it flows to strong coupling under RG
and gaps the charge sector when $\lambda^{\,}_{u(1)}>0$.

To gap out the color (orbital) sector, we add to the free 
Lagrangian density~\eqref{eq: single non-abelian wire} 
the interaction term
\begin{align}\label{eq: su(N_c) current-current}
\mathcal{L}^{\,}_{\mathrm{int}}[su(N^{\,}_{\mathrm{c}})^{\,}_{2}]\:=
-
\lambda^{\,}_{su(N^{\,}_{\mathrm{c}})^{\,}_{2}}
\sum^{N^{2}_{\mathrm{c}}-1}_{\mathsf{a}=1}
\mathsf{J}^{\mathsf{a}}_{\mathrm{L}}\,
\mathsf{J}^{\mathsf{a}}_{\mathrm{R}},
\end{align}
where the currents $\mathsf{J}^{\mathsf{a}}_{\mathrm{M}}$
are defined in Eqs.~\eqref{eq: non-Abelian currents}.  This
current-current interaction is also marginally relevant,
flowing to strong coupling for $\lambda^{\,}_{su(N^{\,}_{\mathrm{c}})^{\,}_{2}}>0$.

At the strong-coupling fixed point dominated by the interactions
\eqref{eq: umklapp} and \eqref{eq: su(N_c) current-current}, the
effective Hamiltonian density for the low-energy sector becomes
\begin{align}
\mathcal{H}^{\,}_{0,\mathrm{eff}}
\:=
2\pi
\left(
T^{\,}_{\mathrm{L}}[su(2)^{\,}_{N^{\,}_{\mathrm{c}}}]
+
T^{\,}_{\mathrm{R}}[su(2)^{\,}_{N^{\,}_{\mathrm{c}}}]
\right).
\end{align}
This is nothing but the Hamiltonian description of the 
$su(2)^{\,}_{N^{\,}_{\mathrm{c}}}$
Wess-Zumino-Witten (WZW) CFT
\cite{Wess71,Witten84}
with the central charge
\begin{align}
c[su(2)^{\,}_{N^{\,}_{\mathrm{c}}}]=
\frac{3\,N^{\,}_{\mathrm{c}}}{2+N^{\,}_{\mathrm{c}}}.
\end{align}
Thus, by adding the interactions \eqref{eq: umklapp} and
\eqref{eq: su(N_c) current-current} to the free theory 
\eqref{eq: single non-abelian wire}, we can convert a quantum
wire containing $N^{\,}_{\mathrm{c}}$ colors (orbitals) of spinful fermions
into a highly nontrivial CFT.
The coupled-wire constructions presented in this paper
use arrays of these $su(2)^{\,}_{N^{\,}_{\mathrm{c}}}$ WZW theories
as building blocks for non-Abelian topological phases.

\section{Non-Abelian topological order in two dimensions}
\label{sec: Warm-up: Non-Abelian topological order in two dimensions}

In this section
we construct a class of $su(2)^{\,}_{k}$ topological quantum
liquids in two spatial dimensions and show,
for the case of $k=2$, how to compute their
topological degeneracy on the torus.
This analysis yields new insights for the description of
non-Abelian topological phases with coupled wires.

\subsection{Definition of the class of models} 
\label{subsec: Definition of the class of models 2D}

\begin{figure}[tb]
\includegraphics[width=0.49\textwidth]{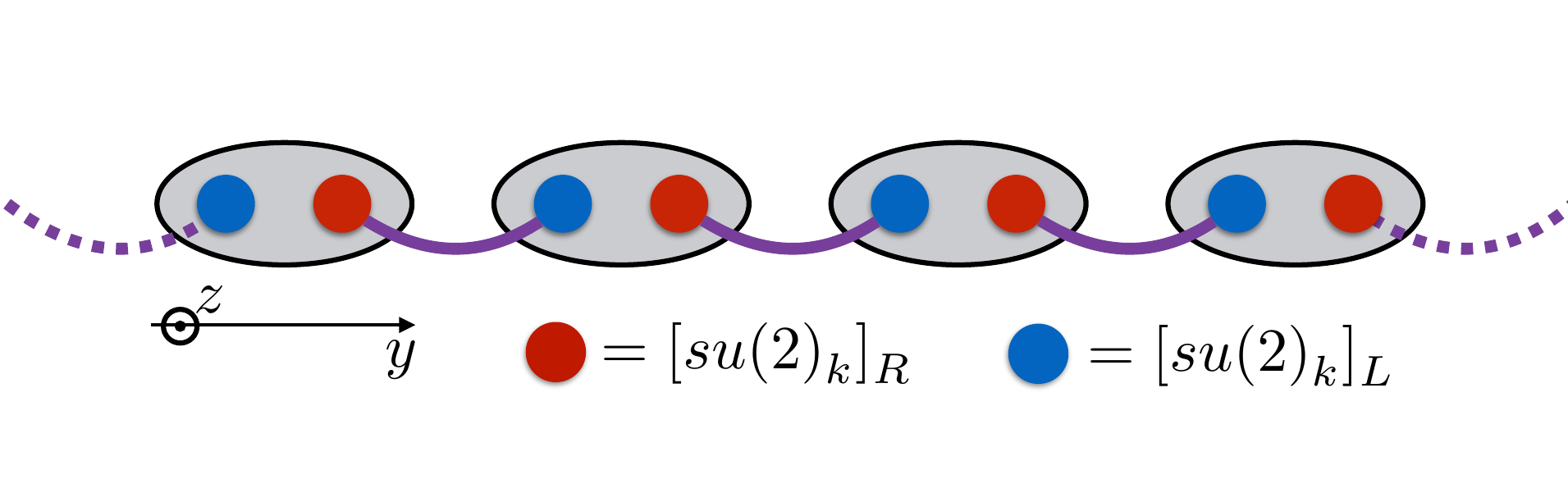}
\caption{
(Color online)
Schematic of the coupled-wire construction for $su(2)^{\,}_{k}$ non-Abelian
topological orders in two spatial dimensions.  Grey ovals represent quantum
wires, while red and blue circles represent chiral $su(2)^{\,}_{k}$ currents.
        }
\label{fig: 2D system}
\end{figure}

We begin with a one-dimensional array $\Lambda$
of parallel nonchiral spinful fermionic
quantum wires aligned along the $z$-direction, each of which is 
described by the Lagrangian density
\eqref{eq: single non-abelian wire} (see Fig.\ \ref{fig: 2D system}).
The cardinality of the one-dimensional lattice $\Lambda$ is
\begin{equation}
|\Lambda|\equiv
L^{\,}_{y}+1.
\end{equation}
We set $N^{\,}_{\mathrm{c}}=k$, where $N^{\,}_{\mathrm{c}}$ is the number of
``colors'' (orbitals) of fermions in each wire.
Each wire has an internal
symmetry $U(2k)^{\,}_{\mathrm{L}}\times U(2k)^{\,}_{\mathrm{R}}$, with
respect to which we carry out the bosonization procedure of 
Sec.~\ref{sec: Non-Abelian bosonization of a single wire}.
We then gap the $u(1)$ and $su(k)^{\,}_{2}$ sectors with the
intrawire interactions discussed in Sec.~\ref{subsec: Free-fermion wire},
leaving behind an $su(2)^{\,}_{k}$ current algebra for
each of the left- and right-moving chiral sectors in every wire.
In the Heisenberg picture and in two-dimensional  Minkowski space,
we denote the chiral $su(2)^{\,}_{k}$ currents by
$\widehat{J}^{a}_{\mathrm{M},y}(z^{\,}_{\mathrm{M}})$
where $\mathrm{M}=\mathrm{L},\mathrm{R}$ labels the chirality, $a=1,2,3$ labels
the $SU(2)$ generators, $y$ labels the wire,
and $z^{\,}_{\mathrm{M}}$ is defined in Eq.~\eqref{eq: def z of M}.

We couple nearest-neighbor wires with the
$su(2)^{\,}_{k}$
interaction (see Fig.~\ref{fig: 2D system})
\begin{subequations}
\label{eq2Dcase: 2D interaction}
\begin{align}
\widehat{\mathcal{L}}^{\,}_{\mathrm{bs}}\equiv
-
\widehat{\mathcal{H}}^{\,}_{\mathrm{bs}}
\:=
-
\frac{\lambda}{2}
\sum_{y=0}^{L^{\,}_{y}-\sigma^{\,}_{\mathrm{BC}}}
\left(
\widehat{J}^{+}_{\mathrm{L},y+1}\,
\widehat{J}^{-}_{\mathrm{R},y}
+
\mathrm{H.c.}
\right),
\label{eq2Dcase: 2D interaction a}
\end{align}
where $\sigma^{\,}_{\mathrm{BC}}=0,1$
for periodic and open boundary conditions,
respectively.
In Eq.~\eqref{eq2Dcase: 2D interaction}, we have introduced
the linear combinations
\begin{align}
\widehat{J}^{\pm}_{\mathrm{M},y}\:=
\widehat{J}^{1}_{\mathrm{M},y}\pm\mathrm{i}\,\widehat{J}^{2}_{\mathrm{M},y}.
\label{eq2Dcase: 2D interaction b}
\end{align}
\end{subequations}
When
periodic boundary conditions are imposed in the $y$-direction,
i.e., when $\sigma^{\,}_{\mathrm{BC}}=0$,
each chiral current is paired with exactly one current of the opposite
chirality in a neighboring wire, and, hence,
the full array of quantum wires
may become gapped in the strong-coupling limit $|\lambda|\gg0$.
Indeed, similar interactions were used in Ref.~\onlinecite{Huang16a}
to construct a large class of topological phases, including the class
of $su(2)^{\,}_{k}$ phases discussed here.  These interactions are marginally
relevant under RG, and their flow to strong coupling is associated with the
opening of a bulk gap in the array of coupled wires when
$\sigma^{\,}_{\mathrm{BC}}=0$. When open boundary conditions are
imposed in the $y$-direction, i.e., when $\sigma^{\,}_{\mathrm{BC}}=1$,
there is a left-moving $su(2)^{\,}_{k}$ current at
$y=0$ and a right-moving $su(2)^{\,}_{k}$ current at $y=L^{\,}_{y}$ that are
fully decoupled from the bulk.  This edge structure is reminiscent of
that of the $su(2)^{\,}_{k}$ non-Abelian Chern-Simons 
theories~\cite{Witten89,Cabra00} and that of the
$\mathbb{Z}^{\,}_{k}$ Read-Rezayi quantum Hall states~\cite{Read99}.

\subsection{Parafermion representation of the interwire interactions}
\label{subsec: Parafermion representation of the interwire interactions 2D}

The interaction \eqref{eq2Dcase: 2D interaction}
can be better understood by rewriting
the $su(2)^{\,}_{k}$ currents in terms of auxiliary degrees of freedom.
This rewriting must preserve the $su(2)^{\,}_{k}$ current algebra,
which is encoded in the operator product expansion (OPE)~\cite{DiFrancesco97}
\begin{align}\label{eq: current-current OPEs}
\widehat{J}^{a}_{\mathrm{L},y}(v)\,\widehat{J}^{\tilde{a}}_{\mathrm{L},\tilde{y}}(w)\sim 
\delta^{\,}_{y,\tilde{y}}\,
\(
\frac{
(k/2)\,\delta^{a\tilde{a}}
     }
     {
v^{2}-w^{2}}
+
\frac{\mathrm{i}\,
\epsilon^{a\tilde{a}b}\,
\widehat{J}^{b}_{\mathrm{L},y}(w)
     }
     {
v-w
     }
\),
\end{align}
for the holomorphic sector $\mathrm{M}=\mathrm{L}$,
and similarly for the antiholomorphic sector $\mathrm{M}=\mathrm{R}$.
(Here, we employ complex coordinates 
$v\equiv t+\mathrm{i}\,z$, obtained from the chiral coordinate
$z^{\,}_{\mathrm{L}}$
defined in Eq.~(\ref{eq: def z of M})
by the analytic continuation $z\to\mathrm{i}\,z$,
and $\bar{v}\equiv t-\mathrm{i}\,z$, obtained from the chiral coordinate
$z^{\,}_{\mathrm{R}}$ also defined in Eq.~(\ref{eq: def z of M})
by the same analytic continuation.)
The group indices $a,\tilde{a}=1,2,3$, and summation over the repeated index
$b=1,2,3$ 
is implied. The symbol $``\sim"$ denotes equality up to nonsingular terms
in the limit $v\to w$.

As shown by Zamolodchikov and Fateev~\cite{Zamolodchikov85}
(see Appendix \ref{appsec: The parafermion current algebra}),
the current algebra \eqref{eq: current-current OPEs}
can be represented in terms of $\mathbb{Z}^{\,}_{k}$
parafermion and chiral boson operators
by [see Eq.\ (5.5) from Ref.~\onlinecite{Zamolodchikov85}]
\begin{subequations}
\label{eq: def chiral parafermion and boson rep su(2)k}
\begin{align}
&
\widehat{J}^{+}_{\mathrm{M},y}\=:
\sqrt{k}\
\widehat{\Psi}^{\,}_{\mathrm{M},y}\, 
\bm{:} 
e^{+\mathrm{i}\sqrt{1/k}\,\widehat{\phi}^{\,}_{\mathrm{M},y}}
\bm{:},
\label{eq: def chiral parafermion and boson rep su(2)k a}
\\
&
\widehat{J}^{-}_{\mathrm{M},y}\=:
\sqrt{k}\
\bm{:}
e^{-\mathrm{i}\sqrt{1/k}\,\widehat{\phi}^{\,}_{\mathrm{M},y}}
\bm{:}\,
\widehat{\Psi}^{\dag}_{\mathrm{M},y},
\label{eq: def chiral parafermion and boson rep su(2)k b}
\\
&
\widehat{J}^{3}_{\mathrm{M},y}\=:
\mathrm{i}\, 
\frac{\sqrt{k}}{2}\,
\partial^{\,}_{\mathrm{M}}\widehat{\phi}^{\,}_{\mathrm{M},y},
\label{eq: def chiral parafermion and boson rep su(2)k c}
\end{align}
\end{subequations}
where $\bm{:}\cdot\bm{:}$ denotes normal ordering with respect to the
many-body ground state of $\widehat{H}^{\,}_{0,\mathrm{eff}}$ within
each wire.
Here, the $\mathbb{Z}^{\,}_{k}$ parafermions
$\widehat{\Psi}^{\,}_{\mathrm{M},y}$ satisfy the equal-time algebra
\begin{subequations}
\label{eq: 2D equal time algebra}
\begin{widetext}
\begin{equation}
\begin{split}
&
\widehat{\Psi}^{\,}_{\mathrm{M},y}(t,z)\, 
\widehat{\Psi}^{\,}_{\mathrm{M}^{\prime},y^{\prime}}(t,z^{\prime})=
\widehat{\Psi}^{\,}_{\mathrm{M}^{\prime},y^{\prime}}(t,z^{\prime})\, 
\widehat{\Psi}^{\,}_{\mathrm{M},y}(t,z)\,
e^{
-\mathrm{i}\,\frac{2\pi}{k}\,\delta^{\,}_{y,y^{\prime}}
\[
(-1)^{\mathrm{M}}\,\delta^{\,}_{\mathrm{M},\mathrm{M}^{\prime}}\,\mathrm{sgn}(z-z^{\prime})
+
\epsilon^{\,}_{\mathrm{M},\mathrm{M}^{\prime}}
\]
+
\mathrm{i}\,\frac{2\pi}{k}\,\mathrm{sgn}(y-y^{\prime})
  },
\end{split}
\label{eq: 2D equal time algebra a}
\end{equation}  
\begin{equation}
\begin{split}
&
\widehat{\Psi}^{\dag}_{\mathrm{M},y}(t,z)\, 
\widehat{\Psi}^{\dag}_{\mathrm{M}^{\prime},y^{\prime}}(t,z^{\prime})=
\widehat{\Psi}^{\dag}_{\mathrm{M}^{\prime},y^{\prime}}(t,z^{\prime})\, 
\widehat{\Psi}^{\dag}_{\mathrm{M},y}(t,z)\,
e^{
-\mathrm{i}\,\frac{2\pi}{k}\,\delta^{\,}_{y,y^{\prime}}
\[
(-1)^{\mathrm{M}}\,\delta^{\,}_{\mathrm{M},\mathrm{M}^{\prime}}\,\mathrm{sgn}(z-z^{\prime})
+
\epsilon^{\,}_{\mathrm{M},\mathrm{M}^{\prime}}
\]
+
\mathrm{i}\,\frac{2\pi}{k}\,\mathrm{sgn}(y-y^{\prime})},
\end{split}
\label{eq: 2D equal time algebra b}
\end{equation}  
\begin{equation}
\begin{split}
&
\widehat{\Psi}^{\,}_{\mathrm{M},y}(t,z)\, 
\widehat{\Psi}^{\dag}_{\mathrm{M}^{\prime},y^{\prime}}(t,z^{\prime})=
\widehat{\Psi}^{\dag}_{\mathrm{M}^{\prime},y^{\prime}}(t,z^{\prime})\, 
\widehat{\Psi}^{\,}_{\mathrm{M},y}(t,z)\,
e^{
+\mathrm{i}\,\frac{2\pi}{k}\,\delta^{\,}_{y,y^{\prime}}
\[
(-1)^{\mathrm{M}}\,\delta^{\,}_{\mathrm{M},\mathrm{M}^{\prime}}\,\mathrm{sgn}(z-z^{\prime})
+
\epsilon^{\,}_{\mathrm{M},\mathrm{M}^{\prime}}
\]
-
\mathrm{i}\,\frac{2\pi}{k}\,\mathrm{sgn}(y-y^{\prime})}.
\end{split}
\label{eq: 2D equal time algebra c}
\end{equation}  
The sign function above is defined such that $\mathrm{sgn}(0)=0$.
The left- and right-moving labels $\mathrm{M}=\mathrm{L},\mathrm{R}$
are defined with
the convention that
$\epsilon^{\,}_{\mathrm{M},\mathrm{M}^{\prime}}$ is the antisymmetric
Levi-Civita symbol obeying
$\epsilon^{\,}_{\mathrm{L},\mathrm{R}}=-\epsilon^{\,}_{\mathrm{R},\mathrm{L}}=1$.
Moreover, $(-1)^{\mathrm{R}}=-(-1)^{\mathrm{L}}\equiv1$.
The algebra of the $su(2)^{\,}_{k}$ currents holds so long as the
equal-time algebra
\begin{align}
\label{eq: 2D equal time algebra d}
\begin{split}
&
\[
\widehat{\phi}^{\,}_{\mathrm{M},y}(t,z),
\widehat{\phi}^{\,}_{\mathrm{M}^{\prime},y^{\prime}}(t,z^{\prime})
\]=
-\mathrm{i}\,
2\pi
\[
(-1)^{\mathrm{M}}\,
\delta^{\,}_{y,y^{\prime}}\,
\delta^{\,}_{\mathrm{M},\mathrm{M}^{\prime}}\, 
\mathrm{sgn}(z-z^{\prime})
+
\delta^{\,}_{y,y^{\prime}}\,
\epsilon^{\,}_{\mathrm{M},\mathrm{M}^{\prime}}
-
\mathrm{sgn}(y-y^{\prime})
\],
\end{split}
\end{align}
\end{widetext}
\end{subequations}
is imposed in the chiral bosonic sector.  In particular, one verifies
that currents defined in different wires commute with one another at
equal times when the definitions
\eqref{eq: def chiral parafermion and boson rep su(2)k} are imposed.
Furthermore, one can show that all
equal-time commutators between $su(2)^{\,}_{k}$ currents differing by
their $\mathrm{L}$ and $\mathrm{R}$ labels also vanish. Finally, the chiral
parafermions commute with the chiral bosons at equal times.

The representation \eqref{eq: def chiral parafermion and boson rep su(2)k}
of the $su(2)^{\,}_{k}$ current algebra provides a convenient interpretation of
the interactions \eqref{eq2Dcase: 2D interaction}
in terms of fractionalized degrees of freedom, as we discuss below.
However, there are several caveats to keep in mind. Chief among these is
the fact that the factorization
\eqref{eq: def chiral parafermion and boson rep su(2)k}--%
\eqref{eq: def chiral parafermion and boson rep su(2)k}
of the $su(2)^{\,}_{k}$ currents re-expresses a set of local operators
(the currents) in terms of products of auxiliary degrees of freedom
(the parafermions and the chiral bosons).
While the $su(2)^{\,}_{k}$ currents admit a local expression 
[Eq.~\eqref{eq: non-Abelian currents b}] in terms of
the original degrees of freedom used to define the theory (the electrons)
these auxiliary degrees of freedom do not.
This fact will be important
when we construct the nonlocal string operators that allow us to
calculate the topological degeneracy in Sec.~\ref{subsec: Case study: su(2)_{2}}.
Furthermore, we note that this parafermion representation
is not unique in two ways.
First, as it factorizes a local (observable) operator
into the product of two operators,
there is an ambiguity with the choice of the phase
assigned to each operator-valued factor.
(This is an explicit manifestation of the nonlocality of
the auxiliary degrees of freedom.)
The choice for this phase cannot have observable consequences.
Second, the dependence on the labels $y\neq y'$
of the equal-time algebra is not unique since
many distinct choices accommodate the fact that any two currents
belonging to two distinct wires $y$ and $y'$ must always commute.
Hence, the dependence
on the labels $y\neq y'$ of the parafermion equal-time algebra
cannot have observable consequences.
We demonstrate that this is true for the case of $su(2)^{\,}_{2}$ in Appendix\
\ref{sec: Independence of string operator algebra on arbitrary phase factors}.

We work with the normalization convention for which   
the operator $\exp(\mathrm{i}\,a\,\widehat{\phi}^{\,}_{\mathrm{M}})$, 
for $a$ any real-valued number,
has
the conformal weights  
$(a^{2},0)$ if $\mathrm{M}=\mathrm{L}$
or $(0,a^{2})$ if $\mathrm{M}=\mathrm{R}$.
With this convention, the chiral vertex operator 
$\exp(\mathrm{i}\,\sqrt{1/k}\,\widehat{\phi}^{\,}_{\mathrm{M}})$, 
which annihilates a chiral Abelian quasiparticle, 
has
the conformal weights  
$(1/k,0)$ if $\mathrm{M}=\mathrm{L}$
or $(0,1/k)$ if $\mathrm{M}=\mathrm{R}$.
In turn, the chiral parafermion operator
$\widehat{\Psi}^{\,}_{\mathrm{M}}$ must have 
the conformal weights  
$(1-1/k,0)$ if $\mathrm{M}=\mathrm{L}$
or $(0,1-1/k)$ if $\mathrm{M}=\mathrm{R}$,
as the current operators have
the conformal weights  
$(1,0)$ if $\mathrm{M}=\mathrm{L}$
or $(0,1)$ if $\mathrm{M}=\mathrm{R}$.
The expressions
\eqref{eq: def chiral parafermion and boson rep su(2)k} 
for the currents are equivalent to the identity~\cite{DiFrancesco97}
\begin{subequations}
\begin{align}
\label{eq: su2 level k decomposition}
su(2)^{\,}_{k}\simeq u(1)^{\,}_{k}\oplus\mathbb{Z}^{\,}_{k},
\end{align}
where
\begin{align}
\mathbb{Z}^{\,}_{k}\equiv \frac{su(2)^{\,}_{k}}{u(1)^{\,}_{k}},
\end{align}
\end{subequations}
which states that an $SU(2)$ WZW theory at level $k$ can be
interpreted as a direct product of a chiral boson and a 
$\mathbb{Z}^{\,}_{k}$ parafermion CFT.

With these definitions, the interactions~\eqref{eq2Dcase: 2D interaction}
take the form
\begin{widetext}
\begin{align}
\widehat{\mathcal L}^{\,}_{\mathrm{bs}}\equiv
-
\widehat{\mathcal{H}}^{\,}_{\mathrm{bs}}&=\,
-\lambda\,
\frac{k}{2}\,
\sum^{L^{\,}_{y}}_{y=0}\,     
\left(
\widehat{\Psi}^{\,}_{\mathrm{L},y}\,
\bm{:}
e^{+\mathrm{i}\sqrt{\frac{1}{k}}\,\widehat{\phi}^{\,}_{\mathrm{L},y}}
\bm{:}
\,
\bm{:}
e^{-\mathrm{i}\sqrt{\frac{1}{k}}\,\widehat{\phi}^{\,}_{\mathrm{R},y+1}}
\bm{:}\,
\widehat{\Psi}^{\dag}_{\mathrm{R},y+1}\,
+
\mathrm{H.c.}
\right),
\label{eq2Dcase: parafermion representation su(2)_k}
\end{align}
\end{widetext}
(We employ periodic boundary conditions for the remainder of this
section.)  Written this way, the current-current
interactions~\eqref{eq2Dcase: 2D interaction} can be reinterpreted as
correlated hoppings of (nonlocal) fractionalized degrees of freedom
between wires.  Indeed, viewing $\widehat{\Psi}^{\dag}_{\mathrm{M},y}$
as the creation operator for a parafermion with chirality $\mathrm{M}$
in wire $y$, and viewing the vertex operator $\bm{:}
e^{-\mathrm{i}\sqrt{\frac{1}{k}}\,\widehat{\phi}^{\,}_{\mathrm{M},y}}
\bm{:}$ as the creation operator for an Abelian quasiparticle, we can
interpret Eq.~\eqref{eq2Dcase: parafermion representation su(2)_k} as
allowing parafermions to hop between wires so long as an Abelian
quasiparticle hops at the same time.  Since the composite of these two
fractionalized excitations is a boson, per Eqs.\
\eqref{eq: def chiral parafermion and boson rep su(2)k},
this correlated hopping process
forbids isolated fractionalized degrees of freedom from hopping
between wires.

When periodic boundary conditions are imposed, the interaction
\eqref{eq2Dcase: parafermion representation su(2)_k} gaps out the
array of wires if the current-current coupling on each bond in the
lattice $\Lambda$ acquires a finite vacuum expectation value.  Such a
scenario is possible in the limit $\lambda\to\infty$.  We will see an
explicit example of this gapping mechanism in the next section.

\subsection{Case study: $su(2)^{\,}_{2}$}
\label{subsec: Case study: su(2)_{2}}

In this section, we work through the example of $k=2$ in detail.
First, we will examine more closely how the interaction
\eqref{eq2Dcase: parafermion representation su(2)_k}
leads to a gapped state of matter.  Next, we
will characterize the topological order in this gapped state of matter
by imposing periodic boundary conditions in the $y$- and
$z$-directions and constructing nonlocal \textit{string operators}
that commute with the interaction
$\widehat{\mathcal{H}}^{\,}_{\mathrm{bs}}$ defined by
Eq.~(\ref{eq2Dcase: 2D interaction}).  These string operators will
label the topologically degenerate ground states in the limit
$\lambda\to\infty$.

The Lagrangian density in this case is (omitting the normal ordering of the
vertex operators)
\begin{widetext} 
\begin{subequations}
\label{eq2Dcase: su(2)_{2} backscattering}
\begin{align}
\widehat{\mathcal{L}}^{\,}_{\mathrm{bs}}\equiv
-
\widehat{\mathcal{H}}^{\,}_{\mathrm{bs}}
&\:=
-\lambda
\sum_{y=0}^{L^{\,}_{y}}
\(
e^{+\mathrm{i}\sqrt{1/2}\,\(\widehat{\phi}^{\,}_{\mathrm{L},y}-\widehat{\phi}^{\,}_{\mathrm{R},y+1}\)}\,
\widehat{\psi}^{\,}_{\mathrm{L},y}\,
\widehat{\psi}^{\,}_{\mathrm{R},y+1}
+
\mathrm{H.c.}
\)
\label{eq2Dcase: su(2)_{2} backscattering a}
\\
&=
-
2\lambda
\sum_{y=0}^{L^{\,}_{y}}
\(
\mathrm{i}\,
\widehat{\psi}^{\,}_{\mathrm{L},y}\,
\widehat{\psi}^{\,}_{\mathrm{R},y+1}
\)
\sin
\(
\sqrt{\frac{1}{2}}
\(
\widehat{\phi}^{\,}_{\mathrm{L},y}
-
\widehat{\phi}^{\,}_{\mathrm{R},y+1}
\)
\),
\label{eq2Dcase: su(2)_{2} backscattering b}
\end{align}
\end{subequations}
\end{widetext}
which should be compared with 
Eq.~\eqref{eq2Dcase: parafermion representation su(2)_k}. 
The chiral operators
\begin{subequations} 
\begin{equation}
\widehat{\psi}^{\,}_{\mathrm{M},y}(t,z)\equiv
\widehat{\Psi}^{\,}_{\mathrm{M},y}(t,z)\equiv
\widehat{\Psi}^{\dag}_{\mathrm{M},y}(t,z)
\end{equation}
with $\mathrm{M}=\mathrm{L},\mathrm{R}$
are Majorana operators
(i.e., $\mathbb{Z}^{\,}_{2}$ parafermions).
Their equal-time exchange algebra is given by 
Eq.~\eqref{eq: 2D equal time algebra a}
with $k=2$.  We also impose the normalization
\begin{equation}\label{eq2Dcase: normalization}
\lim_{z'\to z}
\widehat{\psi}^{\,}_{\mathrm{M},y}(t,z)\,
\widehat{\psi}^{\,}_{\mathrm{M},y}(t,z')\equiv
\lim_{z'\to z}\delta(z-z')\:=\mathcal{N}^{\,}_{\delta},
\end{equation}
\end{subequations}
where $\mathcal{N}^{\,}_{\delta}$ is a constant with
dimension [1/length].
The chiral bosons $\widehat{\phi}^{\,}_{\mathrm{M},y}$
obey the equal-time algebra
\eqref{eq: 2D equal time algebra d},
as before.
Furthermore, the
chiral Majorana operators and the chiral bosons commute at equal
times:
\begin{align}
\[
\widehat{\psi}^{\,}_{\mathrm{M},y}(t,z),
\widehat{\phi}^{\,}_{\mathrm{M}^{\prime},y^{\prime}}(t,z^{\prime})
\]=
0.
\end{align}

The rewriting of the interaction \eqref{eq2Dcase: su(2)_{2} backscattering a}
presented in Eq.~\eqref{eq2Dcase: su(2)_{2} backscattering b} provides
an intuitive illustration of the discussion in 
Sec.~\ref{subsec: Parafermion representation of the interwire interactions 2D}
of how the interaction \eqref{eq2Dcase: parafermion representation su(2)_k}
leads to a gap when periodic boundary conditions are imposed. In this case,
when the bosonic field
$\widehat{\phi}^{\,}_{\mathrm{L},y}-\widehat{\phi}^{\,}_{\mathrm{R},y+1}$
becomes locked to an extremum of the sine potential, a Majorana mass term is
induced for the fermionic degrees of freedom. The simultaneous gapping of the
Majorana modes
and locking of the bosonic fields is consistent due to the
independence of the $u(1)^{\,}_{2}$ and $\mathbb{Z}^{\,}_{2}$ sectors of the
$su(2)^{\,}_{2}$ theory.

\subsubsection{Quasilocal chirality-resolved $\mathbb{Z}^{\,}_{2}$ gauge
symmetry}

Observe that the interaction
\eqref{eq2Dcase: su(2)_{2} backscattering}
is invariant under the M- and $y$-resolved
$\mathbb{Z}^{\,}_{2}$ gauge transformation
\begin{subequations}
\label{eq2Dcase: quasilocal Z_{2} symmetry}
\begin{align}
&
\widehat{\psi}^{\,}_{\mathrm{M},y}(t,z)\mapsto
e^{\mathrm{i}\alpha^{\,}_{\mathrm{M},y}}\,
\widehat{\psi}^{\,}_{\mathrm{M},y}(t,z),
\label{eq2Dcase: quasilocal Z_{2} symmetry a}
\\
&
\widehat{\phi}^{\,}_{\mathrm{M},y}(t,z)\mapsto
\widehat{\phi}^{\,}_{\mathrm{M},y}(t,z)
+
\sqrt{2}\,\alpha^{\,}_{\mathrm{M},y},
\label{eq2Dcase: quasilocal Z_{2} symmetry b}
\end{align}
where the assignments
\begin{equation}
\alpha^{\,}_{\mathrm{M},y}\in\{0,\pi\}
\label{eq2Dcase: quasilocal Z_{2} symmetry c}
\end{equation}
for all chiralities $\mathrm{M}=\mathrm{L},\mathrm{R}$
and all wires $y$ define the map
\begin{equation}
\alpha:
\{\mathrm{M}=\mathrm{L},\mathrm{R}\}
\times
\{y=0,\cdots,L^{\,}_{y}\}\to
\{0,\pi\}.
\label{eq2Dcase: quasilocal Z_{2} symmetry d}
\end{equation}
\end{subequations}
This transformation is implemented by the operator
\begin{equation}
\widehat{\Gamma}^{\,}_{\alpha}(t)\equiv
\prod_{\mathrm{M}=\mathrm{L},\mathrm{R}} 
\prod_{y=0}^{L^{\,}_{y}}
\widehat{\Gamma}^{\,}_{\alpha^{\,}_{\mathrm{M},y}}(t)\:=
\widehat{\mathcal{U}}^{\,}_{\alpha}(t)\,
\widehat{\mathcal{Z}}^{\,}_{\alpha}(t),
\label{eq: def generator M and y resolved Z2 gauge trsf}
\end{equation}
where the operator
\begin{widetext}
\begin{align}
\widehat{\mathcal{U}}^{\,}_{\alpha}(t)\equiv
\prod_{\mathrm{M}=\mathrm{L},\mathrm{R}}\,
\prod_{y=0}^{L^{\,}_{y}}\,
\widehat{\mathcal{U}}^{\,}_{\alpha^{\,}_{\mathrm{M},y}}(t)
\:=
\prod_{\mathrm{M}=\mathrm{L},\mathrm{R}}\,
\prod_{y=0}^{L^{\,}_{y}}\,
\exp
\left(
(-1)^{\mathrm{M}}\,
\frac{
\mathrm{i}\alpha^{\,}_{\mathrm{M},y}
     }
     {
2\pi\sqrt{2}
     }
\int\limits^{L^{\,}_{z}}_{0}\mathrm{d}z\,
\partial^{\,}_{z}\widehat{\phi}^{\,}_{\mathrm{M},y}(t,z)
\right)
\label{eq2Dcase: U_alpha definition}
\end{align}
\end{widetext}
acts only on the chiral boson sector of the theory and
implements the transformation \eqref{eq2Dcase: quasilocal Z_{2} symmetry b},
and where the operator 
\begin{align}
\label{eq2Dcase: Z_alpha def}
\widehat{\mathcal{Z}}^{\,}_{\alpha}(t)
&=
\prod_{\mathrm{M}=\mathrm{L},\mathrm{R}}
\prod_{y=0}^{L^{\,}_{y}}
\widehat{\mathcal{Z}}^{\,}_{\alpha^{\,}_{\mathrm{M},y}}(t)
\end{align}
acts only
on the Ising (i.e., $\mathbbm Z^{\,}_{2}$) sector and implements
the transformation \eqref{eq2Dcase: quasilocal Z_{2} symmetry a}.  The action
of the operator $\widehat{\mathcal{U}}^{\,}_{\alpha}(t)$ on
the chiral bosons follows from the fact that
\begin{equation}
\begin{split}
\widehat{\mathcal{U}}^{\,}_{\alpha^{\,}_{\mathrm{M},y}}(t)\,
\widehat{\phi}^{\,}_{\mathrm{M}',y'}(t,z)\,
\widehat{\mathcal{U}}^{\dag}_{\alpha^{\,}_{\mathrm{M},y}}(t)=&\,
\widehat{\phi}^{\,}_{\mathrm{M}',y'}(t,z)
\\
&\,
+
\sqrt{2}\,
\alpha^{\,}_{\mathrm{M},y}\,
\delta^{\,}_{y,\pri y}\,
\delta^{\,}_{\mathrm{M},\pri M}
\end{split}
\label{eq2Dcase: action widehat mathcal U on phi}
\end{equation}
holds for any pair of chiralities
$\mathrm{M},\mathrm{M}'=\mathrm{L},\mathrm{R}$,
for any pair of wires $y,y'$, and for any
$t$ and $z$ [see Eq.~\eqref{eq: 2D equal time algebra d}].
The action of the operator $\widehat{\mathcal{Z}}^{\,}_{\alpha}(t)$ follows from
the definition of $\widehat{\mathcal{Z}}^{\,}_{\alpha^{\,}_{\mathrm{M},y}}(t)$
in terms of the fermion parity operator in the wire $y$, which is somewhat
involved and will not be presented here.

\subsubsection{$su(2)^{\,}_{2}$ primary fields}
\label{subsec: twist operators}

To construct the excitations of the coupled-wire theory, we will use
the primary operators of the underlying $su(2)^{\,}_{2}$ CFT
defined on each quantum wire in Fig.\
\ref{fig: 2D system}.
Any primary field is labeled by a pair of conformal weights
$(\Delta,\overline{\Delta})$
owing to the underlying Virasoro algebra obeyed by the energy-momentum
tensor. The conformal dimension and spin of this primary field are then
defined to be the linear combinations
$\Delta+\overline{\Delta}$
and
$\Delta-\overline{\Delta}$
of the conformal weights, respectively.
However, the $su(2)^{\,}_{k}$ CFTs have more structure
than the Virasoro algebra alone: any primary field can be chosen to
transform according to an irreducible representation of
the global symmetry
group $SU(2)^{\,}_{\mathrm{L}}\times SU(2)^{\,}_{\mathrm{R}}$.
This means that we can choose the primary fields of the
$su(2)^{\,}_{k}$ CFT to be labeled by the pair of quantum numbers
$(s,m)$ with $m=s,s-1,\dots,-s+1,-s$ and
$s=0,\frac{1}{2},1,\dots,\frac{k}{2}$
delivering the dimension $2s+1$ of an irreducible representation of
$SU(2)$. We shall call the quantum number $s$
the ``spin'' quantum number,
even though the $SU(2)$ symmetry could have originated from
orbital degrees of freedom
instead of electronic spin-1/2 degrees of freedom.
The ``spin'' quantum number $s$
should not be confused with the conformal spin quantum number
$\Delta^{(s)}-\overline{\Delta}^{(s)}$
associated to the Virasoro algebra.
The three primaries of $su(2)_{2}$ are denoted
$I\equiv\widehat{\Phi}^{(0)}(t,z)$,
$\widehat{\Phi}^{(1/2)}_{m,\overline{m}}(t,z)$ with $m,\overline{m}=\pm1/2$,
and
$\widehat{\Phi}^{(1)}_{m,\overline{m}}(t,z)$ with $m,\overline{m}=-1,0,+1$.
They carry the conformal weights
\begin{subequations}
\begin{equation}
\left(
\Delta^{(s)},
\overline{
\Delta^{(s)}
         }
\right)=
\left(
\frac{s(s+1)}{k+2},
\frac{s(s+1)}{k+2}
\right),
\end{equation}
i.e.,
\begin{align}
&
\left(
\Delta^{(0)},
\overline{
\Delta^{(0)}
         }
\right)=
\left(
0,
0
\right),
\\
&
\left(
\Delta^{(1/2)},
\overline{
\Delta^{(1/2)}
         }
\right)=
\left(
\frac{3}{16},
\frac{3}{16}
\right),
\\
&
\left(
\Delta^{(1)},
\overline{
\Delta^{(1)}
         }
\right)=
\left(
\frac{1}{2},
\frac{1}{2}
\right),
\end{align}
\end{subequations}
respectively. As shown by
Zamolodchikov and Fateev (see Eq.~(5.10) in \cite{Zamolodchikov85}),
the primary fields
$\widehat{\Phi}^{(s)}_{m,\overline{m};y}$
with $s=0,1/2,1$ and $m,\overline{m}=-s,-s+1,\cdots,s-1,s$
of the $su(2)^{\,}_{2}$ CFT in any wire $y$
can be represented by
\begin{equation}
\begin{split}
\widehat{\Phi}^{(s)}_{m,\overline{m};y}(t,z)\propto&\,
\widehat{\phi}^{(2s)}_{2m,2\overline{m};y}(t,z)
\\
&\,
\times
\bm{:} 
e^{+\mathrm{i}m           \sqrt{1/k}\,\widehat{\phi}^{\,}_{\mathrm{L},y}(t,z)}\,
e^{-\mathrm{i}\overline{m}\sqrt{1/k}\,\widehat{\phi}^{\,}_{\mathrm{R},y}(t,z)}
\bm{:}
\end{split}
\end{equation}
in the spirit of Eq.~(\ref{eq: def chiral parafermion and boson rep su(2)k}).
The operator $\widehat{\phi}^{(2s)}_{2m,2\overline{m};y}(t,z)$ for $s=0,1/2,1$
is the identity, a continuum analog of the Ising order parameter,
and the identity, respectively.
This means that the primary field
$\widehat{\Phi}^{(1/2)}_{m,\overline{m};y}(t,z)$ with $m,\overline{m}=-1/2,+1/2$
cannot be factorized into a product of holomorphic and antiholomorphic
operators, unlike the primary field
$\widehat{\Phi}^{(1)}_{m,\overline{m};y}(t,z)$ with $m,\overline{m}=-1,0,+1$.

For each $y$, it is convenient to introduce the chiral twist fields
$\widehat{\sigma}^{\,}_{\mathrm{M},y}(t,z)$
with $\mathrm{M}=\mathrm{L},\mathrm{R}$.
They are defined so that they change the periodic boundary conditions
obeyed by the Majorana operator $\widehat{\psi}^{\,}_{\mathrm{M},y}(t,z)$
from periodic to antiperiodic
[see Eqs.\ (\ref{eq2Dcase: Z_{2} psi sigma OPE})].
The chiral twist field
$\widehat{\sigma}^{\,}_{\mathrm{M},y}(t,z)$
has the conformal weight $(1/16,0)$ if $\mathrm{M}=\mathrm{L}$
or $(0,1/16)$ if $\mathrm{M}=\mathrm{R}$.
We then define the auxiliary operator
\begin{subequations}
\begin{align}
\label{eq: spin-1/2 primary}
\widehat{\Upsilon}^{(\frac{1}{2})}_{\mathrm{M},y}(t,z)\:=
\widehat{\sigma}^{\,}_{\mathrm{M},y}(t,z)\, 
\bm{:}
e^{+\mathrm{i}\frac{1}{2\sqrt{2}}\,\widehat{\phi}^{\,}_{\mathrm{M},y}(t,z)}
\bm{:}.
\end{align}
Adding the conformal weights of the chiral twist fields
to those of the vertex operators
$
\bm{:}
e^{+\mathrm{i}\frac{1}{2\sqrt{2}}\,\widehat{\phi}^{\,}_{\mathrm{M},y}}
\bm{:}
$
with $\mathrm{M}=\mathrm{L},\mathrm{R}$
gives the conformal weights
$(3/16,0)$ if $\mathrm{M}=\mathrm{L}$
or $(0,3/16)$ if $\mathrm{M}=\mathrm{R}$
(c.f.\ Appendix~\ref{appsubsec: Gaussian algebra}).
Similarly, we introduce the auxiliary ``spin-$1$'' chiral
operators with the conformal weight $(1/2,0)$ if $\mathrm{M}=\mathrm{L}$
or $(0,1/2)$ if $\mathrm{M}=\mathrm{R}$,~\cite{Zamolodchikov85}
\begin{align}
\label{eq: spin-1 primary}
\widehat{\Upsilon}^{(1)}_{\mathrm{M},y}(t,z)\:= 
\bm{:}
e^{+\mathrm{i}\frac{1}{\sqrt{2}}\,\widehat{\phi}^{\,}_{\mathrm{M},y}(t,z)}
\bm{:}\, .
\end{align}
\end{subequations}
The auxiliary composite chiral operators
$\widehat{\Upsilon}^{(1/2)}_{\mathrm{M},y}(t,z)$
and
$\widehat{\Upsilon}^{(1)}_{\mathrm{M},y}(t,z)$
transform according to the rules
\begin{equation}
\widehat{\Upsilon}^{(1/2)}_{\mathrm{M},y}(t,z)\mapsto
e^{\mathrm{i}\alpha^{\,}_{\mathrm{M},y}/2}\,\widehat{\Upsilon}^{(1/2)}_{\mathrm{M},y}(t,z)  
\end{equation}
and
\begin{equation}
\widehat{\Upsilon}^{(1)}_{\mathrm{M},y}(t,z)\mapsto
e^{\mathrm{i}\alpha^{\,}_{\mathrm{M},y}}\,\widehat{\Upsilon}^{(1)}_{\mathrm{M},y}(t,z)  
\end{equation}
respectively,
under the $y$- and $\mathrm{M}$-resolved $\mathbb{Z}^{\,}_{2}$
gauge transformation
(\ref{eq2Dcase: quasilocal Z_{2} symmetry}).
A such, they are not in the physical sector of the enlarged Hilbert space introduced
by the parton construction of Zamolodchikov and Fateev. However, suitable products thereof will
be gauge invariant.

The pair of operators
\begin{align}
\widehat{\mathcal{O}}^{(1/2)}_{y}(t,z)\propto
\widehat{\Upsilon}^{(1/2)\dag}_{\mathrm{L},y}(t,z)\,
\widehat{\Upsilon}^{(1/2)}_{\mathrm{R},y}(t,z)\sim
\widehat{\Phi}^{(1/2)}_{m,\overline{m};y}(t,z)
\end{align}
and
\begin{align}
\widehat{\mathcal{O}}^{(1)}_{y}(t,z)\propto
\widehat{\Upsilon}^{(1)\dag}_{\mathrm{L},y}(t,z)\,
\widehat{\Upsilon}^{(1)}_{\mathrm{R},y}(t,z)\sim
\widehat{\Phi}^{(1)}_{m,\overline{m};y}(t,z)
\end{align}
will play a fundamental role in the following.
Invariance of
$
\widehat{\mathcal{O}}^{(1/2)}_{y}(t,z)
$
and
$
\widehat{\mathcal{O}}^{(1)}_{y}(t,z)
$
under the $y$- and $\mathrm{M}$-resolved $\mathbb{Z}^{\,}_{2}$
gauge transformation
(\ref{eq2Dcase: quasilocal Z_{2} symmetry})
require that
\begin{equation}
\alpha^{\,}_{\mathrm{M},y}\equiv\alpha^{\,}_{y}\in\{0,\pi\}
\end{equation}
is not $\mathrm{M}$-resolved. We will make this assumption from now on.
Operators
$
\widehat{\mathcal{O}}^{(1/2)}_{y}(t,z)
$
and
$
\widehat{\mathcal{O}}^{(1)}_{y}(t,z)
$
are products of holomorphic and antiholomorphic operators with
the conformal weights $(3/16,3/16)$ and $(1/2,1/2)$, respectively, 
have vanishing conformal spin and, as such, are local~\cite{Knizhnik84}.
For example, if the $su(2)^{\,}_{2}$ CFT describes a quantum spin chain at
criticality, then the operator
$
\widehat{\mathcal{O}}^{(1/2)}_{y}(t,z)\sim
\widehat{\Phi}^{(1/2)}_{m,\overline{m};y}(t,z)
$ 
is related to the continuum limit of the staggered magnetization,
while the operator
$
\widehat{\mathcal{O}}^{(1)}_{y}(t,z)\sim
\widehat{\Phi}^{(1)}_{m,\overline{m};y}(t,z)
$ 
is related to fermion bilinears
that can be constructed from the physical spins~\cite{Tsvelik90}.
We will use these local building blocks to construct the nonlocal
string operators that encode the ground-state degeneracy of the
coupled-wire theory. The relation $\sim$ between
$
\widehat{\mathcal{O}}^{(s)}_{y}(t,z)
$
and
$
\widehat{\Phi}^{(s)}_{m,\overline{m};y}(t,z)
$
means that one can replace the latter (after suitable contraction
of its lower indices) by the former in
correlation functions even though the latter need not factorize
into the product of holomorphic and antiholomorphic
pieces.~\cite{Ardonne10}

\begin{figure}[t]
\includegraphics[width=0.4\textwidth]{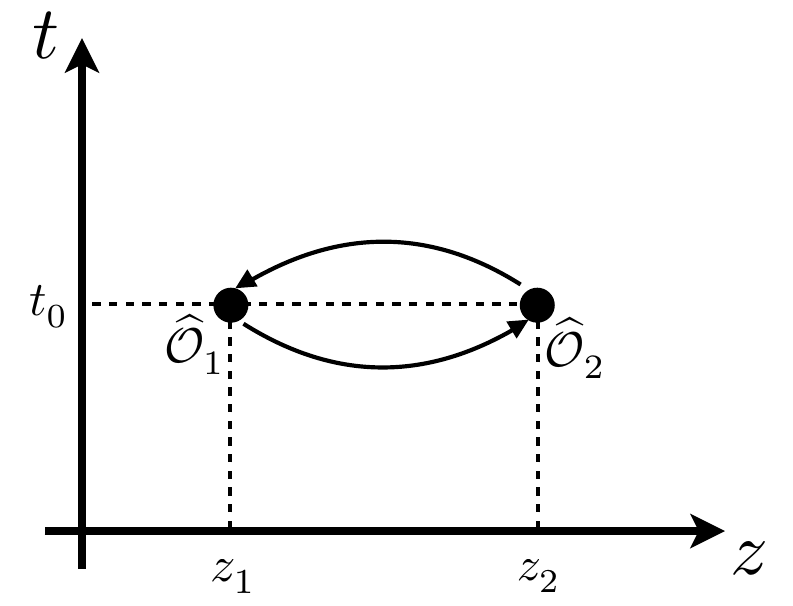}
\caption{
Counterclockwise monodromy of two operators
$\widehat{\mathcal O}^{\,}_{1}(t^{\,}_{0},z^{\,}_{1})$ and 
$\widehat{\mathcal O}^{\,}_{2}(t^{\,}_{0},z^{\,}_{2})$ in the complex plane.
When the operators 
$\widehat{\mathcal O}^{\,}_{1}$ and $\widehat{\mathcal O}^{\,}_{2}$
are evaluated at equal times,
their exchange is related to their monodromy in the complex plane, provided
that the handedness of the monodromy is specified. 
In particular, when $\widehat{\mathcal O}^{\,}_{1}=\widehat{\mathcal O}^{\,}_{2}$,
we adopt the convention that the (holomorphic)
operator with the larger value of $z$ is passed counterclockwise around the
operator with the smaller value of $z$, resulting in the factors of
$\sgn(z-z^{\prime})$
that appear in the exchange algebras for the primary operators in this section.}
\label{fig: monodromy}
\end{figure}

In order to compute commutators of the string operators
that we seek
with the Hamiltonian \eqref{eq2Dcase: su(2)_{2} backscattering}
and with each other, we need to establish the algebra of the primary operators
\eqref{eq: spin-1/2 primary} and \eqref{eq: spin-1 primary}. We can obtain
this by considering the $u(1)^{\,}_{2}$ and $\mathbb{Z}^{\,}_{2}$ sectors
separately.  The algebra of the $u(1)^{\,}_{2}$ vertex operators is obtained
directly from Eq.~\eqref{eq: 2D equal time algebra d}.
The algebra of operators in the $\mathbbm Z^{\,}_{2}$ sector is determined
by considering their monodromy in the complex plane, see Fig.~\ref{fig: monodromy}.

As a first example, we consider the algebra of the Majorana and twist operators.
For any pair of wires $y$ and $y'$, we posit the
OPEs (using the complex coordinates
$v\equiv t+\mathrm{i}\, z$
and
$v^{\prime}\equiv t^{\prime}+\mathrm{i}\, z^{\prime}$)
\begin{subequations}
\label{eq2Dcase: Z_{2} psi sigma OPE}
\begin{align}
&
\widehat{\psi}^{\,}_{\mathrm{L},y}(v)\,
\widehat{\sigma}^{\,}_{\mathrm{L},y^{\prime}}(v^{\prime})=
\delta^{\,}_{y,y^{\prime}}\,
\frac{C^{\sigma}_{\psi\sigma}}
     {(v-v^{\prime})^{1/2}}\,\widehat{\sigma}^{\,}_{\mathrm{L},y}(v)
+
\cdots,
\label{eq2Dcase: Z_{2} psi sigma OPE a}
\\
&
\widehat{\psi}^{\,}_{\mathrm{R},y}(\bar{v})\,
\widehat{\sigma}^{\,}_{\mathrm{R},y^{\prime}}(\bar{v}^{\prime})=
\delta^{\,}_{y,y^{\prime}}\,
\frac{C^{\sigma}_{\psi\sigma}}{(\bar{v}-\bar{v}^{\prime})^{1/2}}\,
\widehat{\sigma}^{\,}_{\mathrm{R},y}(\bar{v})
+
\cdots,
\label{eq2Dcase: Z_{2} psi sigma OPE b}
\\
& 
\widehat{\psi}^{\,}_{\mathrm{L},y}(v)\,
\widehat{\sigma}^{\,}_{\mathrm{R},y^{\prime}}(\bar{v}^{\prime})=
\widehat{\psi}^{\,}_{\mathrm{R},y}(\bar{v})\,
\widehat{\sigma}^{\,}_{\mathrm{L},y^{\prime}}(v^{\prime})=
0
+
\cdots, 
\label{eq2Dcase: Z_{2} psi sigma OPE c}
\end{align}
where the structure constants obey the symmetry condition
\begin{align}
C^{\sigma}_{\psi\sigma}=
C^{\sigma}_{\sigma\psi},
\label{eq2Dcase: Z_{2} psi sigma OPE d}
\end{align}
and $\cdots$ stands for nonsingular terms.
\end{subequations}
It is apparent from Eqs.~\eqref{eq2Dcase: Z_{2} psi sigma OPE a}
that the clockwise or counterclockwise winding of $v$ around $v^\prime$
by an angle $2\pi$ yields an overall minus sign.
Inferring an equal-time exchange algebra from this monodromy is
ambiguous, since the operators $\widehat{\psi}^{\,}_{\mathrm{L},y}$
and $\widehat{\sigma}^{\,}_{\mathrm{L},y}$ are not identical.  We make
the choice
\begin{equation}
\begin{split}
&
\widehat{\psi}^{\,}_{\mathrm{M},y}(t,z)\,
\widehat{\sigma}^{\,}_{\mathrm{M}^{\prime},y^{\prime}}(t,z^{\prime})=
\widehat{\sigma}^{\,}_{\mathrm{M}^{\prime},y^{\prime}}(t,z^{\prime})\,
\widehat{\psi}^{\,}_{\mathrm{M},y}(t,z)
\\
&
\hphantom{
\widehat{\psi}^{\,}_{\mathrm{L},y}(t,z)\,
\widehat{\sigma}^{\,}_{\mathrm{L},y^{\prime}}(t,z^{\prime})=
         }
\times
e^{\mathrm{i}\pi\,(-1)^{\mathrm{M}}\, \delta^{\,}_{y,y^{\prime}}\,\delta^{\,}_{\mathrm{M},\mathrm{M}'}\,\Theta(z-z^{\prime})},
\end{split}
\label{eq2Dcase: psi sigma algebra}
\end{equation}
for any pair of wires $y$ and $y'$ and for any
$z\neq z'$.
This choice amounts to a choice of gauge in which the entirety of the
phase of $\pi$ arising from winding the
$\widehat{\psi}^{\,}_{\mathrm{L},y}$
and
$\widehat{\sigma}^{\,}_{\mathrm{L},y}$
operators around one another comes from the first
``half" of the exchange.  Restricting this half-monodromy to the real
line yields the equal-time algebra. The algebra
\eqref{eq2Dcase: psi sigma algebra}
is consistent with explicit derivations
of the equal-time exchange algebra between the Majorana operators and
the Ising order parameter in the
two-dimensional classical Ising model at criticality, see
e.g.,~\cite{Schroer78},
where the product of twist fields
$
\widehat{\sigma}^{\,}_{\mathrm{L},y^{\prime}}(t,z^{\prime})\,
\widehat{\sigma}^{\,}_{\mathrm{R},y^{\prime}}(t,z^{\prime})
$
is interpreted as representing the local Ising order parameter.

The equal-time algebra of two twist operators is more subtle.
For any pair of wires $y$ and $y'$,
the OPE of two twist fields in the complex plane is given by
(see, e.g., \cite{Ardonne10})
\begin{subequations}
\label{eq2Dcase: sigma sigma OPE}
\begin{align}
\widehat{\sigma}^{\,}_{\mathrm{L},y}(v)\,
\widehat{\sigma}^{\,}_{\mathrm{L},y^{\prime}}(v^{\prime})=&\,
\delta^{\,}_{y,y^{\prime}}\,
\frac{C^{\mathbbm{1}}_{\sigma\sigma}}{(v-v^{\prime})^{1/8}}
\nonumber\\
&
+
\delta^{\,}_{y,y^{\prime}}\,
C^{\psi}_{\sigma\sigma}\,(v-v^{\prime})^{3/8}\,\psi^{\,}_{\mathrm{L},y}(v),
\label{eq2Dcase: sigma sigma OPE a}
\\
\widehat{\sigma}^{\,}_{\mathrm{R},y}(\bar{v})\,
\widehat{\sigma}^{\,}_{\mathrm{R},y^{\prime}}(\bar{v}^{\prime})=&\,
\delta^{\,}_{y,y^{\prime}}\,
\frac{C^{\mathbbm{1}}_{\sigma\sigma}}{(\bar{v}-\bar{v}^{\prime})^{1/8}}
\nonumber\\
&\,
+
\delta^{\,}_{y,y^{\prime}}\,
C^{\psi}_{\sigma\sigma}\,
(\bar{v}-\bar{v}^{\prime})^{3/8}\,
\psi^{\,}_{\mathrm{R},y}(\bar{v}),
\label{eq2Dcase: sigma sigma OPE b}
\\
\widehat{\sigma}^{\,}_{\mathrm{L},y}(v)\,
\widehat{\sigma}^{\,}_{\mathrm{R},y^{\prime}}(\bar{v}^{\prime})=&\,
\widehat{\sigma}^{\,}_{\mathrm{R},y}(\bar{v})\,
\widehat{\sigma}^{\,}_{\mathrm{L},y^{\prime}}(v^{\prime})=
0
+
\cdots.
\label{eq2Dcase: sigma sigma OPE c}
\end{align}
\end{subequations}
Since there are two singular terms appearing on the right-hand side of
Eqs.\ (\ref{eq2Dcase: sigma sigma OPE a}) and
(\ref{eq2Dcase: sigma sigma OPE b}),
the product of two chiral twist fields must be defined with care.
In particular, correlation functions involving multiple
chiral twist fields are not well-defined unless the fusion channel
$\mathbbm{1}$ or $\psi$ is specified~\cite{Fendley07}.
We choose an equal-time operator algebra that reflects this ambiguity 
in the definition of chiral correlation functions involving the twist field.
Thus, we define the equal-time algebra
\begin{subequations}
\label{eq2Dcase: sigma sigma algebra}
\begin{align}
&
\widehat{\sigma}^{\,}_{\mathrm{L},y}(t,z)\,
\widehat{\sigma}^{\,}_{\mathrm{L},y^{\prime}}(t,z')=
\widehat{\sigma}^{\,}_{\mathrm{L},y^{\prime}}(t,z')\,
\widehat{\sigma}^{\,}_{\mathrm{L},y}(t,z)
\nonumber\\
&\qquad
\times 
\begin{cases}
e^{-\mathrm{i}\,\frac{\pi}{8}\,\delta^{\,}_{y,y^{\prime}}\,\mathrm{sgn}(z-z^{\prime})},
&
\mathrm{if}\ \sigma\times\sigma=\mathbbm{1},
\\
e^{+\mathrm{i}\,\frac{3\pi}{8}\,\delta^{\,}_{y,y^{\prime}}\,\mathrm{sgn}(z-z^{\prime})},
&
\mathrm{if}\ \sigma\times\sigma=\psi,
\end{cases}
\\
&
\widehat{\sigma}^{\,}_{\mathrm{R},y}(t,z)\,
\widehat{\sigma}^{\,}_{\mathrm{R},y^{\prime}}(t,z')=
\widehat{\sigma}^{\,}_{\mathrm{R},y^{\prime}}(t,z')\,
\widehat{\sigma}^{\,}_{\mathrm{R},y}(t,z)
\nonumber\\
&\qquad
\times 
\begin{cases}
e^{+\mathrm{i}\,\frac{\pi}{8}\,\delta^{\,}_{y,y^{\prime}}\,\mathrm{sgn}(z-z^{\prime})},
&
\mathrm{if}\ \sigma\times\sigma=\mathbbm{1},
\\
e^{-\mathrm{i}\,\frac{3\pi}{8}\,\delta^{\,}_{y,y^{\prime}}\,\mathrm{sgn}(z-z^{\prime})},
&
\mathrm{if}\ \sigma\times\sigma=\psi,
\end{cases}
\\
&
\widehat{\sigma}^{\,}_{\mathrm{L},y}(t,z)\,
\widehat{\sigma}^{\,}_{\mathrm{R},y^{\prime}}(t,z')=
\widehat{\sigma}^{\,}_{\mathrm{R},y^{\prime}}(t,z')\,
\widehat{\sigma}^{\,}_{\mathrm{L},y}(t,z),
\end{align}
\end{subequations}
in two-dimensional Minkowski space
for any pair of wires $y$ and $y'$ and for any $z\neq z'$.
We have used the shorthand notation $\sigma\times\sigma=\mathbbm{1}$
and $\sigma\times\sigma=\psi$ to distinguish the two possible fusion outcomes.
The appearance of the phases $\pm\pi/8$ and $\mp3\pi/8$
(and the correlation between their
signs) is fixed by the OPE
\eqref{eq2Dcase: sigma sigma OPE a}
and
\eqref{eq2Dcase: sigma sigma OPE b}
and the fusion channel $\mathbbm{1}$ or $\mathbbm{\psi}$,
and the sign $\mathrm{sgn}(z-z^{\prime})$
is used to keep track of the handedness of
the exchange. The choice of the overall sign convention
for the angles $\pm\pi/8$ and $\mp3\pi/8$
is equivalent to a choice of analytic continuation into the complex plane
in order to regularize the equal-time exchange of the two operators.
It is important to stress here that this equal-time algebra is not well-defined
unless one specifies a fusion channel.
This ambiguity is essential. Its origin is physical,
and it reflects the non-Abelian nature of the twist field.
We will see in the next section that this ambiguity has
important consequences for the topological degeneracy.

\subsubsection{String operators and topological degeneracy on the two-torus}
\label{subsec: String operators and topological degeneracy}

We shall consider two distinct wires $y$ and $y'$
and a coordinate $z$ along any one of these wires.
Periodic boundary conditions are imposed both along the $y$-direction and
along the $z$-direction. Hence, the one-dimensional array of wires
has the topology of a torus.

We are going to construct the equal-time algebra
\begin{align}
\left\{
\widehat{\Gamma}^{(\frac{1}{2})}_{1},
\widehat{\Gamma}^{(1)}_{2}
\right\}=0 
\label{eq2Dcase: Gamma (1/2) 1 anticommutes with Gamma (1) 2} 
\end{align}
for a first pair of nonlocal operators
$\widehat{\Gamma}^{(\frac{1}{2})}_{1}$
and
$\widehat{\Gamma}^{(1)}_{2}$.
This pair will be shown to commute with the interaction 
\eqref{eq2Dcase: su(2)_{2} backscattering}.
The nonlocal  operator
$\widehat{\Gamma}^{(\frac{1}{2})}_{1}$
can be thought of as creating a pair of pointlike
``spin-$1/2$'' excitations,
transporting them in opposite directions around
a noncontractible cycle of the torus along the $y$-direction,
and then annihilating them.
Likewise, the nonlocal operator $\widehat{\Gamma}^{(1)}_{2}$
can be thought of as implementing a similar process for a
pair of pointlike ``spin-$1$'' excitations
around a noncontractible cycle of the torus along the $z$-direction.

Similarly, we are going to construct the equal-time algebra
\begin{align}
\left\{
\widehat{\Gamma}^{(1)}_{1},
\widehat{\Gamma}^{(\frac{1}{2})}_{2}
\right\}=0
\label{eq2Dcase: Gamma psi 1 anticommutes with Gamma sigma 2} 
\end{align}
for a second pair of nonlocal operators
$\widehat{\Gamma}^{(1)}_{1}$
and
$\widehat{\Gamma}^{(\frac{1}{2})}_{2}$.
This pair will also be shown to commute with the interaction 
\eqref{eq2Dcase: su(2)_{2} backscattering},
modulo appropriate regularization
of the operator $\widehat{\Gamma}^{(\frac{1}{2})}_{2}$, as we will discuss.
The nonlocal operator
$\widehat{\Gamma}^{(1)}_{1}$
can be thought of as creating a pair of ``spin-$1$'' excitations,
transporting them in opposite directions around
a noncontractible cycle of the torus along the $y$-direction,
and then annihilating them.
The nonlocal 
operator $\widehat{\Gamma}^{(\frac{1}{2})}_{2}$
can be thought of as implementing the same process for a pair
of ``spin-$1/2$'' excitations
around a noncontractible cycle of the torus along the $z$-direction.

If we denote a ground state of the interaction 
\eqref{eq2Dcase: su(2)_{2} backscattering}
by $|\Omega\rangle$, we shall demonstrate that the three states
\begin{align}
|\Omega\rangle,
\quad
\ket{\widehat{\Gamma}^{(\frac{1}{2})}_{1}}\:=
\widehat{\Gamma}^{(\frac{1}{2})}_{1}\,
|\Omega\rangle,
\quad
&
\ket{\widehat{\Gamma}^{(\frac{1}{2})}_{2}}\:=
\widehat{\Gamma}^{(\frac{1}{2})}_{2}\,
|\Omega\rangle,
\label{eq2Dcase: three ground states}
\end{align}
are linearly independent ground states of the interaction
\eqref{eq2Dcase: su(2)_{2} backscattering}. The proof of this
claim relies on the vanishing equal-time commutators
\begin{align}
&
\left[
\widehat{\Gamma}^{(1)}_{2},
\widehat{\Gamma}^{(1)}_{1}
\right]=0,
\label{eq2Dcase: (1) 2-cyle commutes with (1) 1-cycle}
\end{align}
\begin{align}
&
\left[
\widehat{\Gamma}^{(\frac{1}{2})}_{1},
\widehat{\Gamma}^{(1)}_{1}
\right]=0,
\label{eq2Dcase: spin-1/2 1-cyle commutes with spin-1 1-cycle}
\end{align}
and
\begin{align}
&
\left[
\widehat{\Gamma}^{(1)}_{2},
\widehat{\Gamma}^{(\frac{1}{2})}_{2}
\right]=0.
\label{eq2Dcase: psi 2-cyle commutes with sigma 2-cycle}
\end{align}
Crucially, however, the exchange algebra of the nonlocal operators
$\widehat{\Gamma}^{(\frac{1}{2})}_{1}$
and
$\widehat{\Gamma}^{(\frac{1}{2})}_{2}$
suffers from the same ambiguity as that found on the right-hand side of
Eq.~\eqref{eq2Dcase: sigma sigma algebra}.
This is why one cannot infer from
Eqs.\ (\ref{eq2Dcase: Gamma (1/2) 1 anticommutes with Gamma (1) 2})--%
(\ref{eq2Dcase: psi 2-cyle commutes with sigma 2-cycle})
that the state
\begin{equation}
\widehat{\Gamma}^{(\frac{1}{2})}_{1}\,
\widehat{\Gamma}^{(\frac{1}{2})}_{2}\,
|\Omega\rangle  
\end{equation}
is linearly independent from the states (\ref{eq2Dcase: three ground states}).
(See also Appendix%
~\ref{appendix: Commutation between string operators and the Hamiltonian}.)

\begin{proof}
The proof consists of three steps.
  
\textit{Step 1: ``Spin-$1$'' string operators.}
The first string operators that
we will construct are the ``spin-$1$'' string operators.
We begin with strings running along the $y$-direction,
perpendicular to the wires. These strings are built from the local
bilinears
\begin{align}
\begin{split}
\widehat{\mathcal{O}}^{(1)}_{y}(t,z)\propto&\,
\widehat{\Upsilon}^{(1)\dagger}_{\mathrm{L},y}(t,z)\,
\widehat{\Upsilon}^{(1)}_{\mathrm{R},y}(t,z)
\\
\propto&\,
e^{-\mathrm i\frac{1}{\sqrt{2}}\widehat{\phi}^{\,}_{\mathrm{L},y}(t,z)}\,
e^{+\mathrm i\frac{1}{\sqrt{2}}\widehat{\phi}^{\,}_{\mathrm{R},y}(t,z)},
\end{split}
\end{align}
for any $0<z<L^{\,}_{z}$ (hereafter, we suppress the normal
ordering of the vertex operators).
The constants of proportionality omitted above appear in
Sec.~\ref{An aside on locality and energetics}, and are necessary
for the proper normalization of these operators.
Using Eq.~\eqref{eq: 2D equal time algebra d}
for $k=2$, 
we see that a product of ``spin-$1$'' bilinears in neighboring
wires commutes with the part of the interaction
\eqref{eq2Dcase: su(2)_{2} backscattering}
that connects the two wires, since
\begin{widetext}
\begin{align}
\label{eq2Dcase: psi y-string commutators}
\widehat{\mathcal{O}}^{(1)}_{y}(t,z)\,
\widehat{\mathcal{O}}^{(1)}_{y+1}(t,z)\,
e^{+\mathrm{i}\sqrt{1/2}\,\[\widehat{\phi}^{\,}_{\mathrm{L},y}(t,z^{\prime})
-\widehat{\phi}^{\,}_{\mathrm{R},y+1}(t,z^{\prime})\]}=
e^{+\mathrm{i}\sqrt{1/2}\,\[\widehat{\phi}^{\,}_{\mathrm{L},y}(t,z^{\prime})
-\widehat{\phi}^{\,}_{\mathrm{R},y+1}(t,z^{\prime})\]}\,
\widehat{\mathcal{O}}^{(1)}_{y}(t,z)\,
\widehat{\mathcal{O}}^{(1)}_{y+1}(t,z),
\end{align}
and because $\widehat{\mathcal{O}}^{(1)}_{y}(t,z)$
commutes with any operator from the $\mathbb{Z}^{\,}_{2}$ sector of the theory.
Thus, the nonlocal string operator
\begin{align}
\widehat{\Gamma}^{(1)}_{1}(t,z)\:=
\prod_{y=0}^{L^{\,}_{y}}
\widehat{\mathcal{O}}^{(1)}_{y}(t,z)
\label{eq: def Gamma psi 1 at z}
\end{align}
commutes with the interaction
\eqref{eq2Dcase: su(2)_{2} backscattering}
for any value of $0\leq z<L^{\,}_{z}$
when periodic boundary conditions are imposed in the $y$-direction.
The nonlocal operator (\ref{eq: def Gamma psi 1 at z})
is a member of the family
\begin{align}
\widehat{\Gamma}^{(1)}_{1}(t,z^{\,}_{1},\cdots z^{\,}_{L^{\,}_{y}})\:=
\widehat{\Upsilon}^{(1)\dagger}_{\mathrm{L},1}(t,z^{\,}_{1})\,
\widehat{\Upsilon}^{(1)}_{\mathrm{R},1}(t,z^{\,}_{2})\,
\widehat{\Upsilon}^{(1)\dagger}_{\mathrm{L},2}(t,z^{\,}_{2})\,
\widehat{\Upsilon}^{(1)}_{\mathrm{R},2}(t,z^{\,}_{3})\,
\cdots\,
\widehat{\Upsilon}^{(1)\dagger}_{\mathrm{L},L^{\,}_{y}}(t,z^{\,}_{L^{\,}_{y}})\,
\widehat{\Upsilon}^{(1)}_{\mathrm{R},L^{\,}_{y}}(t,z^{\,}_{1})
\label{eq: general ``spin-1'' string operator}
\end{align}
\end{widetext} 
of operators, which all commute with the Hamiltonian 
defined by Eq.~(\ref{eq2Dcase: 2D interaction})
for any values of $0\leq z^{\,}_{1},\cdots,z^{\,}_{L^{\,}_{y}}<L^{\,}_{z}$
when periodic boundary conditions are imposed in the $y$-direction.
Any ``spin-$1$'' string operator from the family
(\ref{eq: general ``spin-1'' string operator})
can be viewed as creating a pair of ``spin-$1$'' excitations and transporting
one of them around a noncontractible loop that encircles the torus in the
$y$-direction
(a noncontractible cycle along the $y$-direction),
before annihilating it with its partner.

To construct a ``spin-$1$'' string running along the $z$-direction,
parallel to the wires, we consider the  operator
\begin{subequations}
\begin{align}
\begin{split}
&
\widehat{\mathcal{O}}^{(1)}_{\mathrm{M},y}(t,z^{\,}_{1},z^{\,}_{2})\propto
\widehat{\Upsilon}^{(1)\dagger}_{\mathrm{M},y}(t,z^{\,}_{2})\,
\widehat{\Upsilon}^{(1)}_{\mathrm{M},y}(t,z^{\,}_{1})
\\
&\qquad\propto
\exp
\Bigg(
-
\mathrm{i}
\frac{1}{\sqrt{2}}
\int\limits^{z^{\,}_{2}}_{z^{\,}_{1}}\mathrm{d}z\,
\partial^{\,}_{z}\widehat{\phi}^{\,}_{\mathrm{M},y}(t,z)
\Bigg)
\end{split}
\label{eq2Dcase: spin-1 z string operator}
\end{align}
for any $0\leq z^{\,}_{1},z^{\,}_{2}<L^{\,}_{z}$ and
$\mathrm{M}=\mathrm{L},\mathrm{R}$.
\end{subequations}
Hence, $\widehat{\mathcal{O}}^{(1)}_{\mathrm{M},y}(t,z^{\,}_{1},z^{\,}_{2})$
is a bilocal operator that also obeys

\medskip\medskip
\begin{widetext}
\begin{align}
\widehat{\mathcal{O}}^{(1)}_{\mathrm{L},y}(t,z^{\,}_{1},z^{\,}_{2})\,
e^{+\mathrm{i}\sqrt{1/2}\,\[\widehat{\phi}^{\,}_{\mathrm{L},y}(t,z)
-\widehat{\phi}^{\,}_{\mathrm{R},y+1}(t,z)\]}
=&\,
e^{+\mathrm{i}\sqrt{1/2}\,\[\widehat{\phi}^{\,}_{\mathrm{L},y}(t,z)
-\widehat{\phi}^{\,}_{\mathrm{R},y+1}(t,z)\]}\,
\widehat{\mathcal{O}}^{(1)}_{\mathrm {L},y}(t,z^{\,}_{1},z^{\,}_{2})
\times e^{+\mathrm{i}2\pi\int\limits^{z^{\,}_{2}}_{z^{\,}_{1}}\mathrm{d}z^{\prime}\,
\delta(z-z^{\prime})},
\label{eq2Dcase: su(2)_{2} z-string commutator}
\end{align}
\end{widetext} \medskip\medskip
as a result of
Eq.\ \eqref{eq: 2D equal time algebra d}
for $k=2$. (A similar expression holds for $\mathrm{M}=\mathrm{R}$.)
Now define the nonlocal operator
\begin{align}
\widehat{\Gamma}^{(1)}_{2,\mathrm{M},y}(t)\propto
\widehat{\mathcal{O}}^{(1)}_{\mathrm{M},y}(t,0,L^{\,}_{z}),
\label{eq: def Gamma (1) 2 at y}
\end{align}
which commutes with the interaction
\eqref{eq2Dcase: su(2)_{2} backscattering}
by Eq.~\eqref{eq2Dcase: su(2)_{2} z-string commutator}.
This ``spin-$1$'' string operator can be
viewed as transporting a ``spin-$1$'' excitation around a noncontractible
loop that encircles the torus in the $z$-direction
(a noncontractible cycle along the $z$-direction).

The equal-time commutation relation between the string operators
(\ref{eq: def Gamma psi 1 at z})
with $0<z<L^{\,}_{z}$ and
(\ref{eq: def Gamma (1) 2 at y})
is computed using
Eq.~\eqref{eq: 2D equal time algebra d}
for $k=2$.
It is simply the commutative rule
\begin{align}
\widehat{\Gamma}^{(1)}_{1}(t,z)\,
\widehat{\Gamma}^{(1)}_{2,\mathrm{M},y}(t)
=&\,
\widehat{\Gamma}^{(1)}_{2,\mathrm{M},y}(t)\,
\widehat{\Gamma}^{(1)}_{1}(t,z),
\label{eq2Dcase: (1) (1) braiding}
\end{align}
for any $\mathrm{M}=\mathrm{L},\mathrm{R}$.
This result reflects the fact
that the spin-1 primary operator in the $su(2)^{\,}_{2}$ has
trivial self-monodromy.
We have established
Eq.~(\ref{eq2Dcase: (1) 2-cyle commutes with (1) 1-cycle})
provided we make the identifications
\begin{equation}
\widehat{\Gamma}^{(1)}_{1}(t,z)\to
\widehat{\Gamma}^{(1)}_{1},
\quad \mathrm{and} \quad
\widehat{\Gamma}^{(1)}_{2,\mathrm{M},y}(t)\to
\widehat{\Gamma}^{(1)}_{2},
\end{equation}
for some choice of chirality $\mathrm{M}$ and wire $y$.

\textit{Step 2: ``Spin-$1/2$'' string operators.}
We next construct string operators associated with the spin-$1/2$
primary of the $su(2)^{\,}_{k}$ theory.
We proceed according to a strategy similar to the one used for the
``spin-$1$'' strings. To construct a ``spin-$1/2$'' string along
the $y$-direction,
let $0<z,z^{\prime}<L^{\,}_{z}$ and consider the local ``spin-$1/2$''
bilinears 
\begin{subequations}
\begin{align}
\widehat{\mathcal{O}}^{(\frac{1}{2})}_{y}(t,z)
&\propto
\widehat{\Upsilon}^{(\frac{1}{2})\dagger}_{\mathrm{L},y}(t,z)\,
\widehat{\Upsilon}^{(\frac{1}{2})}_{\mathrm{R},y}(t,z)
\nonumber
\\
&\propto
e^{-\mathrm i\frac{1}{2\sqrt{2}}\widehat{\phi}^{\,}_{\mathrm{L},y}(t,z)}
e^{+\mathrm i\frac{1}{2\sqrt{2}}\widehat{\phi}^{\,}_{\mathrm{R},y}(t,z)}\\
&\qquad\times
\widehat{\sigma}^{\,}_{\mathrm{L},y}(t,z)\,
\widehat{\sigma}^{\,}_{\mathrm{R},y}(t,z),
\nonumber
\end{align}
where we have defined the operator
\begin{align}
\label{eq2Dcase: phi (1/2) adjoint def}
\widehat{\Upsilon}^{(\frac{1}{2})\dagger}_{\mathrm{L},y}(t,z)\:=
e^{-\mathrm i\frac{1}{2\sqrt{2}}\widehat{\phi}^{\,}_{\mathrm{L},y}(t,z)}\,
\widehat{\sigma}^{\,}_{\mathrm{L},y}(t,z),
\end{align}
\end{subequations}
in which the adjoint operation pertains only to the $u(1)^{\,}_{k}$ vertex operator.
Using Eqs.~\eqref{eq: 2D equal time algebra d}
and \eqref{eq2Dcase: psi sigma algebra}, we find that
the equal-time product of such bilinears over all wires, namely
\begin{align}
\widehat{\Gamma}^{(\frac{1}{2})}_{1}(t,z)\:=
\prod_{y=0}^{L^{\,}_{y}}
\widehat{\mathcal{O}}^{(\frac{1}{2})}_{y}(t,z),
\label{eq: def Gamma (1/2) 1}
\end{align}
commutes with the interaction \eqref{eq2Dcase: su(2)_{2} backscattering}
for any value $0<z<L^{\,}_{z}$
when periodic boundary conditions are imposed in the
$y$-direction. This nonlocal  operator
is a member of the family
\begin{widetext}
\begin{align}
\widehat{\Gamma}^{(\frac{1}{2})}_{1}(t,z^{\,}_{1},\cdots,z^{\,}_{L^{\,}_{y}})\:=
\widehat{\Upsilon}^{(\frac{1}{2})\dagger}_{\mathrm{L},1}(t,z^{\,}_{1})\,
\widehat{\Upsilon}^{(\frac{1}{2})}_{\mathrm{R},1}(t,z^{\,}_{2})\,
\widehat{\Upsilon}^{(\frac{1}{2})\dagger}_{\mathrm{L},2}(t,z^{\,}_{2})\,
\widehat{\Upsilon}^{(\frac{1}{2})}_{\mathrm{R},2}(t,z^{\,}_{3})\,
\cdots\,
\widehat{\Upsilon}^{(\frac{1}{2})\dagger}_{\mathrm{L},L^{\,}_{y}}(t,z^{\,}_{L^{\,}_{y}})\,
\widehat{\Upsilon}^{(\frac{1}{2})}_{\mathrm{R},L^{\,}_{y}}(t,z^{\,}_{1})
\label{eq: general ``spin-1/2'' string operator}
\end{align}
\end{widetext}
of operators that commute with the Hamiltonian 
defined by Eq.~(\ref{eq2Dcase: 2D interaction})
for any values of $0<z^{\,}_{1},\cdots,z^{\,}_{L^{\,}_{y}}<L^{\,}_{z}$
when periodic boundary conditions are imposed in the $y$-direction.
Any ``spin-$1/2$'' string operator from the family
(\ref{eq: general ``spin-1/2'' string operator})
can be interpreted as creating a pair of ``spin-$1/2$''
excitations and transporting
one of them around a noncontractible cycle along the $y$-direction,
before annihilating it with its partner.

We first observe that the operators
$\widehat{\Gamma}^{(1)}_{1}(t,z)$
and
$\widehat{\Gamma}^{(\frac{1}{2})}_{1}(t,z^{\prime})$
commute with one another for any $z$
and $z^{\prime}$, as one can show using the equal-time algebra
\eqref{eq: 2D equal time algebra d},
\begin{equation}
\widehat{\Gamma}^{(1)}_{1}(t,z)\,
\widehat{\Gamma}^{(\frac{1}{2})}_{1}(t,z^{\prime})=
\widehat{\Gamma}^{(\frac{1}{2})}_{1}(t,z^{\prime})\,
\widehat{\Gamma}^{(1)}_{1}(t,z).
\end{equation}
We have established Eq.\
(\ref{eq2Dcase: spin-1/2 1-cyle commutes with spin-1 1-cycle})
provided we make the identifications
\begin{equation}
\widehat{\Gamma}^{(1)}_{1}(t,z)\to
\widehat{\Gamma}^{(1)}_{1},
\qquad
\widehat{\Gamma}^{(\frac{1}{2})}_{1}(t,z^{\prime})\to
\widehat{\Gamma}^{(\frac{1}{2})}_{1}.
\end{equation}

We claim that the ``spin-$1/2$'' string
$\widehat{\Gamma}^{(\frac{1}{2})}_{1}$ can be
interpreted as an operator that ``twists,'' from
antiperiodic to periodic, the boundary conditions of a
``spin-$1$'' excitation that encircles the torus in the $z$-direction. 
To see that this is the case, we use the chiral boson algebra of
Eq.~\eqref{eq: 2D equal time algebra d}
to show that the equal-time operator algebra
\begin{align}
\widehat{\Gamma}^{(1)}_{2,\mathrm{M},y}(t)\,
\widehat{\Gamma}^{(\frac{1}{2})}_{1}(t,z)
=&\,
-
\widehat{\Gamma}^{(\frac{1}{2})}_{1}(t,z)\,
\widehat{\Gamma}^{(1)}_{2,\mathrm{M},y}(t)
\label{eq2Dcase: Gamma^(1)_{2},y and Gamma^(1/2)_{1} commutator}
\end{align}
holds for any choice of chirality $\mathrm{M}=\mathrm{L},\mathrm{R}$
and wire $y$.  We further recall that the operator
$\widehat{\Gamma}^{(1)}_{2,\mathrm{M},y}(t)$ transports a ``spin-$1$''
excitation around the torus along the $z$-direction.  Thus,
Eq.~\eqref{eq2Dcase: Gamma^(1)_{2},y and Gamma^(1/2)_{1} commutator}
shows that the amplitude for transporting a ``spin-$1$'' excitation
around the torus and then applying the operator
$\widehat{\Gamma}^{(\frac{1}{2})}_{1}(t,z)$ differs by a minus sign
from the amplitude for applying the operator
$\widehat{\Gamma}^{(\frac{1}{2})}_{1}(t,z)$ and then transporting a
``spin-$1$'' excitation around the torus.  This is precisely the
action of an operator that twists the boundary conditions of
a``spin-$1$'' excitation.

In deriving
Eq.\ \eqref{eq2Dcase: Gamma^(1)_{2},y and Gamma^(1/2)_{1} commutator},
we have established
Eq.\ (\ref{eq2Dcase: Gamma (1/2) 1 anticommutes with Gamma (1) 2})
provided that we make the identifications
\begin{equation}
\widehat{\Gamma}^{(1)}_{2,\mathrm{M},y}(t)\to
\widehat{\Gamma}^{(1)}_{2},
\qquad
\widehat{\Gamma}^{(\frac{1}{2})}_{1}(t,z)\to
\widehat{\Gamma}^{(\frac{1}{2})}_{1}.
\end{equation}
for some choice of chirality $\mathrm{M}$ and wire $y$.

Next, we seek an operator that twists the boundary conditions
of a ``spin-$1$'' excitation encircling the torus along the $y$-direction.
We proceed in direct analogy with
Eq.\ \eqref{eq2Dcase: spin-1 z string operator}
by defining the (nonlocal ) operator
\begin{align}
\widehat{\mathcal{O}}^{(\frac{1}{2})}_{\mathrm{M},y^{\prime}}(t,z^{\,}_{1},z^{\,}_{2})
&\propto
\widehat{\Upsilon}^{(\frac{1}{2})\dagger}_{\mathrm{M},y^{\prime}}(t,z^{\,}_{2})\,
\widehat{\Upsilon}^{(\frac{1}{2})}_{\mathrm{M},y^{\prime}}(t,z^{\,}_{1})
\nonumber
\\
&\propto
\exp
\Bigg(
-
\mathrm{i}
\frac{1}{2\sqrt{2}}
\int\limits^{z^{\,}_{2}}_{z^{\,}_{1}}\mathrm{d}z\,
\partial^{\,}_{z}\widehat{\phi}^{\,}_{\mathrm{M},y^{\prime}}(t,z)
\Bigg)
\nonumber
\\
&\qquad 
\times
\widehat{\sigma}^{\,}_{\mathrm{M},y^{\prime}}(t,z^{\,}_{2})\,
\widehat{\sigma}^{\,}_{\mathrm{M},y^{\prime}}(t,z^{\,}_{1}).
\label{eq2Dcase: sigma_M,y' sigma_M,y' definition}
\end{align}
We seek
to define a string operator by taking $z^{\,}_{1}\to 0$
and $z^{\,}_{2}\to L^{\,}_{z}$. However, one must be careful in taking
these limits since Eq.~\eqref{eq2Dcase: sigma_M,y' sigma_M,y' definition}
contains two \textit{chiral} $\mathbb{Z}^{\,}_{2}$ twist fields in the
\textit{same} wire.
Due to the ambiguity of the OPE \eqref{eq2Dcase: sigma sigma algebra},
such a product is ill-defined unless a fusion channel is specified.
[Meanwhile, the product of $u(1)^{\,}_{2}$ vertex operators is unambiguous.]
By analogy with the construction of
$\widehat{\Gamma}^{(1)}_{2}$
in Eq.~\eqref{eq2Dcase: spin-1 z string operator}, we would like to define
the string operator $\widehat{\Gamma}^{(\frac{1}{2})}_{2}$ in such a way as
to leave the system in the vacuum sector. Hence, the natural choice is to
specify that the two $\widehat{\sigma}^{\,}_{M,y^{\prime}}$ operators in
Eq.~\eqref{eq2Dcase: sigma_M,y' sigma_M,y' definition}
fuse to the identity operator $\mathbbm{1}$.
In addition to providing
a sensible parallel with the construction of
$\widehat{\Gamma}^{(1)}_{1}$, this choice agrees with the choice
made in the construction of the operator that tunnels an $e/4$
quasiparticle across a quantum point contact in the Moore-Read
state~\cite{Fendley07}.

This motivates the definition of the ``spin-$1/2$'' string operator
\begin{align}
\begin{split}
\widehat{\Gamma}^{(\frac{1}{2})}_{2,\mathrm{M},y^{\prime}}(t,\epsilon)
&\:=
\exp
\Bigg(
-
\mathrm{i}
\frac{1}{2\sqrt{2}}
\int\limits^{L^{\,}_{z}}_{0}\mathrm{d}z\,
\partial^{\,}_{z}\widehat{\phi}^{\,}_{\mathrm{M},y^{\prime}}(t,z)
\Bigg)
\\
&\qquad
\times
\widehat{\mathcal{P}}^{\,}_{\mathbbm{1}}\,
\widehat{\sigma}^{\,}_{\mathrm{M},y^{\prime}}(t,0)\,
\widehat{\sigma}^{\,}_{\mathrm{M},y^{\prime}}(t,\epsilon)\,
\widehat{\mathcal{P}}^{\,}_{\mathbbm{1}},
\end{split}
\label{eq2Dcase: Gamma^(1/2)_{2},y' definition}
\end{align}
where
$\widehat{\mathcal{P}}^{\,}_{\mathbbm{1}}$
is the projection operator onto the fusion channel
$\sigma\times\sigma=\mathbbm{1}$.
(This projection can also be implemented by an appropriate choice of
normalization, as is done in Sec.~\ref{An aside on locality and energetics}.)
One can show that this projector does not affect the
algebra of twist operators $\widehat{\sigma}^{\,}_{\mathrm{M},y}$
and Majorana operators
$\widehat{\psi}^{\,}_{\mathrm{M},y}$.
We claim that the string operator
$\widehat{\Gamma}^{(\frac{1}{2})}_{2,\mathrm{M},y^{\prime}}(t,\epsilon)$
defined in this way commutes with the interaction
\eqref{eq2Dcase: su(2)_{2} backscattering} in the limit $\epsilon\to 0$.
To see this, note that 
\begin{widetext}
\begin{align}
\lim_{\epsilon\to0}
\widehat{\psi}^{\,}_{\mathrm{L},y}(t,z)\,
\widehat{\psi}^{\,}_{\mathrm{R},y+1}(t,z)\,
\widehat{\sigma}^{\,}_{\mathrm{L},y^{\prime}}(t,0)\,
\widehat{\sigma}^{\,}_{\mathrm{L},y^{\prime}}(t,\epsilon)
=
\lim_{\epsilon\to0}
\widehat{\sigma}^{\,}_{\mathrm{L},y^{\prime}}(t,0)\,
\widehat{\sigma}^{\,}_{\mathrm{L},y^{\prime}}(t,\epsilon)\,
\widehat{\psi}^{\,}_{\mathrm{L},y}(t,z)\,
\widehat{\psi}^{\,}_{\mathrm{R},y+1}(t,z)
\times
\begin{cases}
+1, & y\neq y^{\prime},
\\
-1, & y=y^{\prime},
\end{cases}
\end{align}
follows from the algebra \eqref{eq2Dcase: psi sigma algebra},
while
\begin{align}
\begin{split}
&e^{+\mathrm{i}\sqrt{1/2}\,\[\widehat{\phi}^{\,}_{\mathrm{L},y}(t,z)
-\widehat{\phi}^{\,}_{\mathrm{R},y+1}(t,z)\]}\,
\exp
\Bigg(
-
\mathrm{i}
\frac{1}{2\sqrt{2}}
\int\limits^{L^{\,}_{z}}_{0}\mathrm{d}z\,
\partial^{\,}_{z}\widehat{\phi}^{\,}_{\mathrm{L},y^{\prime}}(t,z)
\Bigg)
\\
&\qquad\qquad\qquad
=
\exp
\Bigg(
-
\mathrm{i}
\frac{1}{2\sqrt{2}}
\int\limits^{L^{\,}_{z}}_{0}\mathrm{d}z\,
\partial^{\,}_{z}\widehat{\phi}^{\,}_{\mathrm{L},y^{\prime}}(t,z)
\Bigg)\,
e^{+\mathrm{i}\sqrt{1/2}\,\[\widehat{\phi}^{\,}_{\mathrm{L},y}(t,z)
-\widehat{\phi}^{\,}_{\mathrm{R},y+1}(t,z)\]}
\times
\begin{cases}
+1, & y\neq y^{\prime},\\
-1, & y=y^{\prime},
\end{cases}
\end{split}
\end{align}
\end{widetext}
follows from the algebra
\eqref{eq: 2D equal time algebra d}.
(Similar expressions hold for $\mathrm{M}=\mathrm{R}$.)
Consequently, $\widehat{\Gamma}^{(\frac{1}{2})}_{2,\mathrm{M},y^{\prime}}(t,\epsilon)$
commutes with the interaction \eqref{eq2Dcase: su(2)_{2} backscattering}
in the limit $\epsilon\to 0$~\cite{foot1}.

Moreover, we can also show that
$\widehat{\Gamma}^{(\frac{1}{2})}_{2,\mathrm{M},y^{\prime}}(t,\epsilon)$
twists the boundary conditions of a ``spin-$1$'' excitation encircling
the torus along the $y$-direction. To do this, we use the algebra
\eqref{eq: 2D equal time algebra d} to compute
the exchange relation (in the limit $\epsilon\to0$)
\begin{align}
\widehat{\Gamma}^{(\frac{1}{2})}_{2,\mathrm{M},y^{\prime}}(t,\epsilon)\,
\widehat{\Gamma}^{(1)}_{1}(t,z)\,
=&\,
-
\widehat{\Gamma}^{(1)}_{1}(t,z)\,
\widehat{\Gamma}^{(\frac{1}{2})}_{2,\mathrm{M},y^{\prime}}(t,\epsilon),
\label{eq2Dcase: Gamma^(1/2)_{2},y' and Gamma^(1)_{1} commutator}
\end{align}
which holds for any chirality $\mathrm{M}$ and wire $y^{\prime}$.
This exchange relation has an interpretation similar to
Eq.~\eqref{eq2Dcase: Gamma^(1)_{2},y and Gamma^(1/2)_{1} commutator}.
Thus, we have established Eq.\
(\ref{eq2Dcase: Gamma psi 1 anticommutes with Gamma sigma 2})
provided we make the identifications
\begin{equation}
\widehat{\Gamma}^{(1)}_{1}(t,z)\to
\widehat{\Gamma}^{(1)}_{1},
\qquad
\widehat{\Gamma}^{(\frac{1}{2})}_{2,\mathrm{M},y^{\prime}}(t,\epsilon)
\to
\widehat{\Gamma}^{(\frac{1}{2})}_{2},
\end{equation}
for infinitesimal $\epsilon>0$.
By assumption $y\neq y^{\prime}$. Hence, the operators
$\widehat{\Gamma}^{\psi}_{2,y}\to\widehat{\Gamma}^{(1)}_{2}$
and
$\widehat{\Gamma}^{\sigma}_{2,\mathrm{M},y^{\prime}}\to
\widehat{\Gamma}^{(\frac{1}{2})}_{2}$
commute with one another in a trivial way. This establishes
Eq.~(\ref{eq2Dcase: psi 2-cyle commutes with sigma 2-cycle}).

\textit{Step 3: The topological degeneracy.}
There exists a many-body ground state
\begin{subequations}
\label{eq: Ansatz for three ground states}
\begin{equation}
\ket{\Omega}\equiv\ket{\mathbbm{1}}
\label{eq: Ansatz for three ground states a}
\end{equation}
of the interaction $\widehat{\mathcal{H}}^{\,}_{\mathrm{bs}}$
defined in Eq.~(\ref{eq2Dcase: su(2)_{2} backscattering})
from which
we can obtain two additional many-body states by acting
with the ``spin-$1/2$'' string operators along the
$y$- and $z$-directions,
respectively,
\begin{align}
\ket{\widehat{\Gamma}^{(\frac{1}{2})}_{1}}\:=
\widehat{\Gamma}^{(\frac{1}{2})}_{1}(t,z)\ket\Omega
\label{eq: Ansatz for three ground states b}
\end{align}
and
\begin{align}
\ket{\widehat{\Gamma}^{(\frac{1}{2})}_{2}}\:=
\lim_{\epsilon\to0}
\widehat{\Gamma}^{(\frac{1}{2})}_{2,\mathrm{R},y^{\prime}}(t,\epsilon)
\ket\Omega,
\label{eq: Ansatz for three ground states c}
\end{align}
\end{subequations}
for any $z$, $y^{\prime}$, and $z^{\,}_{1}$.
It is important to point out that not all
choices of $\ket{\Omega}$ are equal. As argued in Appendix%
~\ref{appendix: Commutation between string operators and the Hamiltonian},
depending on the topological sector in which the state
$\ket{\Omega}$ resides, one or both of the states 
\eqref{eq: Ansatz for three ground states b}
and
\eqref{eq: Ansatz for three ground states c}
could have norm zero or infinity.
We will first prove that the many-body states
$\ket{\widehat{\Gamma}^{(\frac{1}{2})}_{1}}$
and
$\ket{\widehat{\Gamma}^{(\frac{1}{2})}_{2}}$
share the same eigenvalue of $\widehat{\mathcal{H}}^{\,}_{\mathrm{bs}}$
as $\ket{\Omega}$.
Second, we will prove that the many-body states
(\ref{eq: Ansatz for three ground states})
are linearly independent. In doing so, we will have established
that the ground state degeneracy on the torus
of the interaction $\widehat{\mathcal{H}}^{\,}_{\mathrm{bs}}$
is threefold.

First, we recall that
$\widehat{\Gamma}^{(\frac{1}{2})}_{1}(t,z)$
commutes with the interaction $\widehat{\mathcal{H}}^{\,}_{\mathrm{bs}}$
defined in Eq.~(\ref{eq2Dcase: su(2)_{2} backscattering}).
Hence, the many-body state $\ket{\widehat{\Gamma}^{(\frac{1}{2})}_{1}}$ defined in 
Eq.~\eqref{eq: Ansatz for three ground states b}
is a ground state of the interaction $\widehat{\mathcal{H}}^{\,}_{\mathrm{bs}}$.
Making sure to treat
the limit $\epsilon\to0$ with care,
we show in Appendix%
~\ref{appendix: Commutation between string operators and the Hamiltonian}
that the many-body state $\ket{\widehat{\Gamma}^{(\frac{1}{2})}_{2}}$ defined
in Eq.~\eqref{eq: Ansatz for three ground states b}
is also a ground state of the interaction
$\widehat{\mathcal{H}}^{\,}_{\mathrm{bs}}$.
Now, we are going to show that the three many-body states
(\ref{eq: Ansatz for three ground states})
are linearly independent. 

The operators $\widehat{\Gamma}^{(1)}_{1}$ and
$\widehat{\Gamma}^{(1)}_{2}$
commute with the interaction
$\widehat{\mathcal{H}}^{\,}_{\mathrm{bs}}$
and with each other
[recall Eq.~(\ref{eq2Dcase: (1) 2-cyle commutes with (1) 1-cycle})].
They are thus simultaneously diagonalizable.
Consequently, we can choose $\ket{\Omega}$
to be a simultaneous eigenstate of the pair of operators
$\widehat{\Gamma}^{(1)}_{1}$
and 
$\widehat{\Gamma}^{(1)}_{2}$.
We assume that 
$\widehat{\Gamma}^{(1)}_{1}$
and 
$\widehat{\Gamma}^{(1)}_{2}$ have the unimodular eigenvalues
$\omega^{(1)}_{1}\neq0$ and $\omega^{(1)}_{2}\neq0$
such that
\begin{subequations}
\begin{align}
\widehat{\Gamma}^{(1)}_{1}\,\ket{\Omega}=
\omega^{(1)}_{1}\,\ket{\Omega},
\end{align}
and
\begin{align}
\widehat{\Gamma}^{(1)}_{2}\,\ket{\Omega}=
\omega^{(1)}_{2}\,\ket{\Omega},
\end{align}
\end{subequations}
respectively.

Because of the anticommutator
(\ref{eq2Dcase: Gamma (1/2) 1 anticommutes with Gamma (1) 2}),
we find the eigenvalue
\begin{align}
&
\widehat{\Gamma}^{(1)}_{2}\,\ket{\widehat{\Gamma}^{(\frac{1}{2})}_{1}}
=
-
\omega^{(1)}_{2}\,
\ket{\widehat{\Gamma}^{(\frac{1}{2})}_{1}}.
\label{eq2Dcase: proof degeneracy of at least three a}
\end{align}
Hence, $\ket{\Omega}$ and $\ket{\widehat{\Gamma}^{(\frac{1}{2})}_{1}}$
are simultaneous eigenstates of the  operator
$\Gamma^{(1)}_{2}$ with distinct eigenvalues.
As such, $\ket{\Omega}$ and $\ket{\widehat{\Gamma}^{(\frac{1}{2})}_{1}}$
are othogonal. Similarly, because of the anticommutator
(\ref{eq2Dcase: Gamma psi 1 anticommutes with Gamma sigma 2}),
we find the eigenvalue
\begin{align}
&
\widehat{\Gamma}^{(1)}_{1}\,\ket{\widehat{\Gamma}^{(\frac{1}{2})}_{2}}
=
-
\omega^{(1)}_{1}\,
\ket{\widehat{\Gamma}^{(\frac{1}{2})}_{2}}.
\label{eq2Dcase: proof degeneracy of at least three b}
\end{align}
Hence, $\ket{\Omega}$ and $\ket{\widehat{\Gamma}^{(\frac{1}{2})}_{2}}$
are simultaneous eigenstates of the  operator
$\Gamma^{(1)}_{1}$ with distinct eigenvalues.
As such,  $\ket{\Omega}$ and $\ket{\widehat{\Gamma}^{(\frac{1}{2})}_{2}}$
are othogonal.

To complete the proof that
$\ket{\Omega}$,
$\ket{\widehat{\Gamma}^{(\frac{1}{2})}_{1}}$,
and $\ket{\widehat{\Gamma}^{(\frac{1}{2})}_{2}}$
are linearly independent, it suffices to show that
$\ket{\widehat{\Gamma}^{(\frac{1}{2})}_{1}}$
and
$\ket{\widehat{\Gamma}^{(\frac{1}{2})}_{2}}$
are orthogonal.
Because of the commutator
(\ref{eq2Dcase: spin-1/2 1-cyle commutes with spin-1 1-cycle}),
we find the eigenvalue
\begin{align}
&
\widehat{\Gamma}^{(1)}_{1}\,\ket{\widehat{\Gamma}^{(\frac{1}{2})}_{1}}
=
+
\omega^{(1)}_{1}\,
\ket{\widehat{\Gamma}^{(\frac{1}{2})}_{1}}.
\label{eq2Dcase: proof degeneracy of at least three c}
\end{align}
Hence, $\ket{\widehat{\Gamma}^{(\frac{1}{2})}_{1}}$ and
$\ket{\widehat{\Gamma}^{(\frac{1}{2})}_{2}}$
are simultaneous eigenstates of the  operator
$\widehat{\Gamma}^{(1)}_{1}$
with the pair of distinct eigenvalues
$+\omega^{(1)}_{1}$
and
$-\omega^{(1)}_{1}$.
As such,
$\ket{\widehat{\Gamma}^{(\frac{1}{2})}_{1}}$
and
$\ket{\widehat{\Gamma}^{(\frac{1}{2})}_{2}}$
are orthogonal. 

We note that the commutator
(\ref{eq2Dcase: psi 2-cyle commutes with sigma 2-cycle})
could equally well have been used to show that
$\ket{\widehat{\Gamma}^{(\frac{1}{2})}_{1}}$ and
$\ket{\widehat{\Gamma}^{(\frac{1}{2})}_{2}}$
are simultaneous eigenstates of the  operator
$\widehat{\Gamma}^{(1)}_{2}$
with the pair of distinct eigenvalues
$+\omega^{(1)}_{2}$
and
$-\omega^{(1)}_{2}$. 

As promised, we have shown that the ground-state
manifold of the interaction $\widehat{\mathcal{H}}^{\,}_{\mathrm{bs}}$
on the torus
is threefold degenerate.
\end{proof}

It is useful to pause at this stage to interpret this lower
bound on the ground state degeneracy and how it comes about.
Naively, given two pairs of anticommuting nonlocal operators, all of which
commute with the Hamiltonian, [i.e., given 
Eqs.\ \eqref{eq2Dcase: Gamma (1/2) 1 anticommutes with Gamma (1) 2} 
and \eqref{eq2Dcase: Gamma psi 1 anticommutes with Gamma sigma 2}]
there are at most four degenerate ground states.  In the case of
Kitaev's toric code~\cite{Kitaev03}, the dimensionality of the ground
state manifold saturates this upper bound.  However, in the case of the 
two-dimensional state of matter that we have constructed here, we argue
that this is not the case.  The reason for this is intimately related to the
nonunitarity of the string operators $\widehat{\Gamma}^{(\frac{1}{2})}_{1}(t,z)$
and $\widehat{\Gamma}^{(\frac{1}{2})}_{2,\mathrm{R},y^{\prime}}(t,\epsilon)$.

In particular, we assert that neither of the naively-expected
fourth states, namely
\begin{subequations}
\label{eq2Dcase: definition of ket sigma_{1},2 and ket sigma_{2},1}
\begin{align}
\ket{\widehat{\Gamma}^{(\frac{1}{2})}_{1}\,
\widehat{\Gamma}^{(\frac{1}{2})}_{2}}\:=
\lim_{\epsilon\to0}
\widehat{\Gamma}^{(\frac{1}{2})}_{1}(t,z)\,
\widehat{\Gamma}^{(\frac{1}{2})}_{2,\mathrm{R},y^{\prime}}(t,\epsilon)\,
\ket{\Omega},
\label{eq2Dcase: definition of ket sigma_{1},2}
\end{align}
and
\begin{align}
\ket{\widehat{\Gamma}^{(\frac{1}{2})}_{2}\,
\widehat{\Gamma}^{(\frac{1}{2})}_{1}}\:=
\lim_{\epsilon\to0}
\widehat{\Gamma}^{(\frac{1}{2})}_{2,\mathrm{R},y^{\prime}}(t,\epsilon)\,
\widehat{\Gamma}^{(\frac{1}{2})}_{1}(t,z)\,
\ket{\Omega},
\label{eq2Dcase: definition of ket sigma_{2},1}
\end{align}
\end{subequations}
belongs to the ground-state manifold
of the interaction $\widehat{\mathcal{H}}^{\,}_{\mathrm{bs}}$.
Note that the limit $\epsilon\to0$ above is to be taken after forming
the products $\widehat{\Gamma}^{(\frac{1}{2})}_{1}(t,z)\,
\widehat{\Gamma}^{(\frac{1}{2})}_{2,\mathrm{R},y^{\prime}}(t,\epsilon)$
and
$\widehat{\Gamma}^{(\frac{1}{2})}_{2,\mathrm{R},y^{\prime}}(t,\epsilon)\,
\widehat{\Gamma}^{(\frac{1}{2})}_{1}(t,z)$, as discussed in
Footnote~\cite{foot1} and Appendix
\ref{appendix: Commutation between string operators and the Hamiltonian}.
If the operator products
$\widehat{\Gamma}^{(\frac{1}{2})}_{1}(t,z)\,
\widehat{\Gamma}^{(\frac{1}{2})}_{2,\mathrm{R},y^{\prime}}(t,\epsilon)$
and
$\widehat{\Gamma}^{(\frac{1}{2})}_{2,\mathrm{R},y^{\prime}}(t,\epsilon)\,
\widehat{\Gamma}^{(\frac{1}{2})}_{1}(t,z)$ were to commute
with the interaction $\widehat{\mathcal{H}}^{\,}_{\mathrm{bs}}$
in the limit $\epsilon\to0$, as they would in an Abelian topological phase,
then there would be no obstruction
to the states $\ket{\widehat{\Gamma}^{(\frac{1}{2})}_{1}\,
\widehat{\Gamma}^{(\frac{1}{2})}_{2}}$ and
$\ket{\widehat{\Gamma}^{(\frac{1}{2})}_{2}\,
\widehat{\Gamma}^{(\frac{1}{2})}_{1}}$
belonging to the ground-state manifold.
The proof that such an obstruction exists in the present (non-Abelian)
case is undertaken in two complementary ways in the present work.
The first, which we call the ``algebraic'' approach,
relies on diagrammatic techniques developed in Appendix 
\ref{appendix: Diagrammatics for operator algebra in the Ising CFT},
and is presented below.
The second, which we call the ``analytic'' approach,
is carried out in Appendix
\ref{appendix: Commutation between string operators and the Hamiltonian}.
Both the ``algebraic'' and ``analytic'' proofs rely on the fact,
discussed in Appendix
\ref{appendix: Commutation between string operators and the Hamiltonian},
that the
operator products
$\widehat{\Gamma}^{(\frac{1}{2})}_{1}(t,z)\,
\widehat{\Gamma}^{(\frac{1}{2})}_{2,\mathrm{R},y^{\prime}}(t,\epsilon)$
and
$\widehat{\Gamma}^{(\frac{1}{2})}_{2,\mathrm{R},y^{\prime}}(t,\epsilon)\,
\widehat{\Gamma}^{(\frac{1}{2})}_{1}(t,z)$
are not bound to
commute with the interaction $\widehat{\mathcal{H}}^{\,}_{\mathrm{bs}}$
in the limit $\epsilon\to0$.
We now proceed with the ``algebraic'' version of the proof,
and refer the reader to Appendices
\ref{appendix: Diagrammatics for operator algebra in the Ising CFT}
and
\ref{appendix: Commutation between string operators and the Hamiltonian}
for more details.

\begin{proof}[Proof (``algebraic'')]

We introduce the projection operator
\begin{equation}\label{eq2Dcase: projection operator}
\begin{split}
\widehat{\mathcal{P}}^{\,}_{\rm{GSM}}\:=&\,
\mathcal{N}^{-1}_{\mathbbm 1}\,
\ket{\mathbbm 1}\bra{\mathbbm 1}
\\
&\,
+
\mathcal{N}^{-1}_{\widehat{\Gamma}^{(\frac{1}{2})}_{1}}\,
\ket{\widehat{\Gamma}^{(\frac{1}{2})}_{1}}\bra{\widehat{\Gamma}^{(\frac{1}{2})}_{1}}
\\
&\,
+
\mathcal{N}^{-1}_{\widehat{\Gamma}^{(\frac{1}{2})}_{2}}\,
\ket{\widehat{\Gamma}^{(\frac{1}{2})}_{2}}\bra{\widehat{\Gamma}^{(\frac{1}{2})}_{2}}
\\
&\,
+
\cdots
\end{split}
\end{equation}
onto the ground state manifold. Here,
$\mathcal{N}^{\,}_{\mathbbm 1}$ is the squared norm of the state
$\ket{\mathbbm 1}\equiv\ket{\Omega}$,
$\mathcal{N}^{\,}_{\widehat{\Gamma}^{(\frac{1}{2})}_{1}}$
is the squared norm of the state
$\ket{\widehat{\Gamma}^{(\frac{1}{2})}_{1}}$,
$\mathcal{N}^{\,}_{\widehat{\Gamma}^{(\frac{1}{2})}_{2}}$
is the squared norm of the state
$\ket{\widehat{\Gamma}^{(\frac{1}{2})}_{2}}$,
and
$\cdots$ is a sum over any remaining elements from the orthonormal basis
of the ground state manifold.
By definition, any one of the three states
$\ket{\mathbbm 1}$,
$\ket{\widehat{\Gamma}^{(\frac{1}{2})}_{1}}$,
and $\ket{\widehat{\Gamma}^{(\frac{1}{2})}_{2}}$
defined in Eq.~(\ref{eq: Ansatz for three ground states})
is invariant under the action of
\begin{equation}
\widehat{\mathcal{P}}^{\,}_{\mathrm{GSM}}=
\widehat{\mathcal{P}}^{2}_{\mathrm{GSM}}.
\end{equation}
Hence, we may write
\begin{subequations}
\label{eq2Dcase: act with P_GSM on new ground states}
\begin{align}
\ket{\widehat{\Gamma}^{(\frac{1}{2})}_{1}}
=&\,
\widehat{\mathcal{P}}^{\,}_{\mathrm{GSM}}\,
\ket{\widehat{\Gamma}^{(\frac{1}{2})}_{1}}
=
\widehat{\mathcal{P}}^{\,}_{\mathrm{GSM}}\,
\widehat{\Gamma}^{(\frac{1}{2})}_{1}(t,z)\, 
\widehat{\mathcal{P}}^{\,}_{\mathrm{GSM}}\,
\ket{\Omega},
\label{eq2Dcase: act with P_GSM on new ground states a}
\\
\ket{\widehat{\Gamma}^{(\frac{1}{2})}_{2}}
=&\,
\widehat{\mathcal{P}}^{\,}_{\mathrm{GSM}}\,
\ket{\widehat{\Gamma}^{(\frac{1}{2})}_{2}}
=
\widehat{\mathcal{P}}^{\,}_{\mathrm{GSM}}\,
\lim_{\epsilon\to0}
\widehat{\Gamma}^{(\frac{1}{2})}_{2,\mathrm{R},y^{\prime}}(t,\epsilon)\, 
\widehat{\mathcal{P}}^{\,}_{\mathrm{GSM}}\,
\ket{\Omega}.
\end{align}
\end{subequations}
On the other hand,
\begin{align}
\widehat{\mathcal{P}}^{\,}_{\rm{GSM}}\,
\widehat{\mathcal{O}}\,
\widehat{\mathcal{P}}^{\,}_{\rm{GSM}}=0
\end{align}
must hold for any operator $\widehat{\mathcal{O}}$
such that $\widehat{\mathcal{O}}$
returns an excited state when applied to any state from the
ground-state manifold.

We are first going to show that the operators
$\widehat{\Gamma}^{(\frac{1}{2})}_{1}(t,z)$
and
$\widehat{\Gamma}^{(\frac{1}{2})}_{2,\mathrm{R},y^{\prime}}(t,\epsilon)$
do not commute in the limit $\epsilon\to0$.
After that, we will
elaborate on why the state 
$
\ket{\widehat{\Gamma}^{(\frac{1}{2})}_{1}\,
\widehat{\Gamma}^{(\frac{1}{2})}_{2}}
$
does not belong to the ground-state manifold
of the interaction $\widehat{\mathcal{H}}^{\,}_{\mathrm{bs}}$.

We begin by considering the exchange algebra of the string operators
$\widehat{\Gamma}^{(\frac{1}{2})}_{1}(t,z)$
and
$\widehat{\Gamma}^{(\frac{1}{2})}_{2,\mathrm{R},y^{\prime}}(t,\epsilon)$
defined in Eqs.~(\ref{eq: def Gamma (1/2) 1})
and (\ref{eq2Dcase: Gamma^(1/2)_{2},y' definition}),
respectively.
Specifically, we consider the product
\begin{equation}
\begin{split}
&
\widehat{\Gamma}^{(\frac{1}{2})}_{1}(t,z)\,
\widehat{\Gamma}^{(\frac{1}{2})}_{2,\mathrm{R},y^{\prime}}(t,\epsilon)
\propto
\\
&\qquad
\(\prod^{L^{\,}_{y}}_{y=0}
\widehat{\sigma}^{\,}_{\mathrm{L},y}(t,z)\,
\widehat{\sigma}^{\,}_{\mathrm{R},y}(t,z) \)
\\
&\qquad\qquad
\times 
\widehat{\mathcal{P}}^{\,}_{\mathbbm{1}}\,
\widehat{\sigma}^{\,}_{\mathrm{R},y^{\prime}}(t,0)\,
\widehat{\sigma}^{\,}_{\mathrm{R},y^{\prime}}(t,\epsilon)\,
\widehat{\mathcal{P}}^{\,}_{\mathbbm{1}},
\end{split}
\label{eq2Dcase: Gamma^sigma_{1} Gamma^sigma_{2} product a}
\end{equation}
where $\epsilon>0$ is infinitesimal and
we have also omitted the operators in the $u(1)^{\,}_{2}$
sector appearing in the
definition \eqref{eq2Dcase: Gamma^(1/2)_{2},y' definition}, 
as these operators commute with all operators in the
$\mathbb{Z}^{\,}_{2}$ sector. Using the fact that
twist operators in different wires
(and in different chiral sectors of the same wire) commute,
we deduce that
\begin{equation}
\begin{split}
&
\widehat{\Gamma}^{(\frac{1}{2})}_{1}(t,z)\,
\widehat{\Gamma}^{(\frac{1}{2})}_{2,\mathrm{R},y^{\prime}}(t,\epsilon)
\propto
\\
&\quad
\(\prod^{\,}_{y\neq y^{\prime}}
\widehat{\sigma}^{\,}_{\mathrm{L},y}(t,z)\,
\widehat{\sigma}^{\,}_{\mathrm{R},y}(t,z) \)
\widehat{\sigma}^{\,}_{\mathrm{L},y^{\prime}}(t,z)
\\
&\quad\quad
\times
\widehat{\sigma}^{\,}_{\mathrm{R},y^{\prime}}(t,z)\,
\widehat{\mathcal{P}}^{\,}_{\mathbbm{1}}\,
\widehat{\sigma}^{\,}_{\mathrm{R},y^{\prime}}(t,0)\,
\widehat{\sigma}^{\,}_{\mathrm{R},y^{\prime}}(t,\epsilon)\,
\widehat{\mathcal{P}}^{\,}_{\mathbbm{1}}.
\end{split}
\label{eq2Dcase: Gamma^sigma_{1} Gamma^sigma_{2} product b}
\end{equation}
Since all operators in the first line of the right-hand side above
commute with all operators in the second line, computing the exchange algebra
of the operators $\widehat{\Gamma}^{(\frac{1}{2})}_{1}$
and
$\widehat{\Gamma}^{(\frac{1}{2})}_{2}$
boils down to considering the following product of operators,
\begin{align}
\lim_{\substack{z^{\,}_{2}\to z^{\,}_{1}+\epsilon\\ z^{\,}_{1}\to0}}
\widehat{\sigma}^{\,}_{\mathrm{R},y^{\prime}}(t,z)\,
\widehat{\mathcal{P}}^{\,}_{\mathbbm{1}}\,
\widehat{\sigma}^{\,}_{\mathrm{R},y^{\prime}}(t,z^{\,}_{1})\,
\widehat{\sigma}^{\,}_{\mathrm{R},y^{\prime}}(t,z^{\,}_{2})\,
\widehat{\mathcal{P}}^{\,}_{\mathbbm{1}}.
\end{align}
Using the prescriptions of Appendix 
\ref{appendix: Diagrammatics for operator algebra in the Ising CFT},
we find that the process of
commuting the leftmost operator, $\widehat{\sigma}^{\,}_{\mathrm{R},y^{\prime}}(t,z)$,
past the remaining two operators is represented by the diagram
\begin{align}
\includegraphics[width=.12\textwidth]{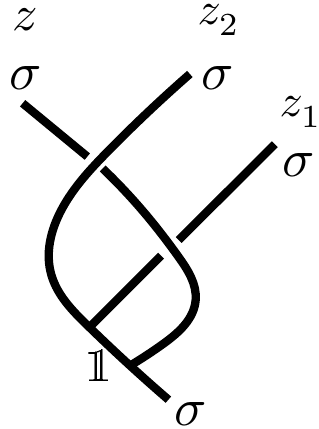}.
\end{align}
Untwisting the legs of this fusion diagram, we find
\\
\\
\begin{widetext}
\begin{align}\label{eq2Dcase: 3-sigma diagram untwisting}
\includegraphics[width=.65\textwidth]{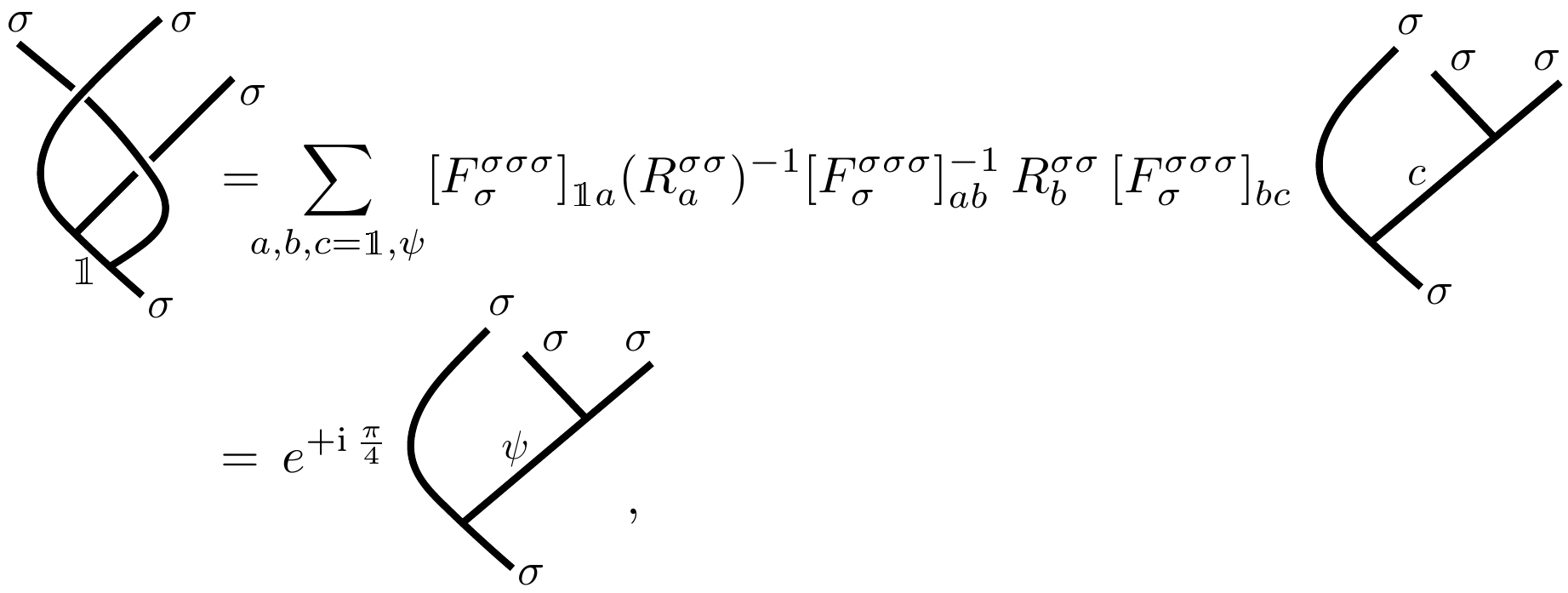}
\end{align}
\end{widetext}
where the $F$- and $R$-symbols are given in Appendix 
\ref{appendix: Diagrammatics for operator algebra in the Ising CFT}.
The diagrammatic relation expressed in Eq.\
\eqref{eq2Dcase: 3-sigma diagram untwisting}
can be rewritten as the algebraic statement
\begin{equation}
\label{eq2Dcase: 3-sigma algebraic untwisting}
\begin{split}
&
\widehat{\sigma}^{\,}_{\mathrm{R},y^{\prime}}(t,z)\,
\widehat{\mathcal{P}}^{\,}_{\mathbbm{1}}\,
\widehat{\sigma}^{\,}_{\mathrm{R},y^{\prime}}(t,z^{\,}_{1})\,
\widehat{\sigma}^{\,}_{\mathrm{R},y^{\prime}}(t,z^{\,}_{2})\,
\widehat{\mathcal{P}}^{\,}_{\mathbbm{1}}=
\\
&\qquad
e^{+\mathrm{i} \frac{\pi}{4}}\,
\widehat{\mathcal{P}}^{\,}_{\psi}\,
\widehat{\sigma}^{\,}_{\mathrm{R},y^{\prime}}(t,z^{\,}_{1})\,
\widehat{\sigma}^{\,}_{\mathrm{R},y^{\prime}}(t,z^{\,}_{2})\,
\widehat{\mathcal{P}}^{\,}_{\psi}\,
\widehat{\sigma}^{\,}_{\mathrm{R},y^{\prime}}(t,z),
\end{split}
\end{equation}
where $\widehat{\mathcal{P}}^{\,}_{\psi}$ is a projection operator
that projects the product
$\widehat{\sigma}^{\,}_{\mathrm{R},y^{\prime}}(t,z^{\,}_{1})\,
\widehat{\sigma}^{\,}_{\mathrm{R},y^{\prime}}(t,z^{\,}_{2})$
into the fusion channel $\sigma\times\sigma=\psi$.  Taking the limits 
$z^{\,}_{2}\to z^{\,}_{1}+\epsilon$
and
$z^{\,}_{1}\to 0$
and restoring the operators
$\widehat{\sigma}^{\,}_{\mathrm{M},y}(t,z)$ present in 
Eq.~\eqref{eq2Dcase: Gamma^sigma_{1} Gamma^sigma_{2} product b}
(as well as the operators from the $u(1)^{\,}_{2}$ sector
that were omitted there), we arrive at the relation
\begin{subequations}
\begin{equation}
\widehat{\Gamma}^{(\frac{1}{2})}_{1}(t,z)\,
\widehat{\Gamma}^{(\frac{1}{2})}_{2,\mathrm{R},y^{\prime}}(t,\epsilon)=
e^{
+\mathrm{i} \frac{3\pi}{4}}\,
\widehat{\widetilde\Gamma}
^{\raisebox{-6pt}{\scriptsize$(\frac{1}{2})$}}_{2,\mathrm{R},y^{\prime}}
(t,\epsilon)\,
\widehat{\Gamma}^{(\frac{1}{2})}_{1}(t,z),
\label{eq2Dcase: Gamma^sigma_{1} Gamma^sigma_{2} exchange algebra}
\end{equation}
in the limit $\epsilon\to0$,
where we have defined the operator
\begin{equation}
\begin{split}
\widehat{\widetilde\Gamma}
^{\raisebox{-6pt}{\scriptsize$(\frac{1}{2})$}}_{2,\mathrm{R},y^{\prime}}
(t,\epsilon)
&\:=
\exp
\Bigg(
-
\mathrm{i}
\frac{1}{2\sqrt{2}}
\int\limits^{L^{\,}_{z}}_{0}\mathrm{d}z\,
\partial^{\,}_{z}\widehat{\phi}^{\,}_{\mathrm{R},y^{\prime}}(t,z)
\Bigg)
\\
&\qquad
\times
\widehat{\mathcal{P}}^{\,}_{\psi}\,
\widehat{\sigma}^{\,}_{\mathrm{R},y^{\prime}}(t,0)\,
\widehat{\sigma}^{\,}_{\mathrm{R},y^{\prime}}(t,\epsilon)\,
\widehat{\mathcal{P}}^{\,}_{\psi},
\end{split}
\end{equation}
\end{subequations}
which is identical to the operator $\widehat{\Gamma}^{(\frac{1}{2})}_{2}$
defined in Eq.~\eqref{eq2Dcase: Gamma^(1/2)_{2},y' definition}, 
except that the product 
$\widehat{\sigma}^{\,}_{\mathrm{R},y^{\prime}}(t,0)\,
\widehat{\sigma}^{\,}_{\mathrm{R},y^{\prime}}(t,\epsilon)$
is evaluated in the fusion channel $\psi$ rather than the fusion channel
$\mathbbm{1}$.  This difference is fundamental.
Since the two twist operators
entering the operator 
$\widehat{\widetilde\Gamma}^{\raisebox{-6pt}{\scriptsize$(\frac{1}{2})$}}_{2}$
fuse to $\psi$, this operator can be interpreted as adding an extra Majorana
fermion to the state on which it acts. Acting with
$\widehat{\widetilde\Gamma}^{\raisebox{-6pt}{\scriptsize$(\frac{1}{2})$}}_{2}$
on any of the states
$\ket{\mathbbm 1},\ket{\widehat{\Gamma}^{(\frac{1}{2})}_{1}},
\ket{\widehat{\Gamma}^{(\frac{1}{2})}_{2}},\cdots$
in the ground-state manifold
of the interaction $\widehat{\mathcal H}^{\,}_{\rm bs}$
can then be viewed as
creating an \textit{excited state}
of the interaction $\widehat{\mathcal H}^{\,}_{\rm bs}$
with one extra fermion.  
In other words, we have
\begin{align}
\label{eq2Dcase: projection of tilde Gamma^sigma_{2} into GSM vanishes}
\widehat{\mathcal{P}}^{\,}_{\mathrm{GSM}}\,
\lim_{\epsilon\to0}
\widehat{\widetilde\Gamma}^{\raisebox{-6pt}{\scriptsize$\sigma$}}_{2,\mathrm{R},y^{\prime}}
(t,\epsilon)\,
\widehat{\mathcal{P}}^{\,}_{\mathrm{GSM}}=
0.
\end{align}
This relation is crucial in what follows. Note also
the difference between the phase on
the RHS of
Eq.~\eqref{eq2Dcase: Gamma^sigma_{1} Gamma^sigma_{2} exchange algebra}
and that on the RHS of 
Eq.~\eqref{eq2Dcase: 3-sigma algebraic untwisting},
which comes from commutators in the $u(1)^{\,}_{2}$
sector.

\begin{table}[t!]
\begin{tabular}{|c||c|c||c|c|}
\hline
\ &
$\widehat{\Gamma}^{(1)}_{1}$ & 
$\widehat{\Gamma}^{(1)}_{2}$ & 
$\widehat{\Gamma}^{(\frac{1}{2})}_{1}$ & 
$\widehat{\Gamma}^{(\frac{1}{2})}_{2}$ % ROW 1
\\
\hline\hline
$\widehat{\Gamma}^{(\frac{1}{2})}_{1}$ &
$+$ & 
$-$ & 
$+$ & 
$\xmark$ % ROW 2
\\
\hline
$\widehat{\Gamma}^{(\frac{1}{2})}_{2}$ &
$-$ & 
$+$ & 
$\xmark$ & 
$+$ % ROW 3
\\
\hline
\end{tabular}
\caption{
Summary of the algebra of the string operators
$\widehat{\Gamma}^{(1)}_{1,2}$ and $\widehat{\Gamma}^{(\frac{1}{2})}_{1,2}$.
Entries corresponding
to a pair of operators that commute are labeled with a $+$.
Entries corresponding
to a pair of operators that anticommute are labeled with a $-$.
Entries corresponding
to a pair of operators that neither commute nor anticommute
are labeled with a $\xmark$.
\label{table2Dcase: summary of algebra}
}
\end{table}

We are now prepared to exclude the state
$\ket{\widehat{\Gamma}^{(\frac{1}{2})}_{1}\,
\widehat{\Gamma}^{(\frac{1}{2})}_{2}}$
from the ground-state manifold of the interaction
$\widehat{\mathcal H}^{\,}_{\rm bs}$.  Applying Eq.\
\eqref{eq2Dcase: Gamma^sigma_{1} Gamma^sigma_{2} exchange algebra}
to the definition
\eqref{eq2Dcase: definition of ket sigma_{1},2} of the state
$\ket{\widehat{\Gamma}^{(\frac{1}{2})}_{1}\,
\widehat{\Gamma}^{(\frac{1}{2})}_{2}}$, we obtain
\begin{align}
\ket{\widehat{\Gamma}^{(\frac{1}{2})}_{1}\,
\widehat{\Gamma}^{(\frac{1}{2})}_{2}}
=
e^{+\mathrm{i}\, \frac{3\pi}{4}}\,
\lim_{\epsilon\to0}
\widehat{\widetilde\Gamma}
^{\raisebox{-6pt}{\scriptsize$(\frac{1}{2})$}}_{2,\mathrm{R},y^{\prime}}
(t,\epsilon)\,
\ket{\widehat{\Gamma}^{(\frac{1}{2})}_{1}}.
\end{align}
If the state $\ket{\widehat{\Gamma}^{(\frac{1}{2})}_{1}\,
\widehat{\Gamma}^{(\frac{1}{2})}_{2}}$
\textit{is} in the ground-state manifold
of the interaction $\widehat{\mathcal H}^{\,}_{\rm bs}$,
then it cannot be a null vector of $\widehat{\mathcal{P}}^{\,}_{\mathrm{GSM}}$.
However, using Eqs.\ \eqref{eq2Dcase: act with P_GSM on new ground states}
and \eqref{eq2Dcase: projection of tilde Gamma^sigma_{2} into GSM vanishes},
we find that
\begin{align}
\begin{split}
&
\widehat{\mathcal{P}}^{\,}_{\mathrm{GSM}}\,
\ket{\widehat{\Gamma}^{(\frac{1}{2})}_{1}\,
\widehat{\Gamma}^{(\frac{1}{2})}_{2}}\\
&\quad=\,
e^{+\mathrm{i}\, \frac{3\pi}{4}}\,
\widehat{\mathcal{P}}^{\,}_{\mathrm{GSM}}\,
\lim_{\epsilon\to0}
\widehat{\widetilde\Gamma}
^{\raisebox{-6pt}{\scriptsize$(\frac{1}{2})$}}_{2,\mathrm{R},y^{\prime}}
(t,\epsilon)\,
\widehat{\mathcal{P}}^{\,}_{\mathrm{GSM}}\,
\ket{\widehat{\Gamma}^{(\frac{1}{2})}_{1}}
\\
&\quad=\,
0.
\end{split}
\label{eq2Dcase: exclude ket sigma_{1},2}
\end{align}
Thus, the state $\ket{\widehat{\Gamma}^{(\frac{1}{2})}_{1}\,
\widehat{\Gamma}^{(\frac{1}{2})}_{2}}$ does not lie in the ground-state
manifold of the interaction $\widehat{\mathcal H}^{\,}_{\rm bs}$.
Similarly, the
state $\ket{\widehat{\Gamma}^{(\frac{1}{2})}_{2}\,
\widehat{\Gamma}^{(\frac{1}{2})}_{1}}$ defined in Eq.\
\eqref{eq2Dcase: definition of ket sigma_{2},1}
is excluded from the ground-state manifold.
We note in passing that a related line of
reasoning was used in Ref.~\onlinecite{Oshikawa07} to exclude certain
states from the ground-state manifold of the gauged $p+\mathrm{i}\, p$
superconductor (see also Ref.~\onlinecite{Read00}).
\end{proof}

In summary, we have shown that the 
$su(2)^{\,}_{2}$ coupled-wire construction
in (2+1)-dimensional spacetime
has a threefold topological degeneracy on the two-torus.
This value of the degeneracy is in agreement with the value $2+1=3$
obtained directly from the $SU(2)$ non-Abelian Chern-Simons theory at
level $2$ \cite{Witten89,Cabra00}.
The proof that this topological degeneracy is threefold and not
fourfold relied on the observation that the ``spin-$1/2$'' string
operators obey the non-Abelian exchange algebra
\eqref{eq2Dcase: Gamma^sigma_{1} Gamma^sigma_{2} exchange algebra}.
We summarize the full algebra of the various string operators in
Table\ \ref{table2Dcase: summary of algebra}.   This algebra,
whereby exchanging the two operators does not simply produce a phase
factor, but instead enacts a nontrivial transformation on the
operators themselves, is the essence of what it means to be a
non-Abelian topological phase.

Although we do not investigate in detail how to compute the ground
state degeneracy of the $su(2)^{\,}_{k}$ family of coupled-wire
theories defined in
Sec.~\ref{subsec: Definition of the class of models 2D}
for $k>2$, the methods of this section can be adapted for
general $k$.  The techniques developed in this section can be used to
demonstrate that a general $su(2)^{\,}_{k}$ coupled wire theory in
this family has a ground state degeneracy on the torus of $k+1$, in
agreement with the value obtained directly from the $SU(2)$
non-Abelian Chern-Simons theory at level $k$~\cite{Witten89}. In all
cases, the primary operators of the $su(2)^{\,}_{k}$ theory are used
as building blocks for the string operators used to calculate the
topological degeneracy on the torus.  The non-Abelian algebra of
string operators that encodes the topological degeneracy is induced by
the algebra of the primary operators used to build the string
operators.

\subsubsection{An aside on locality and energetics}
\label{An aside on locality and energetics}

\begin{widetext}

The four string operators used in Sec.\
\ref{subsec: String operators and topological degeneracy}
to construct the topological degeneracy
are built from the operators
\begin{subequations}
\begin{align}
\widehat{\mathcal{O}}^{(\frac{1}{2})}_{y}(t,z)\equiv&\,
\sqrt{\mathcal{N}^{(\frac{1}{2})}_{\mathrm{LR}}}\times
\widehat{\Upsilon}^{(\frac{1}{2})\dagger}_{\mathrm{L},y}(t,z)\,
\widehat{\Upsilon}^{(\frac{1}{2})}_{\mathrm{R},y}(t,z)
\nonumber\\
\:=&\,
\sqrt{\mathcal{N}^{(\frac{1}{2})}_{\mathrm{LR}}}\times
\bm{:}
e^{-\mathrm i\frac{1}{2\sqrt{2}}\widehat{\phi}^{\,}_{\mathrm{L},y}(t,z)}\,
\widehat{\sigma}^{\,}_{\mathrm{L},y}(t,z)
\bm{:}
\bm{:}
e^{+\mathrm i\frac{1}{2\sqrt{2}}\widehat{\phi}^{\,}_{\mathrm{R},y}(t,z)}\,
\widehat{\sigma}^{\,}_{\mathrm{R},y}(t,z)
\bm{:},
\label{eq: building block 2D string a}
\\
\widehat{\mathcal{O}}^{(1)}_{y}(t,z)\equiv&\,
\sqrt{\mathcal{N}^{(1)}_{\mathrm{LR}}}\times
\widehat{\Upsilon}^{(1)\dagger}_{\mathrm{L},y}(t,z)\,
\widehat{\Upsilon}^{(1)}_{\mathrm{R},y}(t,z)
\nonumber\\
\:=&\,
\sqrt{\mathcal{N}^{(1)}_{\mathrm{LR}}}\times
\bm{:}e^{-\mathrm i\frac{1}{\sqrt{2}}\widehat{\phi}^{\,}_{\mathrm{L},y}(t,z)}\bm{:}\,
\bm{:}e^{+\mathrm i\frac{1}{\sqrt{2}}\widehat{\phi}^{\,}_{\mathrm{R},y}(t,z)}\bm{:},
\label{eq: building block 2D string b}
\\
\widehat{\mathcal{O}}^{(\frac{1}{2})}_{\mathrm{M},y}(t,z^{\,}_{1},z^{\,}_{2})\equiv&\,
\sqrt{\mathcal{N}^{(\frac{1}{2})}_{\mathrm{M}}}\times
\widehat{\Upsilon}^{(\frac{1}{2})\dag}_{\mathrm{M},y}(t,z^{\,}_{2})\,
\widehat{\Upsilon}^{(\frac{1}{2})}_{\mathrm{M},y}(t,z^{\,}_{1})
\nonumber\\
\:=&\,
\sqrt{\mathcal{N}^{(\frac{1}{2})}_{\mathrm{M}}}\times
\bm{:}
e^{-\mathrm i\frac{1}{2\sqrt{2}}\widehat{\phi}^{\,}_{\mathrm{M},y}(t,z^{\,}_{2})}\,
\widehat{\sigma}^{\,}_{\mathrm{M},y}(t,z^{\,}_{2})
\bm{:}
\bm{:}
e^{+\mathrm i\frac{1}{2\sqrt{2}}\widehat{\phi}^{\,}_{\mathrm{M},y}(t,z^{\,}_{1})}\,
\widehat{\sigma}^{\,}_{\mathrm{M},y}(t,z^{\,}_{1})
\bm{:},
\label{eq: building block 2D string c}
\\
\widehat{\mathcal{O}}^{(1)}_{\mathrm{M},y}(t,z^{\,}_{1},z^{\,}_{2})\equiv&\,
\sqrt{\mathcal{N}^{(1)}_{\mathrm{M}}}\times
\widehat{\Upsilon}^{(1)\dag}_{\mathrm{M},y}(t,z^{\,}_{2})\,
\widehat{\Upsilon}^{(1)}_{\mathrm{M},y}(t,z^{\,}_{1})
\nonumber\\
\:=&\,
\sqrt{\mathcal{N}^{(1)}_{\mathrm{M}}}\times
\bm{:}e^{-\mathrm i\frac{1}{\sqrt{2}}\widehat{\phi}^{\,}_{\mathrm{M},y}(t,z^{\,}_{2})}\bm{:}\,
\bm{:}e^{+\mathrm i\frac{1}{\sqrt{2}}\widehat{\phi}^{\,}_{\mathrm{M},y}(t,z^{\,}_{1})}\bm{:}.
\label{eq: building block 2D string d}
\end{align}
\end{subequations}
The need for the normalizations
$\mathcal{N}^{(\frac{1}{2})}_{\mathrm{LR}}$,
$\mathcal{N}^{(1)}_{\mathrm{LR}}$,
$\mathcal{N}^{(\frac{1}{2})}_{\mathrm{M}}$,
and
$\mathcal{N}^{(1)}_{\mathrm{M}}$
will be explained shortly.
Let $|\Omega\rangle$ denote a ground state and denote with
\begin{subequations}
\begin{equation}
\widehat{\mathcal{H}}^{\,}_{\mathrm{bs}\,y'}(t,z')\:=
\mathrm{i}
\widehat{\psi}^{\,}_{\mathrm{L},y'}(t,z')\,
\widehat{\psi}^{\,}_{\mathrm{R},y'+1}(t,z')\,
\bm{:}
\sin
\left(
\sqrt{\frac{1}{2}}
\left(
\widehat{\phi}^{\,}_{\mathrm{L},y'}(t,z')
-
\widehat{\phi}^{\,}_{\mathrm{R},y'+1}(t,z')
\right)
\right)
\bm{:}
\end{equation}
\end{subequations}
the local interaction when $2\lambda=1$.
We claim that
\begin{subequations}
\begin{align}
&
f^{(\frac{1}{2})}_{y|y'}(t,z|t,z')\:=
\left\langle
\Omega
\left|
\left(
\widehat{\mathcal{H}}^{\,}_{\mathrm{bs}\,y'}(t,z')
-
\widehat{\mathcal{O}}^{(\frac{1}{2})\dag}_{y}(t,z)\,
\widehat{\mathcal{H}}^{\,}_{\mathrm{bs}\,y'}(t,z')\,
\widehat{\mathcal{O}}^{(\frac{1}{2})}_{y}(t,z)
\right)
\right|
\Omega
\right\rangle,
\\
&
f^{(1)}_{y|y'}(t,z|t,z')\:=
\left\langle
\Omega
\left|
\left(
\widehat{\mathcal{H}}^{\,}_{\mathrm{bs}\,y'}(t,z')
-
\widehat{\mathcal{O}}^{(1)\dag}_{y}(t,z)\,
\widehat{\mathcal{H}}^{\,}_{\mathrm{bs}\,y'}(t,z')\,
\widehat{\mathcal{O}}^{(1)}_{y}(t,z)
\right)
\right|
\Omega
\right\rangle,
\\
&
f^{(\frac{1}{2})}_{\mathrm{M},y|y'}(t,z^{\,}_{1},z^{\,}_{2}|t,z')\:=
\left\langle
\Omega
\left|
\left(
\widehat{\mathcal{H}}^{\,}_{\mathrm{bs}\,y'}(t,z')
-
\widehat{\mathcal{O}}^{(\frac{1}{2})\dag}_{\mathrm{M},y}(t,z^{\,}_{1},z^{\,}_{2})\,
\widehat{\mathcal{H}}^{\,}_{\mathrm{bs}\,y'}(t,z')\,
\widehat{\mathcal{O}}^{(\frac{1}{2})}_{\mathrm{M},y}(t,z^{\,}_{1},z^{\,}_{2})
\right)
\right|
\Omega
\right\rangle,
\\
&
f^{(1)}_{\mathrm{M},y|y'}(t,z^{\,}_{1},z^{\,}_{2}|t,z')\:=
\left\langle
\Omega
\left|
\left(
\widehat{\mathcal{H}}^{\,}_{\mathrm{bs}\,y'}(t,z')
-
\widehat{\mathcal{O}}^{(1)\dag}_{\mathrm{M},y}(t,z^{\,}_{1},z^{\,}_{2})\,
\widehat{\mathcal{H}}^{\,}_{\mathrm{bs}\,y'}(t,z')\,
\widehat{\mathcal{O}}^{(1)}_{\mathrm{M},y}(t,z^{\,}_{1},z^{\,}_{2})
\right)
\right|
\Omega
\right\rangle.
\end{align}
\end{subequations}
are sharply peaked about $z-z'=0$, $z^{\,}_{1,2}-z'=0$, and $y-y'=0$. This is so because
the equal-time algebra
\begin{equation}
\left[
\widehat{\phi}^{\,}_{\mathrm{M},y}(t,z),
\widehat{\phi}^{\,}_{\mathrm{M}^{\prime},y^{\prime}}(t,z')
\right]=
-\mathrm{i}\,
2\pi
\left[ 
(-1)^{\mathrm{M}}\,
\delta^{\,}_{y,y^{\prime}}\,
\delta^{\,}_{\mathrm{M},\mathrm{M}^{\prime}}\, 
\mathrm{sgn}(z-z')
+
\delta^{\,}_{y,y^{\prime}}\,
\epsilon^{\,}_{\mathrm{M},\mathrm{M}^{\prime}}
-
\mathrm{sgn}(y-y^{\prime})
\right]
\end{equation}
and the equal-time algebra
\begin{equation}
\widehat{\psi}^{\,}_{\mathrm{M},y}(t,z)\,
\widehat{\sigma}^{\,}_{\mathrm{M}^{\prime},y^{\prime}}(t,z')=
\widehat{\sigma}^{\,}_{\mathrm{M}^{\prime},y^{\prime}}(t,z')\,
\widehat{\psi}^{\,}_{\mathrm{M},y}(t,z)\,
e^{
+
\mathrm{i}\,
\pi\,\Theta(z-z')\,
(-1)^{\mathrm{M}}\,\delta^{\,}_{\mathrm{M},\mathrm{M}'}\,\delta^{\,}_{y,y^{\prime}}
  }
\end{equation}
imply that:
\end{widetext}

\noindent (i)
Passing
$\widehat{\mathcal{O}}^{(s)}_{y}(t,z)$ with either $s=1/2$ or $s=1$
through $\widehat{\mathcal{H}}^{\,}_{\mathrm{bs}\,y'}(t,z')$
from the left
creates an infinitely sharp soliton centered at $z$ for
$
\sqrt{1/2}\,\widehat{\phi}^{\,}_{\mathrm{L},y'}(t,z')
$
when $y'=y$ or an infinitely sharp soliton centered at $z$ 
for
$
\sqrt{1/2}\,\widehat{\phi}^{\,}_{\mathrm{R},y'+1}(t,z')
$
when $y'+1=y$.

\noindent (ii)
Passing
$\widehat{\mathcal{O}}^{(s)}_{\mathrm{M},y}(t,z^{\,}_{1},z^{\,}_{2})$
with either $s=1/2$ or $s=1$
through $\widehat{\mathcal{H}}^{\,}_{\mathrm{bs}\,y'}(t,z')$
from the left
creates a pair of infinitely sharp
soliton and antisoliton
centered at $z^{\,}_{1}$ and $z^{\,}_{2}$, respectively.

The normalizations
$\mathcal{N}^{(\frac{1}{2})}_{\mathrm{LR}}$,
$\mathcal{N}^{(1)}_{\mathrm{LR}}$,
$\mathcal{N}^{(\frac{1}{2})}_{\mathrm{M}}$,
and
$\mathcal{N}^{(1)}_{\mathrm{M}}$
are then chosen such that,
upon point splitting,
the operator product expansion of
$
\widehat{\mathcal{O}}^{(s)\dag}_{y}(t,z')\,
\widehat{\mathcal{O}}^{(s)}_{y}(t,z)
$
and
$
\widehat{\mathcal{O}}^{(s)\dag}_{\mathrm{M},y}(t,z^{\prime}_{1},z^{\prime}_{2})\,
\widehat{\mathcal{O}}^{(s)}_{\mathrm{M},y}(t,z^{\,}_{1},z^{\,}_{2})
$
with $s=1/2,1$
deliver the identity operator.

As a result, the operators $\widehat{\mathcal{O}}^{(s)}_{y}(t,z)$
and
$\widehat{\mathcal{O}}^{(s)}_{\mathrm{M},y}(t,z^{\,}_{1},z^{\,}_{2})$
create an energy density with compact support
for both $s=1/2$ and $s=1$.

Creating a single soliton in the Sine-Gordon model costs an energy
that depends on the ratio of potential to kinetic energy. This is the core
energy of the soliton. The width of the soliton is inversely
proportional to the  ratio of potential to kinetic energy.
Hence, in the limit for which the ratio of potential to kinetic energy
diverges, the soliton becomes infinitely sharp while its core energy
diverges. The same is true here, i.e., the four states
$
\widehat{\mathcal{O}}^{(s)}_{y}(t,z)
\left|\Omega\right\rangle
$
and
$
\widehat{\mathcal{O}}^{(s)}_{\mathrm{M},y}(t,z^{\,}_{1},z^{\,}_{2})
\left|\Omega\right\rangle
$
with $s=1/2,1$ are infinitely heavy point-like excitations
in the limit of infinitely strong interaction strength.
Now, the energy cost to opening any one of the four strings
$\widehat{\Gamma}^{(s)}_{1}$,
$\widehat{\Gamma}^{(s)}_{2}$
with $s=1/2,1$
is nothing but the core energy of solitons localized
at either ends of the open strings, i.e., twice the core energies
of the states
$
\widehat{\mathcal{O}}^{(s)}_{y}(t,z)
\left|\Omega\right\rangle
$
and
$
\widehat{\mathcal{O}}^{(s)}_{\mathrm{M},y}(t,z^{\,}_{1},z^{\,}_{2})
\left|\Omega\right\rangle
$
with $s=1/2,1$, respectively.
In the limit for which the ratio of potential to kinetic energy
diverges, the solitons localized at the ends of open strings
are deconfined, although infinitely heavy.
A perturbative treatment of the kinetic energy relative to the
potential energy results in a small decrease of the soliton core energy
and a small string tension. Confinement of the solitons is necessarily
non-perturbative in terms of the ratio of kinetic to potential energy.

\section{Challenges for extensions to 3D}
\label{sec: Challenges for extensions to three-dimensional space}

In Sec.~\ref{sec: Warm-up: Non-Abelian topological order in two dimensions}
we revisited the construction of (2+1)-dimensional non-Abelian topological phases from coupled wires. This discussion advances prior work on the
subject---e.g., in Refs.~\onlinecite{Teo14,Huang16a}---by
providing a methodology for the characterization of such phases using
techniques from conformal field theory.
This methodology will be a crucial ingredient in any
extension of the non-Abelian coupled-wire framework to
(3+1)-dimensional spacetime.

In this section we provide an outlook on the prospects for finding
(3+1)-dimensional generalizations of the $su(2)^{\,}_{k}$ topological
phases constructed in
Sec.~\ref{sec: Warm-up: Non-Abelian topological order in two dimensions}.
We formulate a sharp question informed by the setup studied in
Sec.~\ref{sec: Warm-up: Non-Abelian topological order in two dimensions}:
is it possible to build a gapped
non-Abelian 3D topological phase described by a (3+1)-dimensional
topological quantum field theory using only $su(2)^{\,}_{k}$ CFTs
coupled by bilinear current-current interactions?  We impose the
additional constraint that, like the phases constructed in
Sec.~\ref{sec: Warm-up: Non-Abelian topological order in two dimensions},
the 3D phase in question can be realized starting from
electrons as the fundamental degrees of freedom.  Our conclusion will
be that this question does not appear to have an obvious affirmative
answer.  As we argue below, it seems that the simplest ways of
coupling the constituent $su(2)^{\,}_{k}$ CFTs yield either (1) a
phase that is adiabatically connected to a stack of decoupled 2D
topological phases or (2) a 3D Abelian topological phase.  In case (1)
the phase realized is non-Abelian, but not intrinsically 3D, while in
case (2) the phase realized may be intrinsically 3D, but is not
non-Abelian.  After elaborating on cases (1) and (2) in
Secs.~\ref{subsec: case1} and \ref{subsec: case2}, we will comment in
Sec.~\ref{subsec: workarounds} on possible workarounds for this
problem, the exploration of which we leave for future work.

\subsection{Case (1): Stack of decoupled 2D topological phases}
\label{subsec: case1}

\begin{figure}[t]
\includegraphics[width=\columnwidth]{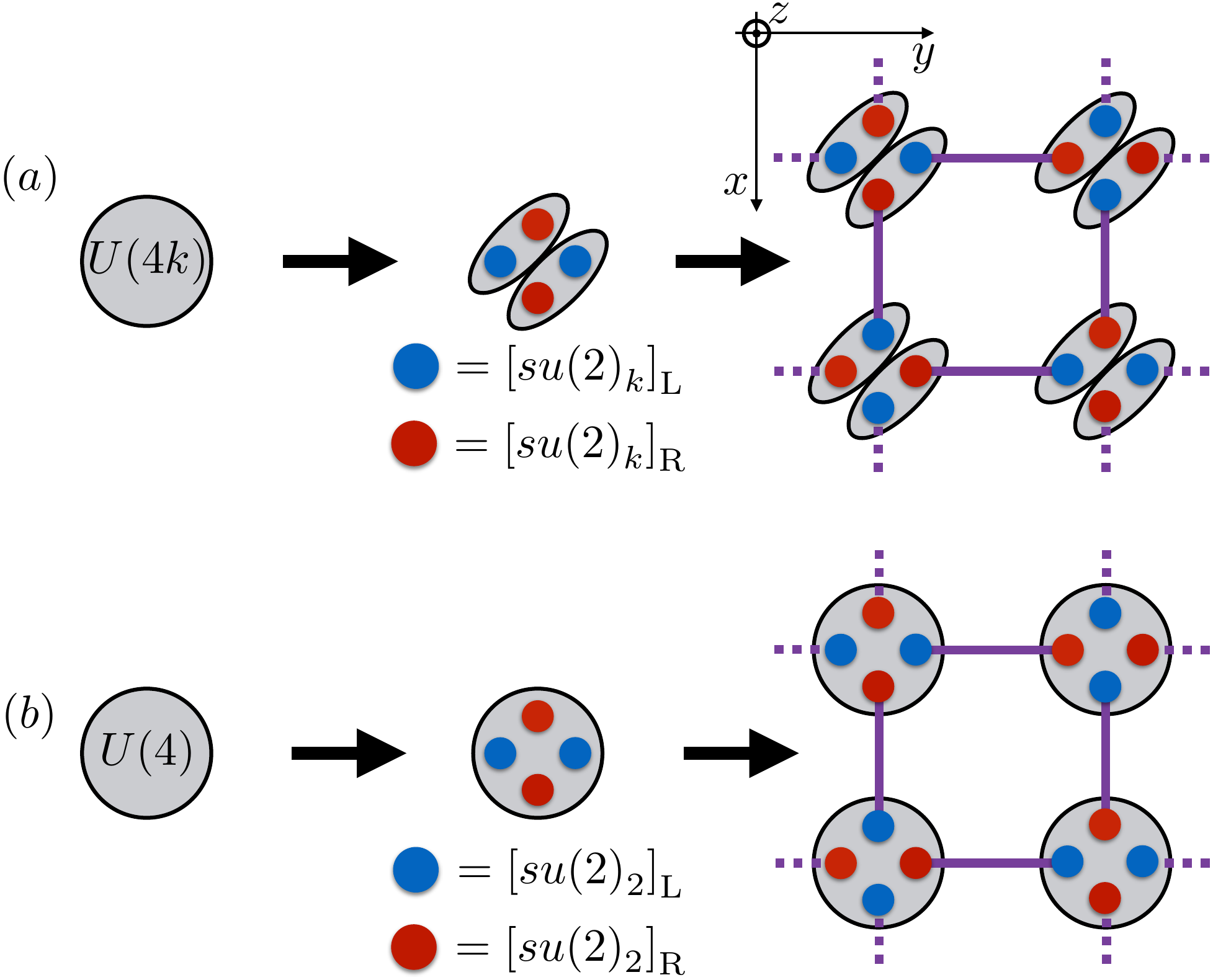}
\caption{
Schematic of two possible generalizations to 3D of the 2D setup
depicted in Fig.~\ref{fig: 2D system}.  (a) The generalization
presented in Sec.~\ref{subsec: case1}.  In this case, the original
wire (gray circle) has a $U(4k)$ symmetry that is broken into two
copies of $U(2k)$, on which the conformal embedding
\eqref{eq: conformal embedding k} is performed. After gapping out the unwanted
gapless sectors, the low-energy theory of a single wire is described
by two decoupled $su(2)^{\,}_k$ CFTs, with each independent copy
depicted as residing within a gray oval.  The remaining low-energy
degrees of freedom are then coupled by interwire current-current
interactions (purple bonds), yielding a 3D phase that is equivalent to
a stack of decoupled 2D topological phases of the type depicted in
Fig.~\ref{fig: 2D system}. (b) An alternative conformal embedding for
$su(2)^{\,}_{2}\oplus su(2)^{\,}_{2}$ into $u(4)^{\,}_{1}$, discussed
in Sec.~\ref{subsec: case2}.  In this case the four chiral
$su(2)^{\,}_{2}$ modes are depicted as residing within a single gray
oval, representing the fact that the two copies of the
$su(2)^{\,}_{2}$ CFT are no longer independent, as discussed in
Sec.~\ref{subsec: case2}.
        }
\label{fig: 3D}
\end{figure}

The simplest way to mimic the construction of
Sec.~\ref{sec: Warm-up: Non-Abelian topological order in two dimensions}
is to generalize the setup depicted in Fig.~\ref{fig: 2D system}
to a set of wires placed on the sites of a square lattice, as depicted in
Fig.~\ref{fig: 3D}.
Since we will ultimately use current-current bilinears to gap
out the array of wires, we need to choose wires that can be broken
into a number of chiral channels that matches the coordination number
4 of the square lattice.  This can be achieved by choosing wires with
an internal symmetry group $U(4k)^{\,}_{\mathrm{L}}\times
U(4k)^{\,}_{\mathrm{R}}$, corresponding to wires of the
form~\eqref{eq: single non-abelian wire} with
$N^{\,}_{\mathrm{c}}=2k$.  To obtain four chiral channels, we consider
couplings symmetric under the diagonal subgroup $U(2k)\times
U(2k)\subset U(4k)$, which effectively breaks each chiral channel
$\mathrm M=\mathrm L,\mathrm R$ into two identical copies.  Then, we
can use the identity
\begin{align}
u(2k)^{\,}_{1}= 
u(1)\oplus 
su(2)^{\,}_{k}\oplus 
su(k)^{\,}_{2}
\label{eq: conformal embedding k}
\end{align}
to define the $\mathrm{M}$-moving chiral currents
$\widehat{j}^{\,}_{\mathrm{M},\gamma}$, $\widehat{J}^{a}_{\mathrm{M},\gamma}$, and 
$\widehat{\mathsf{J}}^{\mathsf{a}}_{\mathrm{M},\gamma}$, where $\gamma=1,2$ labels
the two copies.
These chiral currents, which are given by 
Eqs.\ \eqref{eq: non-Abelian currents} 
with the substitution $N^{\,}_{\mathrm{c}}\to k$,
correspond to the $u(1)$, $su(2)^{\,}_{k}$, and
$su(k)^{\,}_{2}$ sectors, respectively.
We then proceed as in Sec.\
\ref{subsec: Definition of the class of models 2D}.
Namely, we gap out the $u(1)$ and $su(k)^{\,}_{2}$ degrees of freedom
by turning on intrawire interactions of the 
form~\eqref{eq: umklapp} and \eqref{eq: su(N_c) current-current},
respectively, for each $\gamma=1,2$.  We then gap out the
remaining $su(2)^{\,}_{k}$ channels using bilinear current-current interactions
between the wires. This setup is depicted schematically
in Fig.~\ref{fig: 3D}.

While the above scheme yields a gapped state of matter for the same reasons as the
construction presented in Sec.~\ref{sec: Warm-up: Non-Abelian topological order in two dimensions},
this state of matter is not intrinsically 3D.  Rather, it can be described as an array of decoupled 2D
topological phases of the type constructed in
Sec.~\ref{sec: Warm-up: Non-Abelian topological order in two dimensions}.  This fact originates from
the splitting of the degrees of freedom in the original wire into two groups, $U(4k)\supset U(2k)\times U(2k)$, where
the latter two copies of $U(2k)$ are associated with the index $\gamma=1,2$.
By splitting up the wire in this way, one imposes the constraint that any local operator in either of the two $su(2)^{\,}_{2}$ CFTs
originating from the conformal embedding \eqref{eq: conformal embedding k} must be defined exclusively
within the $\gamma=1$ or $\gamma=2$ sector.  This constraint is represented pictorially in Fig.~\ref{fig: 3D}
by the splitting of a gray circle (representing the original wire) into two gray ovals (representing the two channels $\gamma=1,2$).
Thus, the array of coupled wires reduces to a set of decoupled planes, each of which is represented by
a coupled-wire theory of the type constructed in Sec.~\ref{sec: Warm-up: Non-Abelian topological order in two dimensions}.
When the coupled-wire array is defined on a three-torus (i.e., when
periodic boundary conditions are imposed in all directions), each
plane becomes a two-torus that contributes a factor of $k+1$ to the
topological degeneracy, where $k+1$ is the degeneracy of the 2D
$su(2)^{\,}_{k}$ topological phase realized by the class of models
defined in
Sec.~\ref{sec: Warm-up: Non-Abelian topological order in two dimensions}.
The total topological degeneracy is then
$(k+1)^{p}$, where $p$ is the number of independent planes in the
array when periodic boundary conditions are imposed in all directions.

\subsection{Case (2): Reduction to an Abelian phase}
\label{subsec: case2} 

One way to avoid reducing the coupled-wire array to a stack of
decoupled 2D topological phases is to use a different conformal
embedding.  To illustrate this, let us take $su(2)^{\,}_{2}$ as our
working example.  Rather than starting from an array of wires with an
internal $U(8)=U(4\times 2)$ symmetry, as outlined in
Sec.~\ref{subsec: case1}, we will use wires with a $U(4)$ internal
symmetry and the conformal embedding
\begin{align}\label{eq: u(4)_{1} conformal embedding}
u(4)^{\,}_{1}=u(1)\oplus su(2)^{\,}_{2} \oplus su(2)^{\,}_{2}.
\end{align}
At first glance, the low-energy theory obtained from the above conformal embedding
after gapping the $u(1)$ sector is very similar to the one obtained
in Sec.~\ref{subsec: case1} by using the conformal embedding \eqref{eq: conformal embedding k}
on the two copies of the $u(2k)_{1}$ theory with $k=2$.
Namely, in both cases the low-energy theory is described by the affine Lie algebra $su(2)^{\,}_{2} \oplus su(2)^{\,}_{2}$.
Indeed, in both cases one can also obtain a fully gapped state of matter by adding current-current interactions on
the bonds of the square lattice (see Fig.~\ref{fig: 3D}).

However, there is a very important physical difference between the theory arising from the conformal embedding
\eqref{eq: conformal embedding k} and the one arising from the conformal embedding \eqref{eq: u(4)_{1} conformal embedding}.
In the former case, the $su(2)^{\,}_{2} \oplus su(2)^{\,}_{2}$ algebra is embedded in a $u(4)_{1}\oplus u(4)_{1}$ algebra,
so that each copy of $su(2)^{\,}_{2}$ comes from an independent copy of $u(4)_{1}$.  However in the latter case,
which is of interest to us here, \textit{both} copies of $su(2)^{\,}_{2}$ come from the \textit{same} copy of $u(4)_{1}$.
Thus, in the latter case, the two copies of $su(2)^{\,}_{2}$ are not independent but instead nontrivially intertwined.
At a technical level, the difference between these two theories is that they have different partition functions. The partition
function in the case of the conformal embedding $su(2)^{\,}_{2} \oplus su(2)^{\,}_{2}\subset u(4)_{1}\oplus u(4)_{1}$
is known as the ``diagonal" partition function, whereas the partition function in the case of the conformal embedding
$su(2)^{\,}_{2} \oplus su(2)^{\,}_{2}\subset u(4)_{1}$
is known as the ``off-diagonal" partition function. For explicit expressions for these partition functions, we refer
the reader to Sec.~2.2 of Ref.~\onlinecite{Lahtinen17}, which treats the case of two copies of the Ising CFT. The partition
functions appearing there can be translated to the $su(2)^{\,}_{2}$ setting simply by making the substitutions
$(1,\sigma,\psi)\to(0,\frac{1}{2},1)$, where the symbols on the left hand side label primary fields for the Ising CFT
and the symbols on the right-hand side label the primary fields of the $su(2)^{\,}_{2}$ CFT as in Sec.~\ref{subsec: twist operators}.
The off-diagonal partition function for the Ising case is associated with the conformal embedding of two Ising CFTs into a $u(1)^{\,}_4$ CFT,
which is directly analogous to the conformal embedding $su(2)^{\,}_{2} \oplus su(2)^{\,}_{2}\subset u(4)^{\,}_{1}$.
[That $u(1)^{\,}_4$ is relevant to the Ising case while $u(4)^{\,}_{1}$
is relevant to the $su(2)^{\,}_{2}$ case is a consequence of the fact
that the central charge of the Ising~$\times$~Ising CFT is $1$,
matching the central charge of $u(1)^{\,}_4$, while the central charge
of the $su(2)^{\,}_{2} \oplus su(2)^{\,}_{2}$ CFT is $3$, matching the
central charge of $u(4)^{\,}_{1}/u(1)$.]

As explained in Refs.~\onlinecite{Lahtinen17} and
\onlinecite{Neupert16}, the off-diagonal partition function associated
with the conformal embedding of two Ising CFTs into a $u(1)^{\,}_4$
CFT has an interpretation in terms of anyon condensation~\cite{Bais09}
in the Ising~$\times$~Ising topological quantum field theory. (This
interpretation is a specific instance of the more general
correspondence laid out in Ref.~\onlinecite{Bais09} between anyon
condensation and conformal embeddings.) Using this correspondence, one
can argue as is done in Refs.~\onlinecite{Bais09,Neupert16,Lahtinen17}
that the theory described by the off-diagonal partition function is in
fact Abelian (as it should be, since the primaries of the
$u(1)^{\,}_{4}$ CFT have Abelian fusion rules), even though the
underlying Ising~$\times$~Ising theory is non-Abelian.  This reduction
of the non-Abelian theory to an Abelian one arises from constraints
imposed by the branching rules of the conformal embedding, which
ensure that the two copies of the Ising theory are not independent
when embedded within $u(1)^{\,}_4$.  This argument can be directly
extended to the case of the affine Lie algebra $su(2)^{\,}_{2} \oplus
su(2)^{\,}_{2}$ considered here if we replace $u(1)^{\,}_{4}$ by
$u(4)^{\,}_{1}$. Indeed, we have made extensive use in this paper
of the fact that the $su(2)^{\,}_{2}$ CFT is the tensor product of the
Ising CFT and the $u(1)^{\,}_{2}$ CFT. We are thus led to the
conclusion that the gapped phase obtained from the coupled-wire theory
based on the conformal embedding
\eqref{eq: u(4)_{1} conformal embedding}
is Abelian, despite the fact that non-Abelian CFTs were
used in its construction.

\subsection{Possible workarounds}
\label{subsec: workarounds}

In order to circumvent the outcomes discussed in
Secs.~\ref{subsec: case1} and \ref{subsec: case2},
one must go beyond the approach used in this paper.
We now suggest two possible workarounds that could
allow one to construct topological phases that are both (1)
intrinsically 3D (i.e., not adiabatically connected to a stack of
decoupled 2D topological phases) and (2) support non-Abelian pointlike
or stringlike excitations. The approaches we suggest might yield 3D
phases described by topological quantum field theories in
(3+1)-dimensional spacetime, or could yield phases that, like fracton phases%
~\cite{Chamon05,Bravyi11,Haah11,Castelnovo12,Yoshida13,Vijay15,Song18,Prem18},
evade a purely topological field-theoretic description.

\subsubsection{Adding additional intrawire interactions}
\label{subsubsec: Adding additional intrawire interactions}

One approach worth exploring further involves starting from a stack of
decoupled 2D topological phases realized using the conformal embedding
procedure described in Sec.~\ref{subsec: case1}, and then adding
additional intrawire interactions within the $su(2)^{\,}_{k}\oplus
su(2)^{\,}_{k}$ sector.  Adding such intrawire interactions would
amount to adding couplings between the previously decoupled 2D planes.
These couplings should be chosen such that they would not fully gap
out the $su(2)^{\,}_{k}\oplus su(2)^{\,}_{k}$ sector of the array of
quantum wires if their strength was taken to be much larger than
the interwire couplings
(if they were not chosen in this way, then
the resulting phase would be adiabatically connected to a set of
individually gapped, decoupled wires, rendering it topologically
trivial). One class of intrawire interactions that might satisfy
this condition would be a set of interactions that drive an anyon
condensation transition in the case of an isolated bilayer of
$su(2)^{\,}_{k}$ topological phases (which can be viewed as a single
2D system). If the 2D condensation transition driven by these
interactions yields another gapped non-Abelian topological phase, then
there is hope that the 3D phase obtained by coupling more than two
layers would also be non-Abelian.  This approach would make contact
with the coupled-layer construction developed in
Ref.~\onlinecite{Jian14} for Abelian topological phases (see also
Ref.~\onlinecite{Fuji19}).

In order to move in this direction, it will be important to develop a
detailed understanding of how anyon condensation can be implemented in
coupled-wire constructions of 2D topological phases by adding
appropriate intrawire interactions.  This direction is, to our
knowledge, as yet unexplored.  The possible applications to 3D
topological phases mentioned above provide substantial motivation for
such a study.

\subsubsection{Moving beyond bilinear current-current interactions}
\label{subsubsec: Moving beyond bilinear current-current interactions}

Another aspect of our approach that hampers generalizations to 3D is
the fact that we restricted our attention to interwire interactions
that are simple bilinears of currents in neighboring wires, as in
Eq.~\eqref{eq2Dcase: 2D interaction}.  Such interactions have the
advantage of being both mutually commuting and marginally relevant
under RG, and thus they can always be used to open a gap in an array
of coupled wires.  Furthermore, such bilinear interactions are the
most natural ones to use in coupled-wire constructions of 2D
topological phases because they can be viewed as dimerizing a 1D
cross-section of the 2D system (see Fig.~\ref{fig: 2D system}).  The
3D case is more complicated, however. If we generalize the 1D setup in
such a way that a 2D cross-section of the wire array looks like a 2D
lattice (see Fig.~\ref{fig: 3D}), there are many other kinds of
couplings one could imagine adding.  For example, one could use
couplings defined on plaquettes of the 2D lattice that are
\textit{products} of current-current bilinears.  Combining the
plethora of (1+1)-dimensional CFTs with the richer set of 2D lattices
yields a large space of possible 3D coupled-wire theories that has not
yet been explored.

Interactions that cannot be written as bilinears of currents have
already been explored in the context of coupled-wire constructions of
Abelian topological phases in 3D
(see, e.g., Ref.~\onlinecite{Iadecola16}).  In
the non-Abelian case, such interactions suffer from the fact that they
are irrelevant under RG as they involve more than two currents.  This
does not exclude the possibility of using such interactions to open a
gap, however, as one can simply treat the system at strong coupling.
However, such a treatment necessitates the use of nonperturbative
techniques to verify that a gap indeed opens at strong coupling.  This
was done for the Abelian case in Ref.~\onlinecite{Iadecola16}, but the
extension to non-Abelian phases is not obvious.  We plan to explore
the possibility of extending the methods of
Ref.~\onlinecite{Iadecola16} to the non-Abelian case in future work.

\section{Conclusions}
\label{sec: Conclusions}

In this paper, we studied coupled-wire realizations of $su(2)^{\,}_{k}$
topological phases in two spatial dimensions.
These phases inherit their non-Abelian
character from the underlying $su(2)^{\,}_{k}$ CFTs that describe the
constituent interacting fermionic quantum wires in the decoupled limit.
For the special case of $su(2)^{\,}_{2}$, we showed explicitly how to 
construct a set of nonlocal operators that can be used to label a set of
degenerate ground states and to cycle between states in this set,
thus demonstrating how the expected threefold degeneracy arises
in a coupled-wire construction. This calculation relies on the operator
algebra of the underlying CFTs that furnish the low-energy degrees of
freedom for the coupled-wire construction, thus making explicit the
connection between these CFTs and the emergent topological phase.

There are a number of open directions for the study of coupled-wire
constructions that are worth exploring further.  One natural question
is how to extend the methods developed in this paper for calculating
topological degeneracies to the class of 2D topological phases
constructed, e.g., in Refs.~\onlinecite{Teo14,Huang16a}
whose edge states are described by coset conformal field theories.
Another interesting
question raised in
Sec.~\ref{sec: Challenges for extensions to three-dimensional space}
concerns how to describe anyon condensation transitions~\cite{Bais09}
within the coupled-wire framework.  As pointed out in
Sec.~\ref{subsubsec: Adding additional intrawire interactions}, answering
this question could provide a useful path forward for defining interesting
non-Abelian coupled-wire models in 3D.  A related direction of interest
is to study how the gauging of anyonic symmetries in 2D topological
order~\cite{Teo15} can be implemented at the level of coupled-wire
constructions. This gauging procedure is related to the orbifold construction
in CFT~\cite{DiFrancesco97,Ginsparg88,Dijkgraaf89,Moore89b}, which
has been investigated in the context of coupled-wire constructions in
Ref.~\onlinecite{Kane18}.  A final direction worth pursuing is to investigate
whether more complicated current-current interactions like those suggested
in Sec.~\ref{subsubsec: Moving beyond bilinear current-current interactions}
could be used to develop new non-Abelian topological phases in 3D.

\begin{acknowledgments}
We thank D.~Aasen, M.~Barkeshli, P.~Bonderson, F.~Burnell, M.~Cheng,
M.~Metlitski, M.~Oshikawa, Z.~Wang, X.-G.~Wen, and D.~Williamson for
helpful discussions.  T.I. gratefully acknowledges the hospitality of
the KITP, where a significant portion of this work was completed, and
thanks the organizers of the ``Symmetry, Topology, and Quantum Phases
of Matter: From Tensor Networks to Physical Realizations'' and
``Synthetic Quantum Matter'' programs, which were supported by NSF
under Grant No.~NSF PHY11-25915.
Finally, we have benefited from the critical reading of one
heroic referee who painstakingly pointed out inconsistencies in earlier versions
of Sec.\ \ref{sec: Challenges for extensions to three-dimensional space}.
T.I. was supported by a KITP Graduate Fellowship,
the National Science Foundation Graduate Research Fellowship Program
under Grant No.~DGE-1247312, the Laboratory for Physical Sciences,
Microsoft, and a JQI Postdoctoral Fellowship.
C.C. was supported by DOE Grant DE-FG02-06ER46316.
\end{acknowledgments}

\appendix

\section{The parafermion current algebra}
\label{appsec: The parafermion current algebra}

We are going to review how
the affine Lie algebra of level $k=1,2,3,\cdots$
for the compact connected Lie group $SU(2)$
can be represented in terms of parafermions
as was done by  Zamolodchikov and Fateev
in Ref.~\onlinecite{Zamolodchikov85}.

\subsection{Gaussian algebra}
\label{appsubsec: Gaussian algebra}

For any $\kappa>0$,
define the Euclidean action 
\begin{equation}
S\:=
\frac{\kappa}{2}
\int\mathrm{d}^{2}\bm{x}\,
(\bm{\partial}\varphi)^{2}
\end{equation}
for the real-valued scalar field $\varphi$
and the positive number $0<\kappa\in\mathbb{R}$.
Its two-point function is
\begin{equation}
\langle\varphi(\bm{x})\,\varphi(\bm{y})\rangle=
-
\frac{1}{4\pi\kappa}\,
\ln|\bm{x}-\bm{y}|^{2}
\end{equation}
up to an additive dimensionful constant that depends on the boundary 
condition imposed on the Laplacian.
If we trade the complex coordinates $v\in\mathbb{C}$ and $w\in\mathbb{C}$ 
in two-dimensional Euclidean space for the Cartesian coordinates
$\bm{x}\in\mathbb{R}^{2}$ and $\bm{y}\in\mathbb{R}^{2}$, respectively,
then
\begin{equation}
|\bm{x}-\bm{y}|^{2}=
(v-w)\,(\overline{v}-\overline{w})
\end{equation}
and
\begin{subequations}
\begin{align}
&
\langle\varphi(\bm{x})\,\varphi(\bm{y})\rangle=
-
\frac{1}{4\pi\kappa}\,
\left[
\log(v-w)
+
\log(\overline{v}-\overline{w})
\right],
\\
&
\langle\partial^{\,}_{v}\varphi(\bm{x})\,\varphi(\bm{y})\rangle=
-
\frac{1}{4\pi\kappa}\,
\frac{1}{(v-w)},
\\
&
\langle
\partial^{\,}_{v}\varphi(\bm{x})\,
\partial^{\,}_{w}\varphi(\bm{y})
\rangle=
-
\frac{1}{4\pi\kappa}\,
\frac{1}{(v-w)^{2}},
\\
&
\langle
\partial^{\,}_{\overline{v}}\varphi(\bm{x})\,
\varphi(\bm{y})
\rangle=
-
\frac{1}{4\pi\kappa}\,
\frac{1}{(\overline{v}-\overline{w})},
\\
&
\langle
\partial^{\,}_{\overline{v}}\varphi(\bm{x})\,
\partial^{\,}_{\overline{w}}\varphi(\bm{y})
\rangle=
-
\frac{1}{4\pi\kappa}\,
\frac{1}{(\overline{v}-\overline{w})^{2}}.
\end{align}
\end{subequations}
There follows the chiral Abelian OPEs
\begin{subequations}
\begin{align}
&
\partial^{\,}_{v}\varphi(\bm{x})\,
\varphi(\bm{y})=
-
\frac{1}{4\pi\kappa}\,
\frac{1}{(v-w)}
+
\cdots,
\\
&
\partial^{\,}_{v}\varphi(\bm{x})\,
\partial^{\,}_{w}\varphi(\bm{y})=
-
\frac{1}{4\pi\kappa}\,
\frac{1}{(v-w)^{2}}
+
\cdots,
\\
&
\partial^{\,}_{\overline{v}}\varphi(\bm{x})\,
\varphi(\bm{y})=
-
\frac{1}{4\pi\kappa}\,
\frac{1}{(\overline{v}-\overline{w})}
+
\cdots,
\\
&
\partial^{\,}_{\overline{v}}\varphi(\bm{x})\,
\partial^{\,}_{\overline{w}}\varphi(\bm{y})=
-
\frac{1}{4\pi\kappa}\,
\frac{1}{(\overline{v}-\overline{w})^{2}}
+
\cdots,
\\
&
\partial^{\,}_{v}\varphi(\bm{x})\,
\partial^{\,}_{\overline{w}}\varphi(\bm{y})=
0.
\end{align}
\end{subequations}
The conformal weights of the field $\partial^{\,}_{v}\phi$ are 
\begin{equation}
(\Delta^{\,}_{\partial^{\,}_{v}\phi},\overline{\Delta}^{\,}_{\partial^{\,}_{v}\phi})=(1,0).
\end{equation}

Another set of chiral Abelian OPEs follows from making the Ansatz
\begin{subequations}
\begin{align}
&
\varphi(v,\overline{v})\=:
\phi^{\,}_{\mathrm{L}}(v)
+
\phi^{\,}_{\mathrm{R}}(\overline{v}),
\\
&
\left\langle
\partial^{\,}_{v}\phi^{\,}_{\mathrm{L}}(v)\,
\phi^{\,}_{\mathrm{L}}(w)
\right\rangle=
-
\frac{1}{4\pi\,\kappa}
\frac{1}{v-w},
\\
&
\left\langle
\partial^{\,}_{v}\phi^{\,}_{\mathrm{R}}(\overline{v})\,
\phi^{\,}_{\mathrm{R}}(\overline{w})
\right\rangle=
-
\frac{1}{4\pi\,\kappa}
\frac{1}{\overline{v}-\overline{w}},
\\
&
\left\langle
\phi^{\,}_{\mathrm{R}}(v)\,
\phi^{\,}_{\mathrm{L}}(\overline{w})
\right\rangle=
0.
\end{align}
\end{subequations}
The holomorphic, $\phi^{\,}_{\mathrm{L}}$, and antiholomorphic, 
$\phi^{\,}_{\mathrm{R}}$, fields are uniquely defined up to
the addition of holomorphic and antiholomorphic functions,
respectively.
One then deduces from
\begin{subequations} 
\begin{align}
&
\left\langle
e^{+\mathrm{i}a\,\phi^{\,}_{\mathrm{L}}(v)}\,
e^{-\mathrm{i}a\,\phi^{\,}_{\mathrm{L}}(w)}
\right\rangle=
\frac{1}{(v-w)^{\frac{a^{2}}{4\pi\kappa}}},
\\
&
\left\langle
e^{+\mathrm{i}a\,\phi^{\,}_{\mathrm{L}}(v)}\,
e^{+\mathrm{i}a\,\phi^{\,}_{\mathrm{L}}(w)}
\right\rangle=
0,
\\
&
\left\langle
e^{+\mathrm{i}a\,\phi^{\,}_{\mathrm{R}}(\overline{v})}\,
e^{-\mathrm{i}a\,\phi^{\,}_{\mathrm{R}}(\overline{w})}
\right\rangle=
\frac{1}{(\overline{v}-\overline{w})^{\frac{a^{2}}{4\pi\kappa}}},
\\
&
\left\langle
e^{+\mathrm{i}a\,\phi^{\,}_{\mathrm{R}}(\overline{v})}\,
e^{+\mathrm{i}a\,\phi^{\,}_{\mathrm{R}}(\overline{w})}
\right\rangle=
0,
\\
&
\left\langle
e^{\pm\mathrm{i}a\,\phi^{\,}_{\mathrm{L}}(v)}\,
e^{\pm\mathrm{i}a\,\phi^{\,}_{\mathrm{R}}(\overline{w})}
\right\rangle=
0,
\end{align}
\end{subequations}
that
\begin{subequations}\label{appeq: OPE vertex and antivertex operators}
\begin{align}
&
e^{+\mathrm{i}a\,\phi^{\,}_{\mathrm{L}}(v)}\,
e^{-\mathrm{i}a\,\phi^{\,}_{\mathrm{L}}(w)}=
\frac{1}{(v-w)^{\frac{a^{2}}{4\pi\kappa}}}
\nonumber\\
&
\hphantom{
e^{+\mathrm{i}a\,\phi^{\,}_{\mathrm{L}}(v)}\,
e^{-\mathrm{i}a\,\phi^{\,}_{\mathrm{L}}(w)}=
         }
+
\frac{\mathrm{i}a}{(v-w)^{\frac{a^{2}}{4\pi\kappa}-1}}\,
(\partial^{\,}_{w}\phi^{\,}_{\mathrm{L}})(w)
+
\cdots,
\\
&
e^{+\mathrm{i}a\,\phi^{\,}_{\mathrm{R}}(\overline{v})}\,
e^{-\mathrm{i}a\,\phi^{\,}_{\mathrm{R}}(\overline{w})}=
\frac{1}{(\overline{v}-\overline{w})^{\frac{a^{2}}{4\pi\kappa}}}
\nonumber\\
&
\hphantom{
e^{+\mathrm{i}a\,\phi^{\,}_{\mathrm{L}}(v)}\,
e^{-\mathrm{i}a\,\phi^{\,}_{\mathrm{L}}(w)}=
         }
+
\frac{\mathrm{i}a}{(\overline{z}-\overline{w})^{\frac{a^{2}}{4\pi\kappa}-1}}\,
(\partial^{\,}_{\overline{w}}\phi^{\,}_{\mathrm{R}})(\overline{w})
+
\cdots,
\end{align}
\end{subequations}
are the only chiral Abelian OPEs between
the vertex fields
$e^{\pm\mathrm{i}a\,\phi^{\,}_{\mathrm{L}}(v)}$
and 
$e^{\pm\mathrm{i}a\,\phi^{\,}_{\mathrm{R}}(\overline{v})}$
that are proportional to the identity operator to leading order.

At last, we shall need the OPEs
\begin{subequations}\label{appeq: OPE partial phi with vertex operator}
\begin{align}
&
\partial^{\,}_{v}\phi^{\,}_{\mathrm{L}}(v)\,
e^{+\mathrm{i}a\phi^{\,}_{\mathrm{L}}(w)}=
-
\frac{\mathrm{i}a}{4\pi\kappa}
\frac{1}{(v-w)}\,
e^{+\mathrm{i}a\phi^{\,}_{\mathrm{L}}(w)}
+
\cdots,
\\
&
\partial^{\,}_{\overline{v}}\phi^{\,}_{\mathrm{R}}(\overline{v})\,
e^{+\mathrm{i}a\phi^{\,}_{\mathrm{R}}(\overline{w})}=
-
\frac{\mathrm{i}a}{4\pi\kappa}
\frac{1}{(\overline{v}-\overline{w})}\,
e^{+\mathrm{i}a\phi^{\,}_{\mathrm{R}}(\overline{w})}
+
\cdots.
\end{align}
\end{subequations}

In the following, we make the choice
\begin{equation}
\kappa=\frac{1}{8\pi}.
\end{equation}
With this choice, the conformal weights of the vertex fields
$\exp\big(\mathrm{i}a\phi^{\,}_{\mathrm{L}}\big)$
and $\exp\big(\mathrm{i}a\phi^{\,}_{\mathrm{R}}\big)$
are
\begin{equation}
(\Delta^{\,}_{a},\overline{\Delta}^{\,}_{a})\equiv (a^{2},0),
\qquad
(\Delta^{\,}_{\overline{a}},\overline{\Delta}^{\,}_{\overline{a}})\equiv (0,a^{2}),
\end{equation}
respectively.
Moreover, the proportionality constant
on the right-hand side of
Eq.~(\ref{appeq: OPE partial phi with vertex operator})
is $-2a\mathrm{i}$.

\subsection{Parafermion algebra}
\label{appsubsec: Parafermion algebra}

Let $k=0,1,2,\cdots$ be a positive integer.
Define the holomorphic conformal weights
\begin{subequations}
\label{appeq: parafermions OPEs}
\begin{equation}
\label{appeq: parafermions OPEs a}
\Delta^{\,}_{l}\:=
\frac{l(k-l)}{k},
\qquad l=0,\cdots,k-1.
\end{equation}
We posit the family of $k$ local parafermion fields
\begin{equation}\label{appeq: parafermions OPEs b}
I,
\Psi^{\,}_{1}(v),
\cdots,
\Psi^{\,}_{k-1}(v),
\end{equation}
where $I$ is the identity operator with the
conformal weights
\begin{equation}
(\Delta^{\,}_{I},\overline{\Delta}^{\,}_{I})\equiv
(\Delta^{\,}_{0},\overline{\Delta}^{\,}_{0})=(0,0).
\end{equation}
For any $m,n=0,\cdots,k-1$, we impose the OPEs
\cite{Zamolodchikov85}
\begin{equation}\label{appeq: parafermions OPEs c}
\begin{split}
\Psi^{\,}_{m}(v)\,
\Psi^{\,}_{n}(v')=&\,
\frac{
C^{\Psi^{\,}_{m+n}}_{\Psi^{\,}_{m}\Psi^{\,}_{n}}\,
\Psi^{\,}_{m+n}(v')
     }
     {
(v-v')^{\Delta^{\,}_{m}+\Delta^{\,}_{n}-\Delta^{\,}_{m+n}}
     }
+
\cdots
\end{split}
\end{equation}
with the understanding that $m+n$ is defined modulo $k$, i.e.,
\begin{equation}\label{appeq: parafermions OPEs d}
\Psi^{\,}_{0}\equiv\Psi^{\,}_{k}\equiv I.
\end{equation}
The complex-valued number $C^{\Psi^{\,}_{m+n}}_{\Psi^{\,}_{m}\Psi^{\,}_{n}}$
is called a structure constant.
Demanding that the OPEs for the parafermions are associative fixes
this structure constant to be the positive roots of
\cite{Zamolodchikov85}
\begin{equation}
\begin{split}
&
\left(C^{\Psi^{\,}_{m+n}}_{\Psi^{\,}_{m}\Psi^{\,}_{n}}\right)^{2}=
\\
&\quad
\frac{
\Gamma(m+n+1)\,
\Gamma(k-m+1)\,
\Gamma(k-n+1)
     }
     {
\Gamma(m+1)\,
\Gamma(n+1)\,
\Gamma(k-m-n+1)\,
\Gamma(k+1)
     },
\end{split}
  \label{appeq: parafermions OPEs e}
\end{equation}
provided the normalization conditions
\begin{equation}\label{appeq: parafermions OPEs f}
C^{\Psi^{\,}_{k}}_{\Psi^{\,}_{m}\Psi^{\,}_{k-m}}=1, 
\qquad
m=0,\cdots,k-1
\end{equation}
\end{subequations}
are imposed. 

An important consequence of (\ref{appeq: parafermions OPEs e}) 
is the symmetry
\begin{equation}
C^{\Psi^{\,}_{m+n}}_{\Psi^{\,}_{m}\Psi^{\,}_{n}}=
C^{\Psi^{\,}_{m+n}}_{\Psi^{\,}_{n}\Psi^{\,}_{m}}
\qquad
m,n=0,\cdots,k-1,
\end{equation}
under interchanging $m$ and $n$. 
This is why
\begin{subequations}
\begin{equation}
\Psi^{\,}_{n}(v')\,
\Psi^{\,}_{m}(v)
=
(-1)^{\Delta^{\,}_{m+n}-\Delta^{\,}_{m}-\Delta^{\,}_{n}}\,
\Psi^{\,}_{m}(v)\,
\Psi^{\,}_{n}(v'),
\end{equation}
where
\begin{equation}
\Delta^{\,}_{m+n}-\Delta^{\,}_{m}-\Delta^{\,}_{n}=
-\frac{2mn}{k}\equiv
S^{(k)}_{m,n}.
\end{equation}
\end{subequations}
We shall call $\pi\, S^{(k)}_{m,n}$ the mutual (self) statistical angle
between the parafermion $m$ and the parafermion $n\neq m$ (when $n=m$).

Because the OPE between $\Psi^{\,}_{m}$ and $\Psi^{\,}_{k-m}$ gives the
identity operator, we shall use the notation
\begin{subequations}\label{appeq: self and mutual stat for adjoint para pair}
\begin{equation}
\Psi^{\dag}_{m}\equiv
\Psi^{\,}_{k-m}
\end{equation}
for $m=1,\cdots,k-1$.
The self statistical angle of the parafermion $m$ is
\begin{equation}
S^{(k)}_{m,m}=-\frac{2m^{2}}{k}.
\end{equation}
The self statistical angle of the parafermion $k-m$ is
\begin{equation}
S^{(k)}_{k-m,k-m}=-\frac{2(k-m)^{2}}{k}=S^{(k)}_{m,m}\hbox{ mod }\mathbb{Z}.
\end{equation}
The mutual statistics between parafermion $m$ and $k-m$ is
\begin{equation}
S^{(k)}_{m,k-m}=-\frac{2m(k-m)}{k}=-S^{(k)}_{m,m}\hbox{ mod }\mathbb{Z}.
\end{equation}
\end{subequations}

\subsection{Parafermion representation of the $su(2)^{\,}_{k}$ current algebra}
\label{appsubsec: Parafermion representation of the su(2) {k} current algebra}

The $su(2)^{\,}_{k}$ current algebra is defined by the holomorphic
current algebra~\cite{DiFrancesco97}
\begin{equation}
J^{a}(v)\,J^{b}(w)=
\frac{(k/2)\,\delta^{ab}}{(v-w)^{2}}
+
\frac{\mathrm{i}\epsilon^{abc}}{(v-w)}\,J^{c}(w)
+
\cdots
\label{appeq: holomorphic su(2) level k current algebra}
\end{equation}
for any $a,b=1,2,3$ together with its antiholomorphic copy.
Without loss of generality, we consider only this
holomorphic current algebra.

In the basis
\begin{subequations}
\label{appeq: ladder basis affine su2 level k}
\begin{equation}
J^{\pm}\:=
J^{1}\pm\mathrm{i}J^{2},
\qquad
J^{3},
\label{appeq: ladder basis affine su2 level k a}
\end{equation}
the holomorphic current algebra
(\ref{appeq: holomorphic su(2) level k current algebra})
reads
\begin{align}
&
J^{\pm}(v)\,J^{\pm}(w)=
0
+\cdots,
\label{appeq: ladder basis affine su2 level k b}
\\
&
J^{+}(v)\,J^{-}(w)=
\frac{k}{(v-w)^{2}}
+
\frac{2}{(v-w)}\,
J^{3}(w)
+
\cdots,
\label{appeq: ladder basis affine su2 level k c}
\\
&
J^{3}(v)\,J^{\pm}(w)=
\pm
\frac{1}{(v-w)}\,J^{\pm}(w)
+
\cdots,
\label{appeq: ladder basis affine su2 level k d}
\\
&
J^{3}(v)\,J^{3}(w)=
\frac{(k/2)}{(v-w)^{2}}
+
\cdots.
\label{appeq: ladder basis affine su2 level k e}
\end{align}
\end{subequations}
We are going to verify that this current
algebra can be represented in terms of the Gaussian boson $\phi$
from Appendix~\ref{appsubsec: Gaussian algebra}
and the pair of parafermions
$\Psi^{\,}_{1}\equiv\Psi$ and $\Psi^{\,}_{k-1}\equiv\Psi^{\dag}$
from Appendix~\ref{appsubsec: Parafermion algebra}.

We make the Ansatz
\begin{subequations}
\label{eq: parafermion Ansatz}
\begin{align}
&
J^{+}(v)=
\mathcal{N}\,
\Psi^{\,}_{1}(v)\,
e^{+\mathrm{i}\sqrt{\frac{1}{k}}\,\phi(v)}
\nonumber\\
&
\hphantom{J^{+}(v)}
\equiv
\mathcal{N}\,
\Psi(v)\,
e^{+\mathrm{i}\sqrt{\frac{1}{k}}\,\phi(v)},
\label{eq: parafermion Ansatz a}
\\
&
J^{-}(v)=
\mathcal{N}\,
e^{-\mathrm{i}\sqrt{\frac{1}{k}}\,\phi(v)}
\Psi^{\,}_{k-1}(v)
\nonumber\\
&
\hphantom{J^{-}(v)}
\equiv
\mathcal{N}\,
e^{-\mathrm{i}\sqrt{\frac{1}{k}}\,\phi(v)}\,
\Psi^{\dag}(v),
\label{eq: parafermion Ansatz b}
\\
&
J^{3}(v)=
\mathrm{i}
\frac{\sqrt{k}}{2}\,
(\partial^{\,}_{v}\phi)(v),
\label{eq: parafermion Ansatz c}
\end{align}
\end{subequations}
where we impose on $\partial^{\,}_{v}\phi$ the Gaussian algebra
\begin{subequations}
\begin{equation}
\partial^{\,}_{v}\phi(v)\,
\partial^{\,}_{w}\phi(w)=
-\frac{2}{(v-w)^{2}}
+
\cdots,
\label{appeq: choice for OPE between grad phi}
\end{equation}
while we impose on $\Psi$ and $\Psi^{\dag}$ the parafermion algebra
\begin{align}
&
\Psi(v)\,\Psi(w)=
\frac{
C^{I}_{\Psi\Psi}
     }
     {
(v-w)^{2(k-1)/k}
     }
+
\cdots,
\\
&
\Psi^{\dag}(v)\,\Psi^{\dag}(w)=
\frac{
C^{I}_{\Psi^{^{\dag}}\Psi^{^{\dag}}}
     }
     {
(v-w)^{2(k-1)/k}
     }
+
\cdots,
\\
&
\Psi^{\,}(v)\,\Psi^{\dag}(w)=
\frac{
1
     }
     {
(v-w)^{2(k-1)/k}
     }
+
\cdots.
\end{align}
\end{subequations}

The OPE (\ref{appeq: ladder basis affine su2 level k e})
follows from the Ansatz (\ref{eq: parafermion Ansatz c})
with the OPE (\ref{appeq: choice for OPE between grad phi}).
Because of the OPE
(\ref{appeq: OPE partial phi with vertex operator}),
we have the OPE
\begin{equation}
\partial^{\,}_{v}\phi(v)\,
e^{\pm\mathrm{i}\sqrt{\frac{1}{k}}\,\phi(w)}=
\mp
\mathrm{i}\sqrt{\frac{1}{k}}\,
\frac{2}{(v-w)}\,
e^{\pm\mathrm{i}\sqrt{\frac{1}{k}}\,\phi(w)}.
\label{appeq: needed for ladder basis affine su2 level k d}
\end{equation}
The OPE
(\ref{appeq: ladder basis affine su2 level k d})
follows from the Ansatz (\ref{eq: parafermion Ansatz})
with the OPE
(\ref{appeq: needed for ladder basis affine su2 level k d}).  
We thus see that the multiplicative factor $\sqrt{1/k}$ 
entering the argument of the vertex fields
$\exp(\pm\mathrm{i}\sqrt{1/k}\,\phi)$
is fixed by the condition that the two currents have the
holomorphic conformal weight one.
In turn, the normalization factor $\mathcal{N}$
is fixed by the following considerations.
Because of the OPEs
(\ref{appeq: parafermions OPEs})
and
(\ref{appeq: OPE vertex and antivertex operators}),
we have the OPE
\begin{widetext}
\begin{align}
J^{+}(v)\,J^{-}(w)=&\,
\mathcal{N}^{2}\,
\Psi^{\,}_{1}(v)\,
\Psi^{\,}_{k-1}(w)\,
e^{+\mathrm{i}\sqrt{\frac{1}{k}}\,\phi(v)}\,
e^{-\mathrm{i}\sqrt{\frac{1}{k}}\,\phi(w)}
\nonumber\\
=&\,
\left(
\frac{\mathcal{N}^{2}}{(v-w)^{1-\frac{1}{k}+1-\frac{1}{k}}}
+
\cdots
\right)
\frac{1}{(v-w)^{\frac{2}{k}}}
\left(
1
+
\mathrm{i}\sqrt{\frac{1}{k}}\,
(v-w)
(\partial^{\,}_{w}\phi)(w)
+
\cdots
\right)
\nonumber\\
=&\,
\frac{
\mathcal{N}^{2}
     }
     {
(v-w)^{2}
     }
+
\frac{
(2\mathcal{N}^{2}/k)
     }
     {
(v-w)
     }\,
J^{3}(w)
+
\cdots.
\end{align}
\end{widetext}
\medskip
The leading singularity on the right-hand side of
this OPE agrees with the one on the right-hand side of
Eq.~(\ref{appeq: ladder basis affine su2 level k c})
if
\begin{equation}\label{appeq: fixing mathcal N}
\mathcal{N}^{2}=k.
\end{equation}
Finally, the vanishing OPE
(\ref{appeq: ladder basis affine su2 level k b})
follows from the fact that the OPE between any two vertex fields
such that the $\mathbb{C}$-valued prefactors
to the fields $\phi(v)$ and $\phi(w)$
in the arguments of the vertex fields
are not of opposite sign, vanishes to leading order.

We close Appendix\
\ref{appsubsec: Parafermion representation of the su(2) {k} current algebra}
by observing that the Ansatz (\ref{eq: parafermion Ansatz})
is not unique. Indeed, the transformation
\begin{subequations}
\begin{align}
&
\Psi(v)\mapsto
\Psi(v)\,
e^{+\mathrm{i}\alpha},
\\
&
\Psi^{\dag}(v)\mapsto
\Psi^{\dag}(v)\,
e^{-\mathrm{i}\alpha},
\\
&
\phi(v)\mapsto
\phi(v)
-
\sqrt{k}\,
\alpha,
\end{align}
\end{subequations}
leaves the $su(2)^{\,}_{k}$ currents
(\ref{appeq: ladder basis affine su2 level k a})
invariant for any choice of the number $\alpha$.
The number $\alpha$ is defined modulo $2\pi$ and takes $k$
inequivalent values $2\pi n/k$, $n=0,\cdots, k-1$.

\section{Commutation of string operators and the Hamiltonian; ``Analytic''
proof of state exclusion}
\label{appendix: Commutation between string operators and the Hamiltonian}
\subsection{Introduction}

We are given the Hamiltonian
\begin{equation}
\widehat{H}^{\,}_{\mathrm{bs}}\:=
\int\limits_{0}^{L^{\,}_{z}}\mathrm{d}z\,
\widehat{\mathcal{H}}^{\,}_{\mathrm{bs}}
\label{eq: def widehat H bs} 
\end{equation}
and we are told that it commutes with two nonlocal
operators
$\widehat{\Gamma}^{(1)}_{1}$ and $\widehat{\Gamma}^{(1)}_{2}$.
Moreover, we are told that
$\widehat{\Gamma}^{(1)}_{1}$ and $\widehat{\Gamma}^{(1)}_{2}$
commute pairwise.
Hence, we can label any eigenstate of the Hamiltonian
$\widehat{H}^{\,}_{\mathrm{bs}}$
by the simultaneous eigenvalues
$\omega^{(1)}_{1}$ and $\omega^{(1)}_{2}$
of the operators
$\widehat{\Gamma}^{(1)}_{1}$ and $\widehat{\Gamma}^{(1)}_{2}$.
In particular, we can label the basis
for the ground-state manifold by
\begin{equation}
\{|\omega^{(1)}_{1},\omega^{(1)}_{2},\cdots\rangle\}
\label{appeq: def omega basis groundstate manifold}
\end{equation}
where the $\cdots$ allow for additional sources of degeneracies.
We shall demand that this basis is orthonormal.

In order to establish the set to which the eigenvalues
$\omega^{(1)}_{1}$ and $\omega^{(1)}_{2}$
belong, we note that we are given two nonlocal operators
\begin{subequations}
\begin{align}
\widehat{\Gamma}^{(\frac{1}{2})}_{1}(t,z)&\:=
\prod_{y=0}^{L^{\,}_{y}}
\widehat{\sigma}^{\,}_{\mathrm{L}, y}(t,z)\, 
\widehat{\sigma}^{\,}_{\mathrm{R},y}(t,z),
\end{align}
and
\begin{align}
\begin{split}
\widehat{\Gamma}^{(\frac{1}{2})}_{\mathrm{R},2,y}(t,\epsilon)&\:=\,
\exp
\Bigg(
-\frac{\mathrm{i}}{2\sqrt{2}}
\int\limits^{L^{\,}_{z}}_{0}
\mathrm{d}z\,
\partial^{\,}_{z}\widehat{\phi}^{\,}_{\mathrm{R},y}(t,z)
\Bigg)
\\
&\qquad
\times\,
\widehat{\mathcal{P}}^{\,}_{\mathbbm{1}}\,
\widehat{\sigma}^{\,}_{\mathrm{R},y}(t,0)\,
\widehat{\sigma}^{\,}_{\mathrm{R},y}(t,\epsilon)\,
\widehat{\mathcal{P}}^{\,}_{\mathbbm{1}}
\end{split}
\\
&\equiv\,
\widehat{\mathcal U}\,\times\,
\widehat{\mathcal{P}}^{\,}_{\mathbbm{1}}\,
\widehat{\sigma}^{\,}_{\mathrm{R},y}(t,0)\,
\widehat{\sigma}^{\,}_{\mathrm{R},y}(t,\epsilon)\,
\widehat{\mathcal{P}}^{\,}_{\mathbbm{1}}.
\end{align}
\end{subequations}
The operator $\widehat{\Gamma}^{(\frac{1}{2})}_{1}$ is
a discrete product of a countable number of operators
acting along a closed $y$-cycle of the two-torus.
It requires no regularization for its definition.
It anticommutes with
$\widehat{\Gamma}^{(1)}_{2}$, and commutes with
$\widehat{\Gamma}^{(1)}_{1}$ and with the Hamiltonian
\eqref{eq: def widehat H bs}.
In contrast, the operator
$\widehat{\Gamma}^{(\frac{1}{2})}_{\mathrm{R},2,y}(t,\epsilon)$
is a nonlocal operator defined within
one chiral channel of the wire $y$.  It 
acts along an open string (along the
$z$-cycle coinciding with wire $y$)
that fails to close by the infinitesimal amount $\epsilon>0$.
It 
anticommutes with $\widehat{\Gamma}^{(1)}_{1}$
in the limit $\epsilon\to 0$.

If both
$\widehat{\Gamma}^{(\frac{1}{2})}_{1}(t,z)$
and
$\lim_{\epsilon\to0}\widehat{\Gamma}^{(\frac{1}{2})}_{\mathrm{R},2,y}(t,\epsilon)$
were to commute with the Hamiltonian, then so would their product.
The ground-state manifold would then be four-dimensional,
with the orthogonal basis
\begin{subequations}
\label{eq: ''wrong'' states}
\begin{align}
&
|\Omega,\cdots\rangle\:=
|\omega^{(1)}_{1},\omega^{(1)}_{2},\cdots\rangle,
\label{eq: ''wrong'' states a}
\\
&
\widehat{\Gamma}^{(\frac{1}{2})}_{1}(t,z)\,|\Omega,\cdots\rangle\equiv 
\mathcal{N}^{\,}_{1}\,
|\omega^{(1)}_{1},-\omega^{(1)}_{2},\cdots\rangle,
\label{eq: ''wrong'' states b}
\\
&
\lim_{\epsilon\to0}
\widehat{\Gamma}^{(\frac{1}{2})}_{\mathrm{R},2,y}(t,\epsilon)\,
|\Omega,\cdots\rangle\equiv
\mathcal{N}^{\,}_{2}\,
|-\omega^{(1)}_{1},\omega^{(1)}_{2},\cdots\rangle,
\label{eq: ''wrong'' states c}
\\
\begin{split}
&
\widehat{\Gamma}^{(\frac{1}{2})}_{1}(t,z)\,
\[\lim_{\epsilon\to0}\widehat{\Gamma}^{(\frac{1}{2})}_{\mathrm{R},2,y}(t,\epsilon)\,
|\Omega,\cdots\rangle\]\\
&\qquad\qquad\qquad
\equiv
\mathcal{N}^{\,}_{12}\,
|-\omega^{(1)}_{1},-\omega^{(1)}_{2},\cdots\rangle,
\end{split}
\label{eq: ''wrong'' states d}
\\
&
\cdots.
\end{align}
\end{subequations}
We demand that the states on the left-hand side can be normalized.
This can only  be achieved if the normalizations
$\mathcal{N}^{\,}_{1}$,
$\mathcal{N}^{\,}_{2}$,
and
$\mathcal{N}^{\,}_{12}$
are neither zero nor infinity,
for the basis (\ref{appeq: def omega basis groundstate manifold})
is orthonormal by assumption.

However, the logical possibility that one or more of these
normalizations are zero or infinity
cannot be excluded.  In this appendix, we
will assume $\mathcal{N}^{\,}_{1}$ and $\mathcal{N}^{\,}_{2}$ to be
nonvanishing and finite. This assumption amounts
to choosing the ``highest-weight state''
(\ref{eq: ''wrong'' states a}) appropriately.
The quantity $\mathcal{N}^{\,}_{12}$ could be determined by
direct calculation, provided that the explicit form of the state
$\ket{\Omega,\cdots}$ is known. Since we do not have this knowledge,
we leave its value unspecified for the moment.

Given that we do not know the value of $\mathcal{N}^{\,}_{12}$, we proceed
by an alternate route. This line of reasoning makes use of 
the fact that it is not correct to think of the operator 
$\lim_{\epsilon\to0}\widehat{\Gamma}^{(\frac{1}{2})}_{\mathrm{R},2,y}(t,\epsilon)$
as commuting with the Hamiltonian \eqref{eq: def widehat H bs}. It is
a nonlocal operator that changes the topological sector of the state
on which it acts, and can potentially exhibit different limiting behavior
as a function of $\epsilon$
when acting on states belonging to different topological sectors.
Thus, the limit $\epsilon\to0$
must be treated carefully
when multiplying the operators
$\widehat{\Gamma}^{(\frac{1}{2})}_{\mathrm{R},2,y}(t,\epsilon)$
and
$\widehat{\Gamma}^{(\frac{1}{2})}_{1}$.
Indeed, instead of the set of states \eqref{eq: ''wrong'' states},
we can also consider the following set of states,
\begin{subequations}
\label{eq: limiting definition of states}
\begin{align}
&
|\Omega,\cdots\rangle\:=
|\omega^{(1)}_{1},\omega^{(1)}_{2},\cdots\rangle,
\label{eq: limiting definition of states a}
\\
&
\ket{\widehat{\Gamma}^{(\frac{1}{2})}_{1},\cdots}\:=
\widehat{\Gamma}^{(\frac{1}{2})}_{1}(t,z)\,|\Omega,\cdots\rangle,
\label{eq: limiting definition of states b}
\\
&
\ket{\widehat{\Gamma}^{(\frac{1}{2})}_{2},\cdots}\:=
\lim_{\epsilon\to0}
\widehat{\Gamma}^{(\frac{1}{2})}_{\mathrm{R},2,y}(t,\epsilon)
\ket{\Omega,\cdots},
\label{eq: limiting definition of states c}
\\
&
\ket{\widehat{\Gamma}^{(\frac{1}{2})}_{1}
\widehat{\Gamma}^{(\frac{1}{2})}_{2},\cdots}\:=
\lim_{\epsilon\to0}
\[
\widehat{\Gamma}^{(\frac{1}{2})}_{1}(t,z)\,
\widehat{\Gamma}^{(\frac{1}{2})}_{\mathrm{R},2,y}(t,\epsilon)
\ket{\Omega,\cdots}
\].
\label{eq: limiting definition of states d}
\end{align}
\end{subequations}
The only difference between the states
\eqref{eq: limiting definition of states}
and the states \eqref{eq: ''wrong'' states} is that the limit
$\epsilon\to 0$ is taken \textit{after}
forming the product $\widehat{\Gamma}^{(\frac{1}{2})}_{1}\,
\widehat{\Gamma}^{(\frac{1}{2})}_{\mathrm{R},2,y}(t,\epsilon)$
in Eq.~\eqref{eq: limiting definition of states d}.
We adopt the point of view that the dimension
of the ground-state manifold of the Hamiltonian 
\eqref{eq: def widehat H bs}
cannot depend on the choice of when [i.e., before or after
forming the product $\widehat{\Gamma}^{(\frac{1}{2})}_{1}\,
\widehat{\Gamma}^{(\frac{1}{2})}_{\mathrm{R},2,y}(t,\epsilon)$]
the limit $\epsilon\to0$ is taken.
Hence, the number of ground states present in
Eqs.~\eqref{eq: ''wrong'' states} and \eqref{eq: limiting definition of states}
must agree with one another.  For this reason, we ask how many of
the states \eqref{eq: limiting definition of states} are indeed ground states
of the interaction \eqref{eq: def widehat H bs}.
This allows us to scrutinize the limiting behavior of operator products
without losing important information related to the nonlocality
of its constituent operators.
We will show that the state \eqref{eq: limiting definition of states d}
cannot be in the ground-state manifold of the interaction
\eqref{eq: def widehat H bs}.
Logical consistency then demands that $\mathcal{N}^{\,}_{12}=0$ or $\infty$ in
Eqs.~\eqref{eq: ''wrong'' states},
as these are the only two possibilities that would
exclude the state \eqref{eq: ''wrong'' states d} from the ground-state manifold.

The operator
$\widehat{\Gamma}^{(\frac{1}{2})}_{\mathrm{R},2,y}(t,\epsilon)$
does not commute with the interaction
$\widehat{H}^{\,}_{\mathrm{bs}}$
defined by Eq.\
(\ref{eq: def widehat H bs}).
The purpose of this appendix is to determine whether the states
\eqref{eq: limiting definition of states c}
and \eqref{eq: limiting definition of states d},
which involve taking the limit $\epsilon\to 0$,
indeed belong to the ground-state manifold of the interaction
\eqref{eq: def widehat H bs}
once this limit is taken.
More precisely, we define
\begin{subequations}
\begin{equation}
\[
\widehat{H}^{\,}_{\mathrm{bs}},
\widehat{\Gamma}^{(\frac{1}{2})}_{\mathrm{R},2,y}(t,\epsilon)
\]\=:
\widehat{\mathcal{D}}^{\,}_{\mathrm{R},2,y}(t,\epsilon),
\label{eq: D_{2} as commutator}
\end{equation}
where the operator $\widehat{\mathcal{D}}^{\,}_{\mathrm{R},2,y}(t,\epsilon)$
is nonlocal, as we shall see below, and nonvanishing in general.
We further define
\begin{align}
\begin{split}
\[
\widehat{H}^{\,}_{\mathrm{bs}},
\widehat{\Gamma}^{(\frac{1}{2})}_{1}(t,z)\,
\widehat{\Gamma}^{(\frac{1}{2})}_{\mathrm{R},2,y}(t,\epsilon)
\]
&=
\widehat{\Gamma}^{(\frac{1}{2})}_{1}(t,z)\,
\widehat{\mathcal{D}}^{\,}_{\mathrm{R},2,y}(t,\epsilon)
\\
&\=:
\widehat{\mathcal{D}}^{\,}_{1\mathrm{R},2,y}(z,\epsilon).
\end{split}
\end{align}
\end{subequations}
We are going to show that
\begin{subequations}
\begin{align}
\lim_{\epsilon\to0}
\widehat{\mathcal{D}}^{\,}_{\mathrm{R},2,y}(t,\epsilon)
\ket{\Omega,\cdots}
&=
0.
\label{eq: D_{2} ket Omega vanishes}
\end{align}
Equation \eqref{eq: D_{2} ket Omega vanishes}
is equivalent to the statement
\begin{align}
\lim_{\epsilon\to 0}
\[
\widehat{H}^{\,}_{\mathrm{bs}},
\widehat{\Gamma}^{(\frac{1}{2})}_{\mathrm{R},2,y}(t,\epsilon)
\]
\ket{\Omega,\cdots}=
(\widehat{H}^{\,}_{\mathrm{bs}}-
E^{\,}_{\Omega})\,
\ket{\widehat{\Gamma}^{(\frac{1}{2})}_{2},\cdots}
=0,
\end{align}
\end{subequations}
where $E^{\,}_{\Omega}$ is the energy eigenvalue
of the state $\ket{\Omega,\cdots}$.
From this it immediately follows that the state
$\ket{\widehat{\Gamma}^{(\frac{1}{2})}_{2},\cdots}$
indeed belongs to the ground-state manifold of the
interaction \eqref{eq: def widehat H bs}.

We are also going to show that the state
\begin{subequations}
\begin{align}
\lim_{\epsilon\to0}
\widehat{\mathcal{D}}^{\,}_{1\mathrm{R},2,y}(z,\epsilon)
\ket{\Omega,\cdots}
\label{eq: D_{1}2 ket Omega diverges}
\end{align}
has infinite norm as $z\to 0$.
Equation \eqref{eq: D_{1}2 ket Omega diverges}
is equivalent to the statement that
\begin{align}
\begin{split}
&\lim_{z\to 0}
\lim_{\epsilon\to0}
\[
\widehat{H}^{\,}_{\mathrm{bs}},
\widehat{\Gamma}^{(\frac{1}{2})}_{1}(t,z)\,
\widehat{\Gamma}^{(\frac{1}{2})}_{\mathrm{R},2,y}(t,\epsilon)
\]
\ket{\Omega,\cdots}
\\
&\qquad\qquad
=
\lim_{z\to 0}
\(\widehat{H}^{\,}_{\mathrm{bs}}\,
-
E^{\,}_{\Omega}
\)\ket{\widehat{\Gamma}^{(\frac{1}{2})}_{1}\widehat{\Gamma}^{(\frac{1}{2})}_{2},\cdots}
\end{split}
\end{align}
\end{subequations}
is a state with infinite norm.
That this divergence occurs as $z\to 0$ is especially
problematic. In order for the product
$\widehat{\Gamma}^{(\frac{1}{2})}_{1}(t,z)
\widehat{\Gamma}^{(\frac{1}{2})}_{\mathrm{R},2,y}(t,\epsilon)$
of string operators
to yield a topologically-degenerate ground state when
acting on the state $\ket{\Omega,\cdots}$, the resulting
state cannot depend on the quantities $z$ and $\epsilon$
in an observable way as $z\to 0$ and $\epsilon\to 0$.
If this were the case, then the states
 $\ket{\Omega,\cdots}$ and 
$\ket{\widehat{\Gamma}^{(\frac{1}{2})}_{1}
\widehat{\Gamma}^{(\frac{1}{2})}_{2},\cdots}$
could be distinguished by simply evaluating the string operator
$\widehat{\Gamma}^{(\frac{1}{2})}_{1}(t,z)$ near the point $z=0$.
Hence, proving that the state defined in
Eq.~\eqref{eq: D_{1}2 ket Omega diverges}
is not normalizable
will allow us to conclude that the state
$\ket{\widehat{\Gamma}^{(\frac{1}{2})}_{1}\,
\widehat{\Gamma}^{(\frac{1}{2})}_{2},\cdots}$
does not belong to the ground-state manifold of the
interaction \eqref{eq: def widehat H bs}.  

We are left with the conclusion of the paper, namely that
the ground-state manifold of the interaction \eqref{eq: def widehat H bs}
includes the states 
\eqref{eq: limiting definition of states a}--%
\eqref{eq: limiting definition of states c},
and excludes the state \eqref{eq: limiting definition of states d}.
From now on, we ignore the $\cdots$ representing additional degeneracies
for the ground-state manifold.

\subsection{Calculation}

We first prove Eq.~\eqref{eq: D_{2} ket Omega vanishes}.  We begin by
calculating $\widehat{\mathcal{D}}^{\,}_{\mathrm{R},2,y}(t,\epsilon)$.
For finite $\epsilon>0$, we have
\begin{align}
\widehat{\mathcal H}^{\,}_{\mathrm{bs}}\,
\widehat{\Gamma}^{(\frac{1}{2})}_{\mathrm{R},2,y}(t,\epsilon)
&=
\widehat{\Gamma}^{(\frac{1}{2})}_{\mathrm{R},2,y}(t,\epsilon)\,
\widehat{\mathcal H}^{\,}_{\mathrm{bs}}\,
\times
\begin{cases}
+1, & z>\epsilon,\\
+\mathrm{i}, & z=\epsilon,\\
-1, & z<\epsilon.
\end{cases}
\end{align}
We now use the definition
\eqref{eq: D_{2} as commutator},
along with the identity
\begin{align}
\widehat{A}\,
\widehat{B}
&=
\widehat{B}\,
\widehat{A}\,
f(z,\epsilon)
\iff
\[\widehat{A},\widehat{B}\]
=
\widehat{B}\,
\widehat{A}\,
\[f(z,\epsilon)-1\],
\end{align}
which gives
\begin{widetext}
\begin{align}
\widehat{\mathcal{D}}^{\,}_{\mathrm{R},2,y}(t,\epsilon)=&\,
-4\mathrm{i}
\int\limits^{\epsilon}_{0}
\!\mathrm{d}z\,
\sin
\(
\frac{1}{\sqrt{2}}
\(\widehat{\phi}^{\,}_{\mathrm{R},y}(t,z)-\widehat{\phi}^{\,}_{\mathrm{L},y+1}(t,z)\)
\)\, \widehat{\mathcal U}\,
\widehat{\psi}^{\,}_{\mathrm{L},y+1}(t,z)\,
\widehat{\psi}^{\,}_{\mathrm{R},y}(t,z)\,
\widehat{\mathcal{P}}^{\,}_{\mathbbm{1}}\, 
\widehat{\sigma}^{\,}_{\mathrm{R},y}(t,0)\,
\widehat{\sigma}^{\,}_{\mathrm{R},y}(t,\epsilon)\,
\widehat{\mathcal{P}}^{\,}_{\mathbbm{1}},
\label{eq: D_{2} def}
\end{align}
up to a contribution from the set of measure zero where $z=\epsilon$,
which we will ignore.

To prove Eq.~\eqref{eq: D_{2} ket Omega vanishes},
we compute the leading contribution to
$\widehat{\mathcal{D}}^{\,}_{2}(t,\epsilon)$ as $\epsilon\to0$.
For $\epsilon$ infinitesimal,
we may replace the integral in Eq.~\eqref{eq: D_{2} def} by the value of the
integrand at the midpoint of the integration domain,
\begin{align}
\label{eq: D_{2} epsilon to 0 limit before OPE}
\begin{split}
\widehat{\mathcal{D}}^{\,}_{\mathrm{R},2,y}(t,\epsilon)
\approx
-4\mathrm{i}\,
\epsilon\,
\sin
\left(
\frac{1}{\sqrt{2}}
\[
\widehat{\phi}^{\,}_{\mathrm{R},y}
\(\frac{\epsilon}{2}\)
-
\widehat{\phi}^{\,}_{\mathrm{L},y+1}
\(\frac{\epsilon}{2}\)
\]
\right)\, \widehat{\mathcal U}\,
\widehat{\psi}^{\,}_{\mathrm{L},y+1}\(\frac{\epsilon}{2}\)\,
\widehat{\psi}^{\,}_{\mathrm{R},y}\(\frac{\epsilon}{2}\)\,
\widehat{\mathcal{P}}^{\,}_{\mathbbm{1}}\, 
\widehat{\sigma}^{\,}_{\mathrm{R},y}(t,0)\,
\widehat{\sigma}^{\,}_{\mathrm{R},y}(t,\epsilon)\,
\widehat{\mathcal{P}}^{\,}_{\mathbbm{1}}.
\end{split}
\end{align}
We now perform the (equal-time) OPE
\begin{align}
\label{eq: OPEs for D_{2} limit}
\sin
\left(
\frac{1}{\sqrt{2}}
\[
\widehat{\phi}^{\,}_{\mathrm{R},y}
\(\frac{\epsilon}{2}\)-\widehat{\phi}^{\,}_{\mathrm{L},y+1}
\(\frac{\epsilon}{2}\)
\]
\right)\,
\widehat{\mathcal U}
=&\,
\sin
\left(
\frac{1}{\sqrt{2}}
\[
\widehat{\phi}^{\,}_{\mathrm{R},y}
\(\frac{\epsilon}{2}\)-\widehat{\phi}^{\,}_{\mathrm{L},y+1}
\(\frac{\epsilon}{2}\)
\]
\right)\,
\exp
\Bigg(
\!-\!
\frac{\mathrm{i}}{2\sqrt{2}}
\int\limits^{L^{\,}_{z}}_{0}
\mathrm{d} z\,
\partial^{\,}_{z}\, \widehat{\phi}^{\,}_{\mathrm{R},y}(t,z)
\Bigg).
\end{align}
Inserting the OPEs
\begin{subequations}
\begin{align}
&\lim_{\epsilon\to0}
e^{+\frac{\mathrm i}{\sqrt{2}}
\widehat{\phi}^{\,}_{\mathrm{R},y}
\(\frac{\epsilon}{2}\)}
e^{-\frac{\mathrm i}{2\sqrt{2}}\widehat{\phi}^{\,}_{\mathrm{R},y}(L^{\,}_{z})}
\sim
\frac{1}{\epsilon^{1/2}}\, 
e^{+\frac{\mathrm i}{2\sqrt 2}\widehat{\phi}^{\,}_{\mathrm{R},y}(\frac{\epsilon}{2})},
\\
&\lim_{\epsilon\to0}
e^{-\frac{\mathrm i}{\sqrt{2}}\widehat{\phi}^{\,}_{\mathrm{R},y}(\frac{\epsilon}{2})}
e^{+\frac{\mathrm i}{2\sqrt{2}}
\widehat{\phi}^{\,}_{\mathrm{R},y}
\(0\)}
\sim
\frac{1}{\epsilon^{1/2}}\,
e^{-\frac{\mathrm i}{2\sqrt{2}}\widehat{\phi}^{\,}_{\mathrm{R},y}(\frac{\epsilon}{2})},
\\
&\lim_{\epsilon\to0}
e^{+\frac{\mathrm i}{2\sqrt{2}}\widehat{\phi}^{\,}_{\mathrm{R},y}(\frac{\epsilon}{2})}
e^{+\frac{\mathrm i}{2\sqrt{2}}
\widehat{\phi}^{\,}_{\mathrm{R},y}
\(0\)}
\sim
\epsilon^{1/4}\, 
e^{+\frac{\mathrm i}{\sqrt 2}\widehat{\phi}^{\,}_{\mathrm{R},y}(\frac{\epsilon}{2})},
\\
&\lim_{\epsilon\to0}
e^{-\frac{\mathrm i}{2\sqrt{2}}\widehat{\phi}^{\,}_{\mathrm{R},y}(L^{\,}_{z})}
e^{-\frac{\mathrm i}{2\sqrt{2}}
\widehat{\phi}^{\,}_{\mathrm{R},y}
\(\frac{\epsilon}{2}\)}
\sim
\epsilon^{1/4}\,
e^{-\frac{\mathrm i}{\sqrt{2}}\widehat{\phi}^{\,}_{\mathrm{R},y}(\frac{\epsilon}{2})},
\end{align}
\end{subequations}
where ``$\sim$'' denotes equality up to constant factors and
nonsingular terms,
and using the fact that $L^{\,}_{z}\sim 0$ by
periodic boundary conditions,
we find
\begin{align}
\sin
\left(
\frac{1}{\sqrt{2}}
\[
\widehat{\phi}^{\,}_{\mathrm{R},y}\(\frac{\epsilon}{2}\)
-
\widehat{\phi}^{\,}_{\mathrm{L},y+1}\(\frac{\epsilon}{2}\)
\]
\right)\,
\widehat{\mathcal U}
\sim&\,
\frac{1}{\epsilon^{1/4}}\,
\sin
\left(
\frac{1}{\sqrt{2}}
\[
\widehat{\phi}^{\,}_{\mathrm{R},y}\(\frac{\epsilon}{2}\)
-
\widehat{\phi}^{\,}_{\mathrm{L},y+1}\(\frac{\epsilon}{2}\)
\]
\right).
\end{align}
Next, we perform the OPE
\begin{align}
\widehat{\mathcal{P}}^{\,}_{\mathbbm{1}}\, 
\widehat{\sigma}^{\,}_{\mathrm{R},y}(t,0)\,
\widehat{\sigma}^{\,}_{\mathrm{R},y}(t,\epsilon)\,
\widehat{\mathcal{P}}^{\,}_{\mathbbm{1}}
&\sim
\frac{1}{\epsilon^{1/8}}.
\end{align}
Inserting this pair of OPEs into Eq.\
\eqref{eq: D_{2} epsilon to 0 limit before OPE}, we find
\begin{align}
\label{eq: D_{2} final expression}
\begin{split}
\lim_{\epsilon\to0}
\widehat{\mathcal{D}}^{\,}_{\mathrm{R},2,y}(t,\epsilon)
\ket{\Omega}
&\sim
\lim_{\epsilon\to0}
\epsilon^{5/8}\,
\sin
\left(
\frac{1}{\sqrt{2}}
\[
\widehat{\phi}^{\,}_{\mathrm{R},y}\(\frac{\epsilon}{2}\)
-
\widehat{\phi}^{\,}_{\mathrm{L},y+1}\(\frac{\epsilon}{2}\)
\]
\right)\,
\widehat{\psi}^{\,}_{\mathrm{L},y+1}\(\frac{\epsilon}{2}\)\,
\widehat{\psi}^{\,}_{\mathrm{R},y}\(\frac{\epsilon}{2}\)
\ket{\Omega}=0.
\end{split}
\end{align}
\end{widetext}
The form of the operator appearing on the RHS above is not important.
All that matters is that its expectation value in the 
state $\ket{\Omega}$
is not singular in the limit $\epsilon\to0$.
Also of crucial
importance is the factor $\epsilon^{5/8}$ that sends
$\lim_{\epsilon\to0}
\widehat{\mathcal{D}}^{\,}_{\mathrm{R},2,y}(t,\epsilon)
\ket{\Omega}\to 0$
as $\epsilon\to 0$. 
Hence, we may
conclude that the state $\ket{\widehat{\Gamma}^{(\frac{1}{2})}_{2}}$,
defined in Eq.~\eqref{eq: limiting definition of states c}, is
a ground state.

We now turn to the state
$\ket{\widehat{\Gamma}^{(\frac{1}{2})}_{1}\, \widehat{\Gamma}^{(\frac{1}{2})}_{2}}$,
defined in Eq.~\eqref{eq: limiting definition of states d},
and ask if it, too, is a ground state.
We will see that it cannot be a ground state by proving
that the state defined in
Eq.~\eqref{eq: D_{1}2 ket Omega diverges}
has infinite norm as $z\to 0$ and $\epsilon\to0$.
We proceed by setting $z=z^{\,}_0=0$ from the outset.
Using Eq.~\eqref{eq: D_{2} epsilon to 0 limit before OPE},
\begin{widetext}
\begin{align}
\begin{split}
\widehat{\Gamma}^{(\frac{1}{2})}_{1}(t,0)\, 
\widehat{\mathcal{D}}^{\,}_{\mathrm{R},2,y}(t,\epsilon)
\approx&\,
-4\mathrm{i}\,
\epsilon\,
\Bigg(
\prod_{y^{\prime}}
\widehat{\sigma}^{\,}_{\mathrm{L},y^{\prime}}(t,0)\, 
\widehat{\sigma}^{\,}_{\mathrm{R},y^{\prime}}(t,0)
\Bigg)
\,
\sin
\left(
\frac{1}{\sqrt{2}}
\[
\widehat{\phi}^{\,}_{\mathrm{R},y}\(\frac{\epsilon}{2}\)
-
\widehat{\phi}^{\,}_{\mathrm{L},y+1}\(\frac{\epsilon}{2}\)
\]
\right)\,
\widehat{\mathcal U}\,
\\
&\,
\times
\widehat{\psi}^{\,}_{\mathrm{L},y+1}\(\frac{\epsilon}{2}\)\,
\widehat{\psi}^{\,}_{\mathrm{R},y}\(\frac{\epsilon}{2}\)\,
\widehat{\mathcal{P}}^{\,}_{\mathbbm{1}}\, 
\widehat{\sigma}^{\,}_{\mathrm{R},y}(t,0)\,
\widehat{\sigma}^{\,}_{\mathrm{R},y}(t,\epsilon)\,
\widehat{\mathcal{P}}^{\,}_{\mathbbm{1}}.
\end{split}
\end{align}
Using the OPEs~\eqref{eq: OPEs for D_{2} limit} in conjunction with the OPEs
\begin{subequations}
\begin{align}
&
\widehat{\sigma}^{\,}_{\mathrm{R}, y}(t,0)\,
\widehat{\psi}^{\,}_{\mathrm{R},y}\(\frac{\epsilon}{2}\)
\sim
\frac{1}{\epsilon^{1/2}}\,
\widehat{\sigma}^{\,}_{\mathrm{R}, y}(t,0),
\\
&
\widehat{\sigma}^{\,}_{\mathrm{L}, y+1}(t,0)\,
\widehat{\psi}^{\,}_{\mathrm{L},y+1}\(\frac{\epsilon}{2}\)
\sim
\frac{1}{\epsilon^{1/2}}\,
\widehat{\sigma}^{\,}_{\mathrm{L}, y+1}(t,0),
\end{align}
\end{subequations}
we find
\begin{align}
&\widehat{\Gamma}^{(\frac{1}{2})}_{1}(t,0)\, 
\widehat{\mathcal{D}}^{\,}_{\mathrm{R},2,y}(t,\epsilon)
\sim
\frac{1}{\epsilon^{3/8}}
\sin
\left(
\frac{1}{\sqrt{2}}
\[
\widehat{\phi}^{\,}_{\mathrm{R},y}\(\frac{\epsilon}{2}\)
-
\widehat{\phi}^{\,}_{\mathrm{L},y+1}\(\frac{\epsilon}{2}\)
\]
\right)\, 
\widehat{\Gamma}^{(\frac{1}{2})}_{1}(t,0).\nonumber
\end{align}
In contrast to the RHS of Eq.~\eqref{eq: D_{2} final expression},
we now have the product between a local operator and
a nonlocal operator on the RHS.
Furthermore, the real-valued prefactor is a function of $\epsilon$
that diverges as $\epsilon\to 0$.
We conclude that
\begin{align}
\lim_{\epsilon\to0}
\widehat{\mathcal{D}}^{\,}_{1\mathrm{R},2,y}(0,\epsilon)
\ket{\Omega}
&=
\widehat{\Gamma}^{(\frac{1}{2})}_{1}(t,0)\, 
\widehat{\mathcal{D}}^{\,}_{\mathrm{R},2,y}(t,\epsilon)
\ket{\Omega}
\sim
\frac{1}{\epsilon^{3/8}}
\sin
\left(
\frac{1}{\sqrt{2}}
\[
\widehat{\phi}^{\,}_{\mathrm{R},y}\(\frac{\epsilon}{2}\)
-
\widehat{\phi}^{\,}_{\mathrm{L},y+1}\(\frac{\epsilon}{2}\)
\]
\right)
\ket{\widehat{\Gamma}^{(\frac{1}{2})}_{1}}
\end{align}
\end{widetext}
is a state with infinite norm, as advertised, provided that the operator
$\sin
\Big(
\frac{1}{\sqrt{2}}
[
\widehat{\phi}^{\,}_{\mathrm{R},y}(\epsilon/2)
-
\widehat{\phi}^{\,}_{\mathrm{L},y+1}(\epsilon/2)
]
\Big)$ does not
annihilate the state $\ket{\widehat{\Gamma}^{(\frac{1}{2})}_{1}}$.
(Determining whether or not this is the case again requires
an explicit expression for the state $\ket{\Omega}$, which
we do not have at our disposal.)
In that case, we conclude that the state
$\ket{\widehat{\Gamma}^{(\frac{1}{2})}_{1}\, \widehat{\Gamma}^{(\frac{1}{2})}_{2}}$
cannot be a ground state of the interaction
$\widehat{H}^{\,}_{\mathrm{bs}}$
defined by Eq.\
(\ref{eq: def widehat H bs}).

\section{Diagrammatics for operator algebra in the Ising CFT}
\label{appendix: Diagrammatics for operator algebra in the Ising CFT}

The discussion surrounding Eqs.\ \eqref{eq2Dcase: psi sigma algebra} in
the main text concerns how to infer the exchange algebra of two chiral
primary operators in the Ising CFT from their operator product
expansion.  This exchange algebra is simple to determine in cases
where the two primary operators have a unique fusion product, as in
the case of the $\sigma$ and $\psi$ operators in Eqs.\
\eqref{eq2Dcase: psi sigma algebra}.  However, when the two primary operators
\textit{do not} have a unique fusion product, as occurs in the case of
two $\sigma$ operators [see the OPEs in Eqs.\
\eqref{eq2Dcase: sigma sigma OPE}],
the exchange algebra depends on the fusion channel in
which the product of the pair of operators is evaluated [see the
exchange algebra in Eqs.\ \eqref{eq2Dcase: sigma sigma algebra}].
This poses a challenge for calculations. It is necessary to keep track
of both fusion and braiding in a way that respects consistency
conditions between the two.  This challenge is the essence of the
difference between Abelian and non-Abelian excitations in quantum
field theory.

To this end, it is expedient to make use of the diagrammatic calculus
developed in, e.g.,
Refs.~\onlinecite{Friedan87,Vafa88,Verlinde88,Moore89a}
to represent chiral algebras associated with rational conformal field
theories (RCFTs).  In this Appendix, we review aspects of this
calculus, as they relate to the wire constructions of non-Abelian topological
phases discussed in this work.  For simplicity, we focus on the example of the
Ising CFT, although generalizations to other RCFTs are straightforward.

We first define the data necessary to compute the exchange algebra of
chiral primary fields in a general RCFT.  These are the fusion rules,
the $R$-symbols, and the $F$-symbols.  The nontrivial fusion rules of the
Ising ($\mathbb Z^{\,}_{2}$) RCFT are
\begin{subequations}
\begin{align}
\label{appeq: Z_{2} fusion rules}
\psi\times \psi &= \mathbbm 1\\
\sigma\times\sigma &= \mathbbm 1+\psi\\
\sigma\times\psi &= \sigma.
\end{align}
\end{subequations}
In general, for chiral primary fields $a$, $b$, and $c$, the fusion rules
take the form
\begin{subequations}
\label{appeq: fusion algebra def}
\begin{align}
a\times b=
\sum_{c}
N^{c}_{ab}\, c\, ,
\label{appeq: fusion algebra def a}
\end{align} 
with $N^{c}_{ab}$ nonnegative integers.
The diagrammatic representation of a product of two chiral primary
fields $a$ and $b$ that fuse to $c$ is
\begin{align}
\includegraphics[width=.09\textwidth]{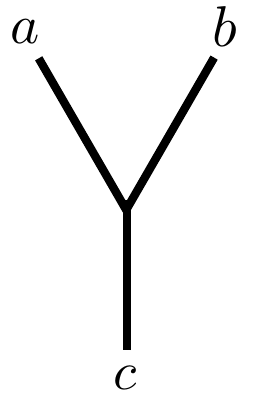}.
\label{appeq: fusion algebra def c}
\end{align}
The requirement that the fusion
algebra \eqref{appeq: fusion algebra def a}
be associative imposes the constraints
\begin{align}
\sum_{d}
N^{d}_{ab}\,
N^{e}_{dc}=
\sum_{f}
N^{e}_{af}\,
N^{f}_{bc}.
\label{appeq: fusion algebra def b}
\end{align}
\end{subequations}
For many interesting RCFTs,
including all of the $\mathbbm Z^{\,}_{k}$ CFTs, the fusion coefficients
$N^{c}_{ab}=0$ or $1$.  For simplicity, we will
restrict ourselves to this class of RCFTs, which is known as the class
of RCFTs without fusion multiplicity since the nonnegative integers
$N^{c}_{ab}=0$ are never larger than one.

Read from bottom to top, diagram (\ref{appeq: fusion algebra def c})
is an element of the vector
space $V^{ab}_{c}$,
which is known as a ``splitting space.'' Read from
top to bottom, it is an element of the vector space $V^{c}_{ab}$,
which is known as a ``fusion space.''  These vector spaces are dual to
one another, and we will use the terms ``fusion'' and ``splitting''
interchangeably unless otherwise noted.
The $R$-symbols are defined to be unitary maps 
\begin{subequations}
\begin{align}
R^{ab}_{c}:V^{ba}_{c}\to V^{ab}_{c}
\end{align}
that implement the diagrammatic braiding operation
\begin{align}\label{appeq: R-symbol def diagram}
\includegraphics[width=.2\textwidth]{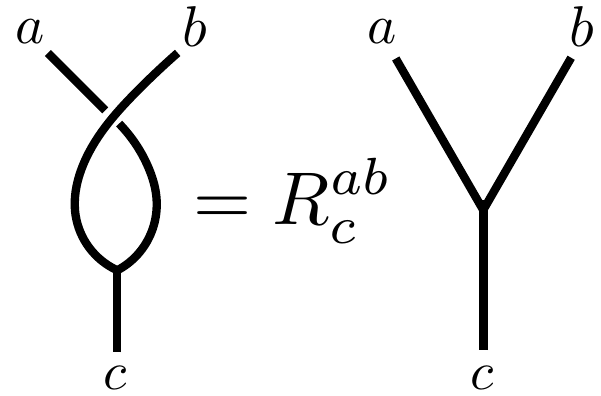}.
\end{align}
\end{subequations}
Note that we have defined the diagrammatic action of the $R$-symbols
in such a way that the left leg of the fusion tree passes over the
right leg.  If instead the right leg passes over the left leg, then
the inverse $R$-symbol $(R^{ab}_{c})^{-1}$ appears.  The $R$-symbols
are essential for determining how primary operators in an RCFT behave
under exchange.

The final data necessary to determine the exchange
algebra of primary operators in an RCFT are the $F$-symbols.
These are required if exchange of chiral primary fields is to
be associative. Associativity of the fusions rules
(\ref{appeq: fusion algebra def a}) is encoded by
Eq.~(\ref{appeq: fusion algebra def b}).
Equation (\ref{appeq: fusion algebra def b})
suggests that one defines the splitting space $V^{abc}_{d}$
that encodes the fusion of three chiral fields
$a$, $b$, $c$ into one chiral field $d$
by demanding that
\begin{subequations}
\label{appeq: def F symbols}
\begin{align}
\sum_{e}V^{ab}_{e}\otimes V^{ec}_{d} =
\sum_{f}V^{af}_{d}\otimes V^{bc}_{f}\equiv V^{abc}_{d}
\label{appeq: def F symbols a}
\end{align}
holds. The $F$-symbols are then defined to be unitary maps
\begin{align}
[F^{abc}_{d}]^{\,}_{ef}:
V^{ab}_{e}\otimes V^{ec}_{d}\to
V^{af}_{d}\otimes V^{bc}_{f}
\label{appeq: def F symbols b}
\end{align}
that implement the diagrammatic operation
\begin{align}
\includegraphics[width=.3\textwidth]{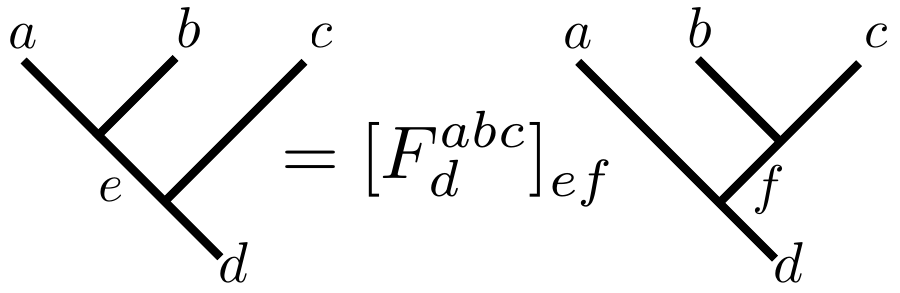}.
\label{appeq: def F symbols c}
\end{align} 
\end{subequations} 
The $F$-symbols $F^{abc}_{d}$ are thus automorphisms
(i.e., changes of basis) of the splitting space $V^{abc}_{d}$.
The fusion rules, $F$-symbols, and
$R$-symbols define a mathematical structure known as a braided fusion
category (BFC).  This structure can be used as a starting point for an
axiomatic formulation of RCFT~\cite{Moore89a}.

For the Ising RCFT, whose fusion rules are given in Eqs.\
\eqref{appeq: Z_{2} fusion rules},
the $R$-symbols are given by~\cite{Rowell09}
\begin{subequations}
\begin{align}
\label{appeq: Ising RCFT R-symbols}
&
R^{\sigma\sigma}_{\mathbbm{1}}=
e^{+\mathrm{i}\,\frac{\pi}{8}},
\\
&
R^{\sigma\sigma}_{\psi}=
e^{-\mathrm{i}\,\frac{3\pi}{8}},
\\
&
R^{\psi\psi}_{\mathbbm{1}}=
-1,
\\
&
R^{\psi\sigma}_{\sigma}=
R^{\sigma\psi}_{\sigma}=
+\mathrm{i},
\end{align}
\end{subequations}
with all other $R$-symbols trivial (i.e., equal to $+1$).  Note that,
up to complex conjugation, these $R$-symbols coincide with the phases
acquired in Eqs.\ \eqref{eq2Dcase: psi sigma algebra} and
\eqref{eq2Dcase: sigma sigma algebra} when the corresponding chiral
primary fields are exchanged.  This is by design. The $R$-symbols
reflect the monodromy of products of chiral primary fields in the
corresponding RCFT.
The $F$-symbols for the Ising RCFT are given by~\cite{Rowell09}
\begin{subequations}
\begin{align}
&
F^{\psi\psi\sigma}_{\sigma}=
F^{\psi\sigma\psi}_{\sigma}=
F^{\sigma\psi\psi}_{\sigma}=
-1,
\\
&
F^{\psi\sigma\sigma}_{\psi}=
F^{\sigma\psi\sigma}_{\psi}=
F^{\sigma\sigma\psi}_{\psi}=
-1,
\\
&
F^{\sigma\sigma\sigma}_{\sigma}=
\frac{1}{\sqrt{2}}
\begin{pmatrix}
1 & 1
\\
1 &-1
\end{pmatrix},
\end{align}
\end{subequations}
with all other $F$-symbols trivial (i.e., equal to $+1$).

We will now demonstrate, using the example of the Ising RCFT, how
to translate diagrams like those appearing in Eqs.\
\eqref{appeq: R-symbol def diagram} and
\eqref{appeq: def F symbols c}
into algebraic statements.  Performing this translation
requires one to fix a chiral sector of the CFT.
We choose to work with the chiral sector
$\mathrm{M} =\mathrm{R}$.
Once this choice is made, the starting point for this
``dictionary'' is to compare the diagram corresponding
to the action of a particular $R$-symbol, say
\begin{subequations}
\label{appeq: dictionary diagram}
\begin{align}
\includegraphics[width=.2\textwidth]{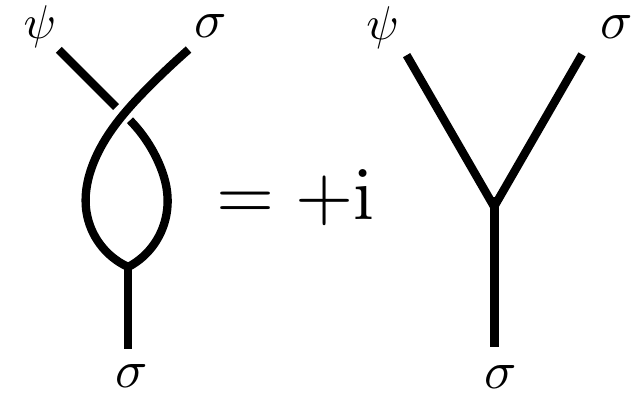},
\label{appeq: dictionary diagram a}
\end{align}
with its algebraic analogue, 
given up to a constant phase factor by
Eq.\ \eqref{eq2Dcase: psi sigma algebra},
\begin{align}
\widehat{\psi}^{\,}_{\mathrm{R}}(z)\,
\widehat{\sigma}^{\,}_{\mathrm{R}}(z^{\prime})=
\widehat{\sigma}^{\,}_{\mathrm{R}}(z^{\prime})\,
\widehat{\psi}^{\,}_{\mathrm{R}}(z)\,
e^{+\mathrm{i}\, \frac{\pi}{2}\,\mathrm{sgn}(z-z^{\prime})},
\label{appeq: dictionary diagram b}
\end{align}
\end{subequations}
where we have suppressed the coordinate $t$ as we assume all operators
to be evaluated at equal times, and where we have suppressed the wire
labels $y,y^{\prime}$ as we are working within a single chiral sector
of a single CFT.
[The exchange algebra Eq.~\eqref{appeq: dictionary diagram b} arises from
a different choice of gauge for the monodromy of a $\widehat{\psi}$ and
a $\widehat{\sigma}$ operator, in which the total phase $-1$ arising upon
a full winding of the operator coordinates by $2\pi$ arises as a product of
two factors $e^{+\mathrm i\, \pi/2}$ arising from the first and second ``halves"
of the winding process. We choose this gauge for consistency with the 
axiomatic RCFT data \eqref{appeq: Ising RCFT R-symbols},
and only use it to consider the braiding of excitations,
which is gauge-invariant.]
Comparing Eqs.\ \eqref{appeq: dictionary diagram a}
and \eqref{appeq: dictionary diagram b}, we see that the phases only
coincide if the diagram \eqref{appeq: dictionary diagram a} is
interpreted such that the coordinate $z$ attached to the $\psi$ branch
is larger than the coordinate $z^{\prime}$ attached to the $\sigma$
branch [i.e., if $\mathrm{sgn}(z-z^{\prime})=+1$].  We thus establish
\begin{align}
\parbox{0.4\textwidth}{
\textit{Rule 1:}
In the operator product corresponding to a fusion tree, the spatial
coordinates $z$, at which the operators are evaluated, are ordered
according to the positions of the corresponding branches of the fusion
tree on the axis pointing \textit{into the page}.
                      }
\end{align}
As a sanity check of this rule, we note that if the $\psi$ branch
instead passed \textit{over} the $\sigma$ branch in the diagram in
Eq.~\eqref{appeq: dictionary diagram a}, we would use
$(R^{\psi\sigma}_{\sigma})^{-1}=-\mathrm{i}$ instead, in accordance
with Eq.~\eqref{appeq: R-symbol def diagram}, but the ordering of the
legs would now dictate that $\sgn(z-z^{\prime})=-1$ in Eq.\
\eqref{appeq: dictionary diagram b}.  Thus, Rule 1 ensures a meaningful
correspondence between the $R$-symbols in the diagrammatics and the
phases acquired under exchanging two operators in the CFT.

Next, we need to establish a convention for ordering the operators in
an algebraic expression based on a fusion tree, and vice versa.  There
are various ways of doing this, but we choose to use
\begin{align}
\parbox{0.4\textwidth}{
\textit{Rule 2:} In the operator product corresponding to a fusion
tree, the operators are ordered from \textit{left to right} according
to the order from \textit{right to left} of the corresponding branches
of the fusion tree, \textit{before} any braiding is performed.
                      }
\end{align}
In Rule 2, the word ``before'' is interpreted under the assumption that
the diagram is read from bottom to top.  In this way, the ordering of
operators in Eq.~\eqref{appeq: dictionary diagram b} agrees with the
ordering of the branches of the fusion tree in Eq.\
\eqref{appeq: dictionary diagram a}.

With Rules 1 and 2 in place, we can now reliably translate fusion
diagrams into equations and vice versa.  For example, the
correspondence
\begin{align}
\includegraphics[width=.3\textwidth]{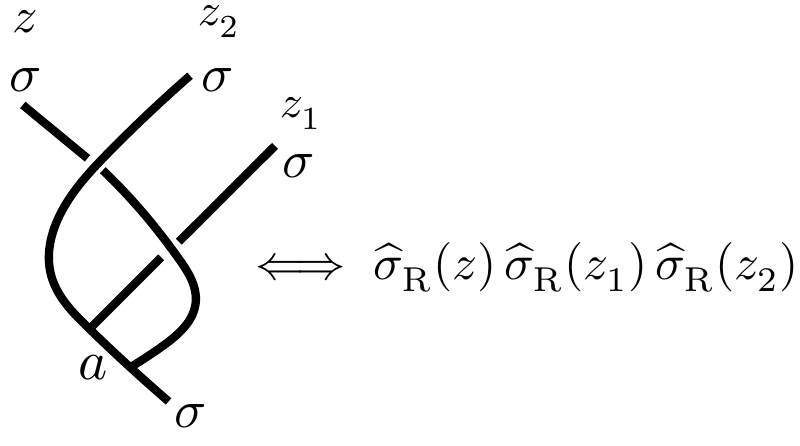}
\end{align}
is used in Eqs.\ \eqref{eq2Dcase: 3-sigma diagram untwisting} and
\eqref{eq2Dcase: 3-sigma algebraic untwisting} of the main text.

\section{Independence of string-operator algebra on arbitrary phase factors}
\label{sec: Independence of string operator algebra on arbitrary phase factors}

We have made extensive use of the
fact that the OPE of two operators in the \textit{same} wire
determines the algebra of these two operators under exchange.
However, in certain situations
[e.g.,\ Eq.\ \eqref{eq: def chiral parafermion and boson rep su(2)k}],
we found it important (on physical grounds) to
modify the exchange algebra between operators in \textit{different}
wires.  We will now show that, despite their importance in calculating
local quantities, these modifications
have no effect on topological features like the ground state
degeneracy.

We proceed with an explicit example that illustrates how this comes
about for the $su(2)^{\,}_{2}$ case in 2D
studied in Sec.~\ref{subsec: Case study: su(2)_{2}}.
We begin by rewriting the exchange algebra
\eqref{eq2Dcase: psi sigma algebra}, but this time allowing for operators in
different wires to have nontrivial commutation with one another.
Hence, we posit that
\\

\begin{widetext}
\begin{equation}
\widehat{\psi}^{\,}_{\mathrm{M},y}(t,z)\,
\widehat{\sigma}^{\,}_{\mathrm{M}^{\prime},y^{\prime}}(t,z^{\prime})=
\widehat{\sigma}^{\,}_{\mathrm{M}^{\prime},y^{\prime}}(t,z^{\prime})\,
\widehat{\psi}^{\,}_{\mathrm{M},y}(t,z)\, 
e^{
+\mathrm{i}\,\pi\,(-1)^{M}\,\delta^{\,}_{\mathrm{M},\mathrm{M}^{\prime}}\,
\delta^{\,}_{y,y^{\prime}}\,\Theta(z-z^{\prime})
  }\,
e^{
+\mathrm{i}\,\epsilon^{\,}_{\mathrm{M},\mathrm{M}^{\prime}}\,
\delta^{\,}_{y,y^{\prime}}\, \varphi
  }\,
e^{
+\mathrm{i}\,\mathrm{sgn}(y-y^{\prime})\, \theta^{\,}_{\mathrm{M},\mathrm{M}^{\prime}}
  },
\label{eq2Dcase: modified psi sigma algebra}
\end{equation}
where $(-1)^{\mathrm{R}}\equiv-(-1)^{\mathrm{L}}\equiv1$,
$\epsilon^{\,}_{\mathrm{L},\mathrm{R}}=-\epsilon^{\,}_{\mathrm{R},\mathrm{L}}=1$,
and $\epsilon^{\,}_{\mathrm{R},\mathrm{R}}=\epsilon^{\,}_{\mathrm{L},\mathrm{L}}=0$.
The reason why the choice (\ref{eq2Dcase: modified psi sigma algebra})
has no effect on the topological
features of the phase is that all of these features depend on the
algebra of string operators, which are constructed from bilinears in
the operators $\widehat{\psi}^{\,}_{\mathrm{M},y}$ and
$\widehat{\sigma}^{\,}_{\mathrm{M},y}$.  In particular, for Majorana
and twist-field operators in the same wire $y$, we have
\begin{align}
\begin{split}
\widehat{\psi}^{\,}_{\mathrm{R},y}(t,z)\,
\widehat{\psi}^{\,}_{\mathrm{L},y}(t,z)\,
\widehat{\sigma}^{\,}_{\mathrm{R},y}(t,z^{\prime})\,
\widehat{\sigma}^{\,}_{\mathrm{L},y}(t,z^{\prime})=&\,
\widehat{\sigma}^{\,}_{\mathrm{R},y}(t,z^{\prime})\,
\widehat{\sigma}^{\,}_{\mathrm{L},y}(t,z^{\prime})\,
\widehat{\psi}^{\,}_{\mathrm{R},y}(t,z)\,
\widehat{\psi}^{\,}_{\mathrm{L},y}(t,z)
\\
&\,
\times
e^{+\mathrm{i}\, \epsilon^{\,}_{\mathrm{L},\mathrm{R}}\, \varphi}\,
e^{+\mathrm{i}\, \pi\, \Theta(y-y^{\prime})}\,
e^{-\mathrm{i}\, \pi\, \Theta(y-y^{\prime})}\,
e^{+\mathrm{i}\, \epsilon^{\,}_{\mathrm{R},\mathrm{L}}\, \varphi}
\\
=&\,
\widehat{\sigma}^{\,}_{\mathrm{R},y}(t,z^{\prime})\,
\widehat{\sigma}^{\,}_{\mathrm{L},y}(t,z^{\prime})\,
\widehat{\psi}^{\,}_{\mathrm{R},y}(t,z)\,
\widehat{\psi}^{\,}_{\mathrm{L},y}(t,z).
\end{split}
\end{align}
For Majorana and twist-field operators in different wires $y\neq y^{\prime}$,
we find that
\begin{align}
\begin{split}
\widehat{\psi}^{\,}_{\mathrm{R},y}(t,z)\,
\widehat{\psi}^{\,}_{\mathrm{L},y}(t,z)\,
\widehat{\sigma}^{\,}_{\mathrm{R},y^{\prime}}(t,z^{\prime})\,
\widehat{\sigma}^{\,}_{\mathrm{L},y^{\prime}}(t,z^{\prime})=&\,
\widehat{\sigma}^{\,}_{\mathrm{R},y^{\prime}}(t,z^{\prime})\,
\widehat{\sigma}^{\,}_{\mathrm{L},y^{\prime}}(t,z^{\prime})\,
\widehat{\psi}^{\,}_{\mathrm{R},y}(t,z)\,
\widehat{\psi}^{\,}_{\mathrm{L},y}(t,z)\,
\\
&\,
\times
e^{
+\mathrm{i}\,\mathrm{sgn}(y-y^{\prime})\,
\left(
\theta^{\,}_{\mathrm{L},\mathrm{R}}
+\theta^{\,}_{\mathrm{L},\mathrm{L}}+\theta^{\,}_{\mathrm{R},\mathrm{R}}
+\theta^{\,}_{\mathrm{R},\mathrm{L}}
\right)
  }
\\
=&
\widehat{\sigma}^{\,}_{\mathrm{R},y^{\prime}}(t,z^{\prime})\,
\widehat{\sigma}^{\,}_{\mathrm{L},y^{\prime}}(t,z^{\prime})\,
\widehat{\psi}^{\,}_{\mathrm{R},y}(t,z)\,
\widehat{\psi}^{\,}_{\mathrm{L},y}(t,z)
\end{split}
\end{align}
\end{widetext}
holds so long as the angles $\theta^{\,}_{\mathrm{M},\mathrm{M}^{\prime}}$ satisfy
\begin{subequations}
\begin{align}
\label{eq2Dcase: Klein factor constraint}
\theta^{\,}_{\mathrm{L},\mathrm{R}}
+
\theta^{\,}_{\mathrm{L},\mathrm{L}}
+
\theta^{\,}_{\mathrm{R},\mathrm{R}}
+
\theta^{\,}_{\mathrm{R},\mathrm{L}}\in2\pi\,\mathbb{Z}.
\end{align}
For general choices of the angles
$\theta^{\,}_{\mathrm{M},\mathrm{M}^{\prime}}$, Eq.\
\eqref{eq2Dcase: Klein factor constraint}
is automatically satisfied if
\begin{align}
\theta^{\,}_{\mathrm{R},\mathrm{R}}=
-\theta^{\,}_{\mathrm{L},\mathrm{L}},
\qquad
\indent\theta^{\,}_{\mathrm{L},\mathrm{R}}=-\theta^{\,}_{\mathrm{R},\mathrm{L}}.
\end{align}
\end{subequations}
Thus, when string operators are built from bilinears like
$
\widehat{\psi}^{\,}_{\mathrm{R},y}\,
\widehat{\psi}^{\,}_{\mathrm{L},y}
$
and
$
\widehat{\sigma}^{\,}_{\mathrm{R},y},
\widehat{\sigma}^{\,}_{\mathrm{L},y}
$,
the additional phases in the
exchange algebra \eqref{eq2Dcase: modified psi sigma algebra} drop out
of all calculations.

The calculations of the previous paragraph generalize readily to other
combinations of primary operators.
The key observation in all cases is that string 
operators are built either from nonchiral bilinears of primary
operators, like the ones studied in the previous paragraph, or from
operators like $\widehat{\mathcal U}^{\,}_{\alpha}(t)$ [defined in
Eq.~\eqref{eq2Dcase: U_alpha definition}]
that act only within one channel of one wire.

\bibliographystyle{apsrev}

\bibliography{bib_non-Abelian_3d_wire}

\end{document}